\documentclass[twocolumn,tighten]{aastex62}
\usepackage{bm}
\usepackage{xspace}
\usepackage{enumitem}
\usepackage{graphicx}
\usepackage{color}
\usepackage{comment}
\usepackage{amsmath}
\usepackage[T1]{fontenc}

\def\red#1{\textcolor{red}{\textbf{#1}}}
\def\blue#1{\textcolor{blue}{\textbf{#1}}}

\def\flx{erg cm$^{-2}$ s$^{-1}$}
\graphicspath{{./}{figures/}}
\shorttitle{Modeling TDE Accretion}
\shortauthors{Wen et al.}

\begin{document}

\title{Continuum-Fitting the X-ray Spectra of Tidal Disruption Events}
\correspondingauthor{Sixiang Wen}
\email{wensx@email.arizona.edu}
\author{Sixiang Wen}
\affiliation{University of Arizona, 933 N. Cherry Ave., Tucson, AZ  85721}
\affiliation{School of Physics and Astronomy, Sun Yat-Sen University, 2 Daxue Road, Tangjia, Zhuhai, China}

\author{Peter G.~Jonker}
\affiliation{Department of Astrophysics/IMAPP, Radboud University, P.O.~Box 9010, 6500 GL, Nijmegen, The Netherlands}
\affiliation{SRON, Netherlands Institute for Space Research, Sorbonnelaan 2, 3584~CA, Utrecht, The Netherlands}
\author{Nicholas C. Stone}
\affiliation{Racah Institute of Physics, The Hebrew University, Jerusalem, 91904, Israel}
\affiliation{Department of Astronomy, University of Maryland, Stadium Drive, College Park, MD, 20742, USA}
\affiliation{Department of Physics, Department of Astronomy, and Columbia Astrophysics Laboratory, Columbia University, New York, NY, 10027, USA}

\author{Ann I. Zabludoff}
\affiliation{University of Arizona, 933 N. Cherry Ave., Tucson, AZ  85721}

\author{Dimitrios Psaltis}
\affiliation{University of Arizona, 933 N. Cherry Ave., Tucson, AZ  85721}

\begin{abstract}
We develop a new model for X-ray emission from tidal disruption events (TDEs), applying stationary general relativistic ``slim disk'' accretion solutions to supermassive black holes (SMBHs) and then ray-tracing the photon trajectories from the image plane to the disk surface, including gravitational redshift, Doppler, and lensing effects self-consistently. We simultaneously and successfully fit the multi-epoch XMM-{\it Newton} X-ray spectra for two TDEs: ASASSN-14li and ASASSN-15oi. We test explanations for the observed, unexpectedly slow X-ray brightening of ASASSN-15oi, including delayed disk formation and variable obscuration by a reprocessing layer. We propose a new mechanism that better fits the data: a ``Slimming Disk'' scenario in which accretion onto an edge-on disk slows, reducing the disk height and exposing more X-rays from the inner disk to the sightline over time. For ASASSN-15oi, we constrain the SMBH mass to $4.0^{+2.5}_{-3.1} \times 10^6M_\odot$. For ASASSN-14li, the SMBH mass is $10^{+1}_{-7}\times 10^6M_\odot$ and the spin is $>0.3$. For both TDEs, our fitted masses are consistent with independent estimates; for ASASSN-14li, application of the external mass constraint narrows our spin constraint to $>0.85$. The mass accretion rate of ASASSN-14li decays slowly, as $\propto t^{-1.1}$, perhaps due to inefficient debris circularization. Over $\approx$1100 days, its SMBH has accreted $\Delta M \approx 0.17 M_\odot$, implying a progenitor star mass of $> 0.34 M_\odot$, i.e., no ``missing energy problem.'' For both TDEs, the hydrogen column density declines to the host galaxy plus Milky Way value after a few hundred days, suggesting a characteristic timescale for the depletion or removal of obscuring gas.
\end{abstract}


\section{Introduction}
\label{int}

Tidal disruption events (TDEs) are one of the most direct and promising routes to studying supermassive black holes (SMBHs). When a star comes close to a SMBH, it will be broken apart by the black hole's tidal force \citep{Hills75, Rees1988}. Part of the debris will then be accreted by the black hole. During the accretion process, the gas will produce a flare of strong electromagnetic radiation. The luminosity and the decay time of the radiation are sensitive to the properties of both the SMBH and the victim star.

The flare's luminosity can decline by orders of magnitude, from highly super-Eddington to significantly sub-Eddington, in less than a year \citep{Evans1989}. Tidal disruption flares have been discovered via quasi-thermal emission in optical \citep{vanVelzen+11, Gezari+12, Arcavi+14, Holoien2016a, vanVelzen+20}, near-ultraviolet \citep[NUV;][]{Gezari+06, Gezari+08}, and soft X-ray \citep{Bade+96, Greiner+00, Komossa+04, Saxton+17} wavelengths. If the debris circularization process is rapid, and the disk accretion rate follows the debris fallback rate, one can use the optical and UV light curves to directly constrain properties of the SMBH and the victim star \citep{Mockler2018}. 

Yet both analytical arguments \citep{Piran+15, Dai+15} and hydrodynamical simulations \citep{Shiokawa+15} suggest that circularization and disk assembly may be inefficient processes, complicating the translation from light curves into parameter constraints.  Because self-intersection shocks forced by general relativistic precession are the most promising dissipation sites for TDE debris streams, the efficiency of circularization and disk assembly should vary strongly depending on the degree to which the debris pericenter is relativistic \citep{Hayasaki+13, Dai+15, Bonnerot+17}. Nodal precession caused by SMBH spin may further complicate matters \citep{GuillochonRamirezRuiz15, Hayasaki+16}.  Parameter estimation becomes even more difficult when looking at relatively long wavelengths, as there are multiple plausible power sources and photospheres for the observed optical/NUV emission in TDEs \citep{Loeb+97, Guillochon+14, Piran+15, Shiokawa+15, Metzger2016, LuBonnerot19, BonnerotLu19}.  

While our theoretical picture of TDE accretion flows remains incomplete, the underlying physics is more straightforward for those bright in quasi-thermal soft X-rays, emission likely to originate in
a standard accretion disk on scales near the innermost stable circular orbit (ISCO). Observations \citep{Miller2015, Gezari2017} and theory \citep{Ulmer99, Lodato2011} agree that, for such TDEs, thermal X-rays are emitted on the Wien tail of a multi-color blackbody accretion disk, whose spectrum peaks in the extreme ultraviolet (EUV). The size of the innermost accretion disk annuli, which emit most of the X-rays, is determined primarily by the SMBH mass and spin. Generally, the more massive the SMBH, the larger the size and the lower the effective temperature of the disk; the faster the (prograde) SMBH spin, the greater the inward extent and the higher the effective temperature of the disk. Studying the X-ray emission is thus a promising way to recover SMBH properties.

There is a long history
of inferring black hole properties by fitting X-ray spectra of stellar-mass (X-ray binary) and supermassive (AGN) systems \citep{Titarchuk1994,Hua1995,McClintock2006,Done2012}. While X-ray binaries exhibit dramatic fluctuations in their accretion rate and corresponding disk ``state changes'' \citep{HasingervanDerKlis89, Fender+04}, 
SMBH disks tend to have a stable accretion rate over long periods of time\footnote{``Changing-look'' AGN \citep{Shappee+14}
are an exception and may represent scaled-up versions of X-ray binary state transitions \citep{NodaDone18, Ruan+19}.}. As a result, different X-ray spectral observing epochs do not generally sample wildly different disk conditions in AGN. This stability
is a challenge for constraining SMBH properties, requiring that priors be imposed on other factors affecting the X-ray spectrum, e.g., disk inclination, HI absorption, spectral hardening factor \citep{ST93,ST95}, and background counts.

Modeling the X-rays from TDEs 
can teach us more about the SMBH,
because the accretion rate 
varies dramatically over even the months spent near peak emission. Later, the mass fallback rate can decline by orders of magnitude, from highly super-Eddington to significantly sub-Eddington, in less than a year \citep{Evans1989}. Multi-epoch fitting can thus break degeneracies in parameter estimation, leading to constraints on the SMBH mass and spin, as well as on the time evolution of the mass accretion rate.




The
X-ray emission is also cleaner to model than the complex optical/NUV photosphere. A
well-known issue in the study of optical/NUV emission from TDEs is the ``missing energy problem'' \citep{Stone&Metzger16, Piran+15}: the total energy emitted by TDEs in their optical/NUV light curves is $\sim 10^{51}~{\rm erg}$, much less than the $10^{52-53}~{\rm erg}$ expected from radiatively efficient accretion of a lower main sequence star.  Simple disk models suggest that the bulk of a TDE's luminosity is emitted in EUV bands \citep{Lodato2011, Lu&Kumar18}, which are challenging to observe directly and may have little correlation with the optical/NUV photosphere. X-ray continuum fitting has the potential to measure mass accretion rates directly, place a limit on the total accreted mass, and thereby test for any missing energy.


While current TDE samples are modest, with only about two dozen strong candidate flares detected \citep{Komossa15,Hung+17}, the recently launched eROSITA satellite will detect up to $\approx 1000$ X-ray TDEs every year \citep{Khabibullin+14,Jonker+19}. Future ``lobster-eye'' X-ray instruments, such as the proposed ISS-TAO and the planned Einstein Probe, are each expected to find $\sim$100 TDEs per year \citep{Yuan+15}. Optically-selected TDEs will be discovered at rates of $\sim$10/yr by the ongoing ZTF survey \citep{vanVelzen+20} and $\sim$1000/yr by LSST \citep{Bricman&Gomboc20}; many of these events may also be observed in the X-rays, either serendipitously by the aforementioned wide-field surveys or in targeted follow up.  Producing science from this impending explosion of data, especially when sensitive soft X-ray follow-ups from {\it Chandra}, {\it XMM-Newton}, and {\it Swift} are still possible, requires accurate models for TDE disks to be developed now. 

In this paper, we apply a general relativistic model of stationary slim disk accretion \citep{Sadowski2009,Sadowski2011} to observations of ASASSN-14li and ASASSN-15oi, two of the TDEs best characterized at soft X-ray wavelengths. 
We first estimate the X-ray spectra emitted locally from each annulus as a multicolor blackbody modified by a spectral hardening factor \citep{DE2018}. We then use a general relativistic ray-tracing code \citep{JP11} to calculate the X-ray flux observed by a distant observer after considering beaming, lensing, and redshift effects. 
We introduce the model in Section \ref{Methodology}. We describe and reduce the archival X-ray spectral data in Section \ref{data}. In Section \ref{RandD}, we present the X-ray spectra and light curves predicted by the model for a wide range of parameter space, the best fits to the multi-epoch X-ray observations of ASASSN-14li and ASASSN-15oi, and the resulting constraints on key TDE parameters, including SMBH mass and spin.
We close with our conclusions in Section \ref{Conclusions}.   Additional details about the effects of the spectral hardening factor assumptions, the stationary slim disk model, and the ray tracing algorithm are in the Appendices.

\section{Methodology}
\label{Methodology}

In this section, we present a step-by-step procedure for continuum fitting the X-ray emission in TDEs.  This includes the details of the stationary slim disks used to model inner regions of a TDE accretion flow, the assumptions concerning the emitted spectrum, our procedure for general relativistic ray tracing, a theoretical prescription for the evolution of mass fallback in TDEs, and the details of building a model library and applying it to X-ray datasets.

\subsection{Relativistic stationary slim disk model}

The accretion rates in disks formed after tidal disruption can vary from highly super-Eddington to quite sub-Eddington. When the accretion rate of the disk is significantly higher than $\sim 0.1 \dot M_{\rm Edd}$, the gas orbits become significantly non-Keplerian, the aspect ratio of the disk approaches unity, and advective heat losses due to radial gas inflow can no longer be neglected \citep{Abramowicz1988}. As a result, the standard thin disk model \citep{Shakura1973,Novikov1973} is not adequate to generally model TDE disks, and a slim disk model is more accurate. Slim disks are optically thick accretion flows, first developed by \cite{Abramowicz1988}, which incorporate advective cooling and do not assume Keplerian angular momentum, allowing them to work in both super- and sub-Eddington regimes.

Since the development of the first slim disk models,
many analytic and numerical studies have further developed the structure and dynamics of slim disks \citep{Lasota1994,Abramowicz1996,Abramowicz1997, Gammie1998, Beloborodov1998}. 
Based on the works of \cite{Abramowicz1996,Abramowicz1997}, \cite{Sadowski2009,Sadowski2011} developed a GR slim disk code to solve the disk equations numerically. However, this model was only used to study accretion disks around small back holes \citep{Straub2011}. Following the work of \cite{Sadowski2009}, we estimate the viscous heating term analytically instead of numerically and rewrite the code by using the shooting method \citep{Press2002} to determine the sonic point. For convenience, we write the underlying stationary disk equations in the appendix. Solving these equations, we find the local flux $F(r)$, the surface density $\Sigma(r)$, the ratio of disk height to radius $H/r$ and the four-velocity $(u_t,u_r,u_\theta,u_\phi)$ of the gas at radius r. There are several important assumptions made in this model, such as the imposition of a zero-torque inner boundary condition, the large assumed optical depth, neglect of self-irradiation due to light bending, an absence of magnetic pressure support, and no angular momentum loss due to radiation or outflows.

\subsection{X-ray emissivity of accretion disk}
Early studies \citep{ST93,ST95} demonstrated that the local X-ray spectrum of the disk does not behave as a perfect multi-color black body. Due to electron scattering and the temperature gradient in the atmosphere, the real X-ray flux at each annulus will be higher than the corresponding black body flux. In principle, one has to solve the full vertical radiation transfer equation to determine the local X-ray flux. The solution is determined by the vertical structure of the disk, which is determined by the effective temperature $T_e=(F(r)/2\sigma)^{1/4}$ ($\sigma$ is Stefan-Boltzmann constant), surface density $\Sigma(r)$, and the strength of vertical gravity $Q(r)$ at each annulus of the disk. Working in the context of thin disks, \citet{ST95} found that the X-ray flux can be well approximated by introducing a color correction factor $f_{\rm c}$,
\begin{equation}
\label{Iv}
 I_{\rm d}(\nu)=\frac{2h\nu^3c^{-2}f_{\rm c}^{-4}}{\exp(h\nu/k_{\rm B} f_{\rm c}T_{\rm e})-1}.
\end{equation}
Here, $h$ is the Plank constant and $k_{\rm B}$ is the Boltzmann constant; $f_{\rm c}$ is also commonly called the spectral hardening factor. Since both slim disk and thin disk sharing the same radiation transfer physics, we assume Eq. \eqref{Iv} is also true for slim disk, but slim disk will adopt a different $f_{\rm c}$ value due to different disk parameters when compare with thin disk. For a thin disk, $f_{\rm c}$ is about 1.7, and is insensitive to disk parameters \citep{ST95}. Later studies \citep{MFR2000,GD04,DBHT05} showed that $f_{\rm c}$ may increase as the accretion rate increases. 
Recently, \citet{DE2018} (hereafter DE19) gave an approximate formula to estimate $f_{\rm c}$ for a non-spinning SMBH,
\begin{eqnarray}
\nonumber 
f_{\rm c}=1.74+&&1.06(\log_{10} T_{\rm e}-7)-0.14[\log_{10} Q(r)-7] \\ 
&&-0.07\{ log_{10}[\Sigma(r)/2]-5 \}, 
\label{fc2}
\end{eqnarray}
This equation holds for case of accretion rate between 0.01 to 1 Eddington accretion. In the super-Eddington regime, $f_{\rm c}$ would not increase to infinity as $T_e$ increases and would instead saturate \citep{Davis2006} at
\begin{equation}
 f_{\rm c}^{\star}\approx\left( \frac{72~ \rm keV}{T_{\rm e}k_{\rm B}}\right)^{1/9}.
\label{fc3}
\end{equation}
This implicates that $f_{\rm c}^\star\approx 2.4$ for an
SMBH accreting near Eddington. Above this value, $f_{\rm c}$ could increase only weakly.  

The spectral hardening factor $f_{\rm c}$ can significantly change the local X-ray flux, and is thus a key input to constrain parameters in X-ray spectral fitting. One may either treat it as a nuisance parameter to be fitted independently or use the prescriptions above to estimate {\it a priori} values for $f_{\rm c}$. In this work, we take the latter approach as our fiducial model, but explore the impact of letting $f_{\rm c}$ float as a free parameter in Appendix \ref{app:14li}.  Given the range of $T_{\rm e}$, $Q(r)$ and $\Sigma(r)$ predicted from the slim disk solution, Eq. \eqref{fc2} is unlikely to exceed the saturation value $f_{\rm c}^\star$ but we also explore this issue more carefully in Appendix \ref{app:14li}.

\subsection{Disk spectrum seen by distant observer}

 We use a general relativistic ray-tracing code \citep{JP10,JP11} to calculate the trajectories of photons reaching the observer from the surface of the disk. With this code, we trace the null geodesics of photons backwards in time to the disk, and calculate their energy $(E_{\rm d})$ and position $(r,\phi)$, given initial (i.e. observed) energies ($E_{\rm obs}$) and positions $(x_{\rm o},y_{\rm o})$ in the image plane of the observer. This ray-tracing calculation self-consistently accounts for gravitational redshift, the Doppler effect, and strong lensing in Schwarzschild or Kerr spacetimes. The observed flux can then be calculated by integrating over the flux on the image plane. For the reader's convenience, we write the underlying ray-tracing equations in Appendix \ref{app:rays}.

A monochromatic flux of disk radiation, as seen by a distant observer, can be calculated as \citep{Baubock2015}
\begin{equation}
F_{\rm obs}(E_{\rm obs})=\frac{1}{d^2}\int I_{\rm obs}(E_{\rm obs},x_{\rm o},y_{\rm o})\,{\rm d}x_{\rm o} {\rm d}y_{\rm o} .
\label{flux_o}
\end{equation}
Here, $d$ is the distance from the observer to the disk and $I_{\rm obs}(E_{\rm obs},x_{\rm o},y_{\rm o})$ is the observed intensity of photons with energy $E_{\rm obs}$ in the image plane.
Since $I/E^3$ is Lorentz invariant, we have
\begin{equation}
\frac{I_{\rm d}}{E_{\rm d}^3}=\frac{I_{\rm obs}}{E_{\rm obs}^3},
\end{equation}
where $I_{\rm d}$ and $E_{\rm d}$ are the specific intensity and photon energy, respectively,
as measured on the surface of the disk. By defining the redshift factor
\begin{equation}
\Upsilon=\frac{E_{\rm obs}}{E_{\rm d}},
\end{equation}
Eq. \eqref{flux_o} can be rewritten as:
\begin{equation}
F_{\rm obs}(E_{\rm obs})=\frac{1}{d^2}\int \Upsilon^3I_{\rm d}(E_{\rm obs}/\Upsilon,r,\phi) {\rm d}x_{\rm o} {\rm d}y_{\rm o},
\label{F_o}
\end{equation}
with an emitted intensity $I_{d}(E_{d}/\Upsilon,r,\phi)$ that is calculated by Eq. \eqref{Iv}. Therefore, we can calculate the observed flux at a single frequency by integrating Eq. \eqref{F_o} numerically.

Since the height of the slim disk cannot be neglected, we trace the ray back to the surface of the disk, a finite height above the midplane. 
We note that when the observer is at a high inclination (observing nearly edge-on), the height of the edge can shield the X-rays that are preferentially emitted from the inner, hottest annuli of the disk\footnote{We assume the outer edge of the disk to emit radiation at the local annular effective temperature.}.  Even a perfectly edge-on disk will not be completely dim in the X-rays, however, due to strong lensing effects.

\subsection{Evolution of TDE accretion disk}

The evolution of the mass fallback rate in TDEs at late times can be expressed as
\begin{equation}
\dot M_{\rm t}=\dot M_{\rm peak} \left( \frac{t}{t_{\rm fall}}\right)^{n},
\label{Mf}
\end{equation}
where $\dot M_{\rm peak}$ is the peak fallback rate, $t$ is the time since the most tightly bound debris has returned to pericenter, $t_{\rm fall}$ is the fallback time of the most tightly bound debris, and $n$ is the power-law decay index, which is equal to $-5/3$ in simple analytic models of the disruption process \citep{Rees1988, Phinney89}. 
While analytic estimates for these variables exist \citep{Rees1988, Stone+13}, they can be more precisely calibrated via hydrodynamic simulations of the disruption process \citep{James13, Mainetti+17, Ryu+20b}. In Newtonian gravity\footnote{The late-time mass fallback rate may become more complicated for highly relativistic TDEs, where the gravitational radius $r_{\rm g}=GM_\bullet / c^2 \sim r_{\rm p}$; in this case, the fallback rate acquires some dependence on $r_{\rm p}/r_{\rm g}$ \citep{Kesden12, ChengBogdanovic14, Ryu+20d} and potentially even on the SMBH spin $a_\bullet$ \citep{Kesden12, Stone+13, Tejeda+17}.  In this paper, we neglect these general relativistic corrections to the fallback rate.}, $\dot M_{\rm peak}$, $t_{\rm fall}$, and $n$ are determined primarily by the SMBH mass $M_\bullet$, the victim star's mass $m_\star$ and radius $r_\star$, and the star's pericenter $r_{\rm p}$. 
\citet{James13} provide fitting formulas for the quantities $\dot M_{\rm peak}$, $t_{\rm fall}$ and $n$ in the aftermath of disruption of a polytropic star:
\begin{eqnarray}
\dot M_{\rm peak}&=&46A_{\gamma}\eta_{-1}M_6^{-3/2} \left(\frac{m_{\star}}{M_\odot}\right)^2 \left(\frac{r_{\star}}{R_\odot}\right)^{-3/2}{\dot M_{\rm Edd}},\\
\label{Mpeak}
t_{\rm fall}&=&B_{\gamma}M_6^{1/2} \left(\frac{m_{\star}}{M_\odot}\right)^{-1} \left(\frac{r_{\star}}{R_\odot}\right)^{3/2} ~{\rm yr},\\
\label{tpeak}
n_\infty&=&D_{\gamma}.
\end{eqnarray}
Here, $\gamma$ is the polytropic index, $\eta$ is the radiative efficiency of accretion, $\eta_{-1}=\eta/0.1$, $M_6=M_\bullet/(10^6M_\odot)$, and $M_\odot$ is the solar mass. 
We have defined the peak fallback rate in terms of the Eddington-limited accretion rate,
\begin{equation}
    \dot M_{\rm Edd}=1.37 \times 10^{21} ~{\rm kg~s}^{-1} \eta^{-1}_{-1} M_6.
\end{equation}
Hereafter, we refer to accretion and fallback rates in dimensionless Eddington units,
i.e., $\dot m= \dot M/ \dot M_{\rm Edd}$. $A_{\gamma}$, $B_{\gamma}$ and $D_{\gamma}$ are fitting functions that depend only on $\gamma$ and the penetration parameter $\beta\equiv r_t/r_p$ (note that tidal disruption radius $r_{\rm t} = r_\star(M_\bullet / m_\star)^{1/3}$)
and the precise form of the $A_\gamma$, $B_\gamma$, and $D_\gamma$ functions can be found in the erratum of \citet{James13}. 

We estimate $m_{\star}$ from $r_{\star}$ using the mass-radius relation of \citet{Tout96}, for $m_{\star}>0.1M_\odot$. If $m_{\star}<0.1M_\odot$, we set $r_{\star}=0.1R_\odot$. The numerical fitting formulae above are valid for $0.6 \le \beta \le 2.5$, and are calibrated for two specific values of $\gamma$: $4/3$ and $5/3$. These adiabatic indices are appropriate for high-mass ($m_\star \gtrsim M_\odot$) and low-mass ($m_\star \lesssim 0.5 M_\odot$) stars, respectively \citep{Lodato2009}.  
For stars in the transition region $(0.5M_{\odot}< m_\star < M_{\odot})$, we use a linear interpolation, 
\begin{equation}
\dot M_{\rm t}=2\dot M_{{\rm t}}^{\gamma={5/3}}(M_\odot-m_{\star})+\dot M_{{\rm t}}^{\gamma={4/3}}(2m_{\star}-M_\odot), 
\end{equation}
to estimate the fallback rate. Thus, in order to calculate the fallback rate at $t$, one need to employ four free parameters, which are $M_\bullet$, $t_{\rm peak}$, $m_{\star}$ and $\beta$.
Since it is difficult to detect the exact date of peak fallback, $t$ is determined by:
\begin{equation}
t=t_{\rm obs}-t_{\rm returned}=t_{\rm obs}-t_{\rm peak}+t_{\rm fall}.
\label{t}
\end{equation}
Here, $t_{\rm obs}$, $t_{\rm returned}$, and $t_{\rm peak}$ are the date of observations (which varies from epoch to epoch), the first material returned date, and the peak luminosity date. In order to calculate the accretion rate from Eq. \ref{Mf}, one must know the 5 parameters, $M_\bullet$, $\beta$, $m_\star$, $t_{\rm obs}$ and $t_{\rm peak}$. 

Eq. \ref{Mf} is approximate in two significant ways.  The first is that the fallback rate is only a power law at late times, and this simple functional form neglects the early rise to peak fallback that characterizes the first days to weeks of a TDE \citep{Lodato2009, James13}.  More important, however, is the nontrivial translation between mass fallback rates $\dot{M}_{\rm t}$ and inner disk accretion rates $\dot{M}$. As the most tightly bound debris returns to pericenter, it can begin to form an accretion flow through dissipational processes. While the circularization and disk formation process in TDEs remains an unsolved problem, there is evidence that circularization rates vary enormously across the parameter space of TDEs.  Gas circularization may be rapid if $r_{\rm p} \sim r_{\rm g}$, in which case relativistic apsidal precession will cause debris streams to self-intersect and thermalize their orbital energy in shocks \citep{Rees1988, Hayasaki+13, Dai+15}. In this regime, one can expect that the accretion rate of the inner disk, $\dot{M}$, follows the fallback rate of the gas, $\dot{M}_{\rm t}$. 

On the other hand, if $r_{\rm p} \gg r_{\rm g}$, apsidal precession will be very weak, and self-intersection shocks will occur far from the SMBH, leading to inefficient circularization \citep{Piran+15, Shiokawa+15, Dai+15, Bonnerot+17}.  In this regime, $\dot{M}$ will probably not track $\dot{M}_{\rm t}$.  Even in the $r_{\rm p} \sim r_{\rm g}$ regime, circularization may be delayed if the SMBH is spinning rapidly at an angle misaligned from the debris orbital angular momentum \citep{GuillochonRamirezRuiz15, Hayasaki+16}.  In this work, we take a primarily agnostic view on these open theoretical questions.  In our fiducial multi-epoch X-ray spectral fitting, we allow $\dot{M}$ to float between epochs as free parameters, but in non-fiducial modeling, we consider how our parameter estimation change if circularization is rapid, and we can make the strong assumption that $\dot{M}$ has the functional form of Eq. \eqref{Mf}.

\subsection{Assumptions and fitting procedure}

Here we summarize the assumptions we have made in the previous subsections, and give a brief description of our actual fitting procedure.

Throughout this paper, we assume that:
\begin{enumerate}
\item The dynamic inner accretion flow can be approximated by a series of stationary, circular and equatorial general relativistic slim disk models. In all cases we assume that the disk, if initially misaligned from the SMBH equatorial plane, has become aligned by the time of observations.
  
\item The disk structure is determined assuming a zero-torque inner boundary condition, large optical depth, no self-irradiation\footnote{For simplicity, as can be seen below, we treat photons intersecting the finite height of the disk as completely lost, and neglect the effect of their heating on disk structure.}, no magnetic pressure support, and no angular momentum losses through winds or radiation.
  
\item The local X-ray emissivity of the disk can be described by Eq.~\eqref{Iv}: a multicolor blackbody modified by a (radius-dependent) spectral hardening factor given in Eq.~\eqref{fc2}.
  
\item The X-rays are generated at the finite-height photospheric surface of the disk, which we assume to be equal to its scale height $H$, and if a photon's trajectory intersects the disk after emission, it will be absorbed completely.
  
\item The inner edge of the disk is set to be the innermost stable circular orbit (ISCO), and the outer edge of the disk is twice the tidal disruption radius ($2r_t$).

\item While we do not usually assume a prior for the disk accretion rate $\dot{M}(t)$, in the non-fiducial cases when we do, we assume that the circularization process is rapid, so that $\dot{M}(t) = \dot{M}_{\rm t}(t)$, as in Eq. \eqref{Mf}.
\end{enumerate}

These assumptions, which allow us to
compare multi-epoch observations with theoretical predictions for the soft X-ray continuum,
represent idealizations to some degree.
Assumption (1) above is probably the most questionable, as TDE disks may form with generic misalignments from the SMBH \citep{Stone2012} and have 
non-axisymmetric initial accretion flows \citep{Shiokawa+15}.
Our approach is likely more accurate at later epochs, as initially tilted TDE disks align over time \citep{Franchini2016, XiangGruess+16, Ivanov+18, ZanazziLai19}, and initially eccentric motions within the accretion flow circularize \citep{Sadowski+16}. The circular disk assumption may also be more valid for the innermost, X-ray producing annuli than for the disk as a whole, as gas must undergo significant dissipation in order to inspiral from scales $\approx r_{\rm t}$ to the ISCO.  
A closely related issue is global disk precession, which may modulate TDE X-ray light curves in a quasi-periodic way at early times \citep{Stone2012, Franchini2016}.  These modulations could become severe if the disk precesses into an edge-on configuration.  Our simplifying assumption that the disk quickly aligns into the equatorial plane excludes the possibility of precession, an issue to address in future work.

The other assumptions listed above
may cause issues in specific cases.  For example, if circularization 
is not radiatively efficient,
the disk outer edge can exceed $2r_{\rm t}$ \citep{Hayasaki+16, BonnerotLu19}.  However, because most X-rays are generated at small radii, our predictions are generally insensitive to the choice of outer edge, unless both (a) the disk is nearly edge-on (so that the disk edge strongly shields the X-rays from the inner disk) and (b) the largest scale height, $H/r$, is achieved at the outer edge. We find that condition (b) is rarely satisfied, and only for $\dot{m} \gtrsim 100$ does $H/r$ rise monotonically with increasing $r$.  Since such severely super-Eddington accretion rates are not expected for main sequence TDEs, our model's X-ray predictions should not be very sensitive to the choice of outer boundary. 
With these caveats in mind, we now describe our specific fitting procedure.

First, we build a library of slim disk structural solutions. The disk structure is completely determined by $M_\bullet$, $a_\bullet$, $\dot m$ and $\alpha$. Throughout our grid, we set $\alpha=0.1$; as we will see later, $\alpha$ has no important effect on the fitting. Given a $\{M_\bullet,a_\bullet\}$ pair, we solve for 70 disk structures, with $\dot m$ varying from 600 to 0.1 (for the case of $a_\bullet=0.998$, we extend the solution of $\dot m$ to 0.06).

Second, we estimate the theoretical spectrum for each disk model in our library. Given $\dot m$, $M_\bullet$, $a_\bullet$, and the observer's inclination angle\footnote{Note that $\theta$ is a colatitude, so that $\theta=0^\circ$ is a face-on observer, and $\theta=90^\circ$ is an edge-on observer.} $\theta$, we use the ray-tracing code to calculate the flux in each bin of photon energy. We use linear interpolation between energy bins with different accretion rate to estimate the model flux. We then fit to observations by using the XSPEC package (version 12.10.1; \citealt{Arnaud1996}).

Finally, for any given epoch of X-ray observations, we search across our model grid and optimize the model parameters using the $\chi^2$ statistic (or Cash statistic) for a given $\{M_\bullet,a_\bullet\}$ pair. Both SMBH mass and spin are assumed constant between epochs.
We then calculate the $\chi^2$ (or Cstat) across the $\{M_\bullet,a_\bullet\}$ plane, determining the significance of each $\{M_\bullet,a_\bullet\}$ pair.

\begin{figure}[ht!]
\includegraphics[height=0.45\textheight]{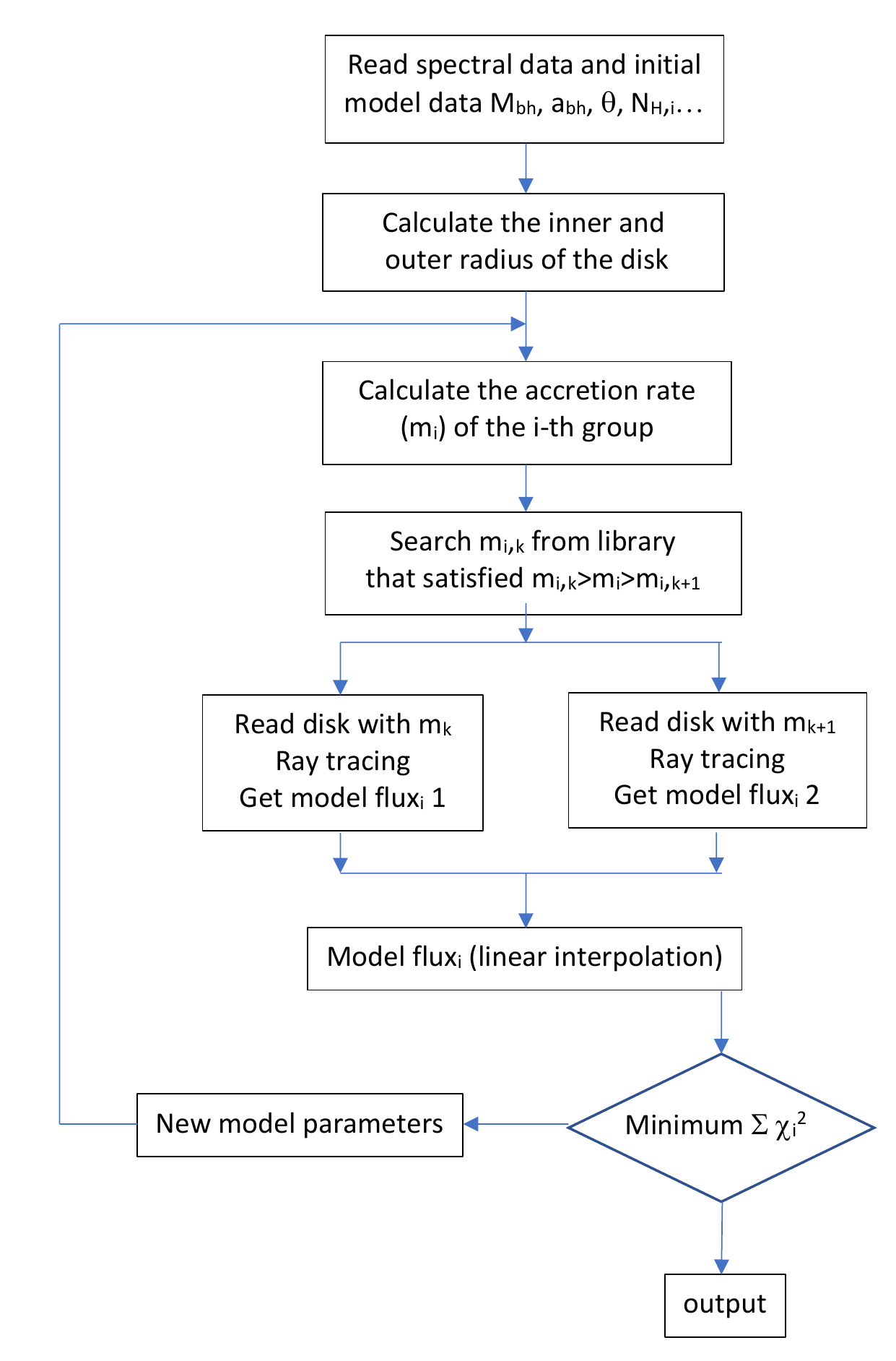}
\caption{Schematic illustration of our continuum fitting procedure for X-ray observations of TDEs.}
\label{procedure1}
\end{figure}

The free parameters for the slim disk are $M_\bullet$, $a_\bullet$, $\dot m_i$ and $\theta$. Here, the subscript index $i$ denotes the $i$-th observational epoch. If we estimate $\dot m_i$ with Eq. \eqref{Mf} (i.e., assume efficient circularization of returning tidal debris), $\dot m$ is replaced by $\beta$, $m_{\star}$, and $t_{\rm peak}$. Given these parameters, as well as $f_{\rm c}$ (determined by Eq.~\eqref{fc2}), we can calculate synthetic X-ray spectra for all observational epochs. The real X-ray spectrum is, however, 
subject to 
interstellar extinction, so we introduce the extinction $N_{\rm H}$. The fitting procedure is shown schematically in Fig. 1. The priors on the free parameters are listed in Table \ref{freep}.

\begin{deluxetable*}{ccccccccccc}
\label{freep}
\tablecaption{Allowed ranges for TDE parameters.}
\tablewidth{0pt}
\tablehead{
\colhead{Name} & \colhead{$M_\bullet^a$} & \colhead{$a_\bullet^a$} & \colhead{$\beta^d$} & \colhead{$m_{\star}^{d}$} & \colhead{$\dot m_i^c$} & \colhead{$\theta$} & \colhead{$N_{\rm H, i}$} &  \\
\colhead{} & \colhead{$[10^6 M_\odot]$} & \colhead{} & \colhead{} & \colhead{$[M_{\odot}]$} & \colhead{$[\rm Edd]$} & \colhead{$[^{\circ}]$} & \colhead{$[10^{22}{\rm cm}^{-2}]$} & 
}
\startdata
ASASSN-14li & (2, 20) & (-0.9, 0.998) & (0.6, 2.5) & (0.1, 100) & (0.1, 600) & (5, 90) & (0.02, 1) &  \\
ASASSN-15oi & (0.8, 10) & (-0.9, 0.9) & (0.6, 2.5) & (0.1, 100)& (0.1, 600) & $(5, 90)^b$ & (0.04, 1) &  \\
\enddata
\tablecomments{$^a$In individual-epoch fits, $M_\bullet$ and $a_\bullet$ are discretely sampled across the given ranges. $^b$For ASASSN-15oi, the inclination prior will be handled more carefully to test a range of hypotheses. $^c$ For the largest spin we consider, $a_\bullet=0.998$, the lower limit of $\dot m$ extends to 0.06. The accretion rate $\dot{m}$ here is calculated by assuming $\eta=0.1$ and listed in dimensionless Eddington units. 
$^d$In our most general fits, we allow $\dot{m}$ to float over time. Else, we assume that
$\dot{m}$ follows the gas fallback rate, which requires us to fit two additional variables: stellar mass $m_\star$ and penetration parameter $\beta$.}
\end{deluxetable*}

\section{X-RAY DATA ANALYSIS}
\label{data}
We will use our slim disk models to fit the soft X-ray data for two TDEs, ASASSN-14li and ASASSN-15oi.  For both events, we will focus on observations obtained with the XMM-{\it Newton} \citep{Jansen+01} satellite.

\subsection{ASASSN-14li data}
\label{sec:AS14li}

ASASSN-14li was first detected by the All Sky Automated Search for Supernova (ASASSN) on MJD $56983.6$ in a post-starburst galaxy with $z=0.0206$ \citep{jose2014}. ASASSN-14li has been observed extensively at different frequencies \citep{Miller2015,van2016,Pasham2017,Bright2018}. We extracted ten XMM--{\it Newton} spectra from archival observations. Table~\ref{tab:xmm14li} lists some details of these observations.

We run the default {\sc SAS} v18 (20190531) tools under the HEASOFT {\sl ftools} software version 6.26.1 to extract source spectra and filter the data. We initially focused on data from the EPIC pn detector \citep{Strüder+01,Turner+01} as that provides the highest count rate. We filtered the pn data for periods of enhanced background radiation, where we require that the 10--12 keV detection rate of pattern 0 events is lower than 0.4 counts s$^{-1}$. The effective exposure time for each observation after filtering is given in Table~\ref{tab:xmm14li}. All these observations are done with the pn detector in Small Window mode providing a time resolution of 5.7 ms, however, for reasons we explain below, we also extracted the Reflection Grating Spectrometer (RGS; \citealt{denHerder2001}) data for the first five epochs of observations.  

If the source flux is very high, the XMM--{\it Newton} (pn) data might be affected by photon pile-up. Pile-up spuriously influences the observed spectral properties of the source under study. However, owing to the high time resolution afforded by the pn in Small Window mode pile-up can be avoided in many cases. The XMM--{\it Newton} handbook states that pile-up is important for point sources with count rates above 25 counts per second. As the issue can be especially important for very soft sources such as typically found for TDEs, we investigate the event pattern distribution. The event pattern is the property that the charge cloud released by the incoming photon can be distributed over one, two, or more detector pixels. For the XMM--{\it Newton} pn detector we use only events detected in one or two pixels. If pile-up is present, the observed number of one and two pixel events deviates from non-piled-up observations. We use the {\sc SAS} command {\sc epatplot} to compare the observed and expected number of single and double event pattern as a function of photon energy. We conclude that pile-up is important for the first four observations in Table~\ref{tab:xmm14li} and probably also for the fifth observation. Furthermore, for a source as bright as ASASSN-14li early-on in the observing campaign, there is no area of the read--out part of the detector that can be used to estimate a reliable background spectrum. This is a limiting factor for the first five pn observations in Table~\ref{tab:xmm14li}, which we therefore do not use further. Instead, for those first five observations, we extracted the RGS data following the standard {\sc SAS} procedure, where we defined as good times those where the background count rate on CCD number 9 on RGS 1 is lower than 1 count s$^{-1}$.

In the pn observations from epochs 6-10, background photons were extracted from a source-free rectangular region towards the edge of the field of view of the Small Window mode. In selecting the location of the rectangular background region, we avoid the wings of the point spread function of the source, including the counts caused by the diffraction spikes, and the area that is affected by so-called out-of-time events. We extracted the source using a circular aperture of 15\arcsec\, radius centered on the optical position of the source.
The extracted spectra are rebinned to yield a minimum of 30 counts per bin.

\begin{deluxetable*}{lrcrc}
\tablecaption{XMM--{$\it Newton$} observations of ASASSN-14li and ASASSN-15oi used in this paper.}
\tablewidth{0pt}
\tablehead{
\colhead{Observing ID \& instrument} & \colhead{Start date \& time} & \colhead{Exp time} & \colhead{Counts$^{\dagger}$} & \colhead{Aperture} \\
\colhead{} & \colhead{[UTC]} & \colhead{[ks]} & \colhead{[phot]} & \colhead{[$\arcsec$]}
}
\startdata
ASASSN-14li-0694651201  RGS &  2014-12-06 23:27:23 & 11.45 & 19078 &  RGS   \\
ASASSN-14li-0722480201  RGS & 2014-12-08 12:53:00 & 30.19 & 85077  & RGS  \\
ASASSN-14li-0694651401  RGS & 2015-01-01 21:29:52 & 16.32 & 17876  & RGS  \\
ASASSN-14li-0694651501  RGS & 2015-07-10 05:55:50 & 15.01 & 4854 & RGS   \\
ASASSN-14li-0770980101  RGS & 2015-12-10 11:36:43 & 59.68 & 20509 & RGS  \\
ASASSN-14li-0770980501 pn & 2016-01-12 04:14:51 & 4.9 & 5225  & 15  \\
ASASSN-14li-0770980601 pn & 2016-06-04 23:28:25 & 9.09 & 2839 & 15   \\
ASASSN-14li-0770980701 pn & 2016-12-04 14:24:45 & 6.50 & 1445  & 15   \\
ASASSN-14li-0770980801 pn & 2017-06-08 01:01:40 & 8.05 & 760  & 15  \\
ASASSN-14li-0770980901 pn & 2017-12-05 08:51:52 & 13.43 & 782 & 15  \\
\hline
ASASSN-15oi-0722160501 pn & 2015-10-29 14:38:02 & 8.68 & 223 &   15  \\
ASASSN-15oi-0722160701 pn & 2016-04-04 15:30:49 & 9.84 & 1115 & 15 \\
\label{tab:xmm14li}
\enddata
\tablecomments{$^{\dagger}$ Counts detected after filtering on photon energy between 0.3--10 keV for the pn and 0.35--1.9 keV for the RGS data (combining the RGS1 and RGS2 data). }
\end{deluxetable*}

\subsection{ASASSN-15oi data}

ASASSN-15oi was discovered on 2015, August 14 in the nucleus of a quiescent early-type galaxy at $z=0.0484$ \citep{Holoien16b}.
XMM-{\it Newton} observed the source ASASSN-15oi twice, namely, on 2015, October 29 and 2016, April 04 with observation IDs 0722160501 and 0722160701, respectively. These ASASSN-15oi observations are done with the pn detector in Full Frame mode providing a time resolution of 73.4 ms. We run the default {\sc SAS} v18 (20190531) tools under the HEASOFT {\sl ftools} software version 6.26.1 to extract source spectra. In particular, we filtered the data for periods of enhanced background radiation, where we require that the 10--12 keV detection rate of pattern 0 events is lower than 0.4 counts s$^{-1}$. The remaining, effective exposure time after filtering out periods of high background is given in Table~\ref{tab:xmm14li}. We also checked the data for the presence of pile--up but we found that pile-up did not affect the observations of ASASSN-15oi. We extracted the source using a circular aperture of a 15\arcsec\, radius centered on the optical position of the source. The background counts and spectrum were extracted from a source free, rectangular region on the same detector.

\section{Results and Discussion}
\label{RandD}

In this section, we present synthetic X-ray spectra and light curves predicted by our TDE model and then show the best model fits to the TDEs ASASSN-14li and ASASSN-15oi.

\subsection{Theoretical X-ray spectra and light curves}
\label{thsl}

\begin{figure}
\includegraphics[height=0.5\textheight]{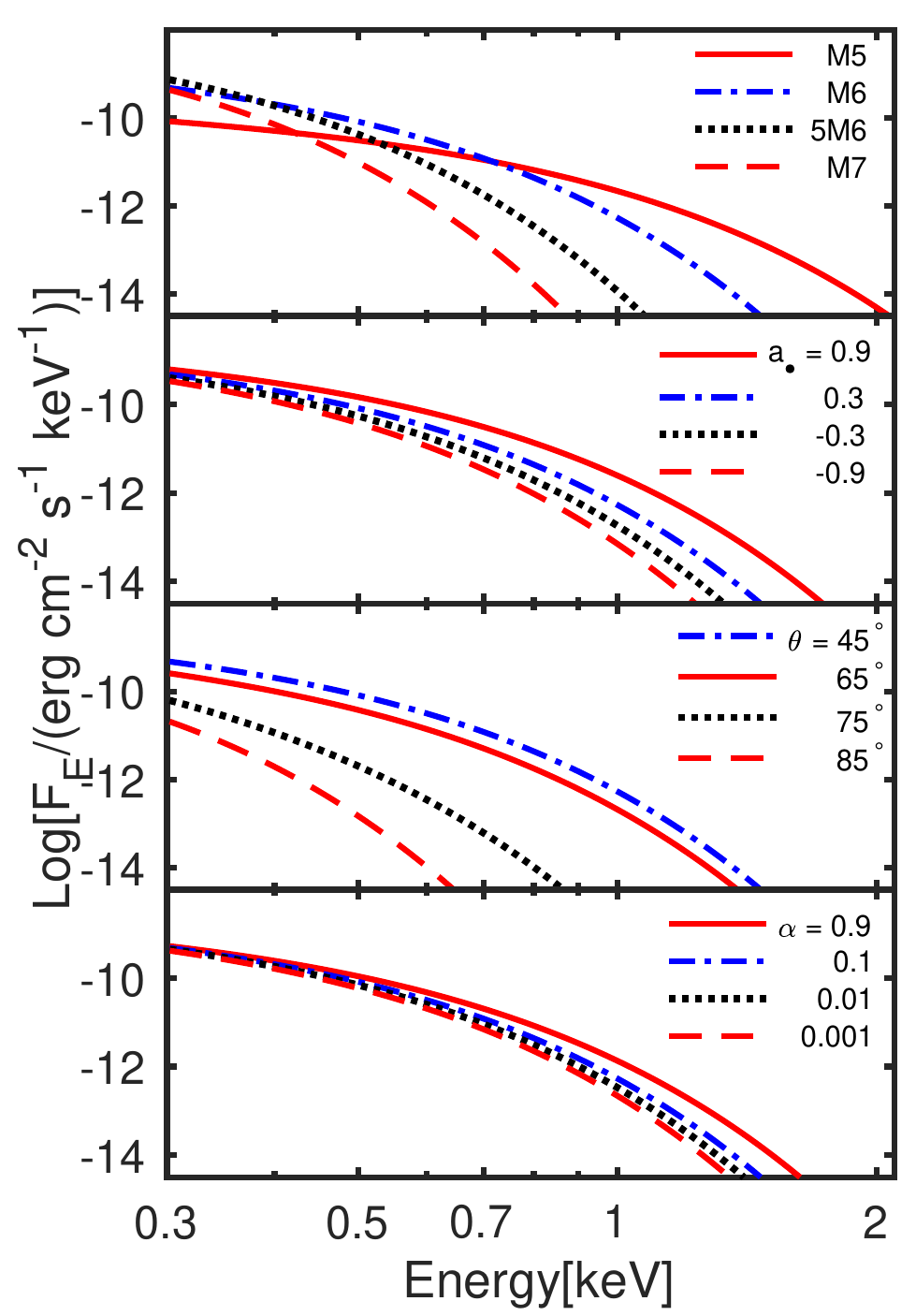}
\caption{Theoretical spectra for our slim disk TDE model, 
assuming that the mass accretion rate follows the fallback rate, 
as seen near the time of peak mass fallback. The fiducial case here
has SMBH mass $M_\bullet=10^6M_\odot$, spin $a_\bullet=0.3$, viewing angle $\theta=45^\circ$, and disk viscosity parameter $\alpha=0.1$ (blue dot-dashed line). In each panel, we vary one of these four parameters while holding the others fixed. We assume a distance $d=100 ~{\rm Mpc}$, penetration parameter $\beta=1.8$, and a progenitor star mass $m_\star=M_\odot$ throughout.
The dimensionless accretion rates 
for the SMBH masses of $10^5$, $10^6$, $5\times 10^6$, and $10^7 M_\odot$ are $2000$, $100$, $10$, and $4$, respectively.
{\it Top}: SMBH mass $M_\bullet$ varied between $10^5$ and $10^7 M_\odot$
(in the figure legend, $Mx=10^xM_\odot$). 
The change in the spectra with $M_\bullet$ 
reflects the variation in the physical size of the disk and in the accretion rate.
%
%
{\it Upper middle}: SMBH spin $a_\bullet$ varied from $-0.9$ to $0.9$.  
%
Positive spins correspond to prograde disks, and negative spins to retrograde ones. 
{\it Lower middle}: Observer's viewing angle $\theta$ varied from $45^\circ$ to $85^\circ$, 
where $\theta=90^\circ$ is an edge-on disk.
%
{\it Bottom}: Dimensionless disk viscosity parameter $\alpha$ varied from $10^{-3}$ to $0.9$.
This figure shows that 1) smaller $M_\bullet$ produces a higher quasi-thermal hard X-ray luminosity, but a lower soft X-ray luminosity; 2) higher $a_\bullet$ produces more luminosity in both hard and soft X-rays; 
3) large $\theta$ strongly screens the inner X-ray emission;  4) $\alpha$ has little effect by itself.}
\label{fig:theorySEDs1}
\end{figure}

\begin{figure}
\includegraphics[height=0.5\textheight]{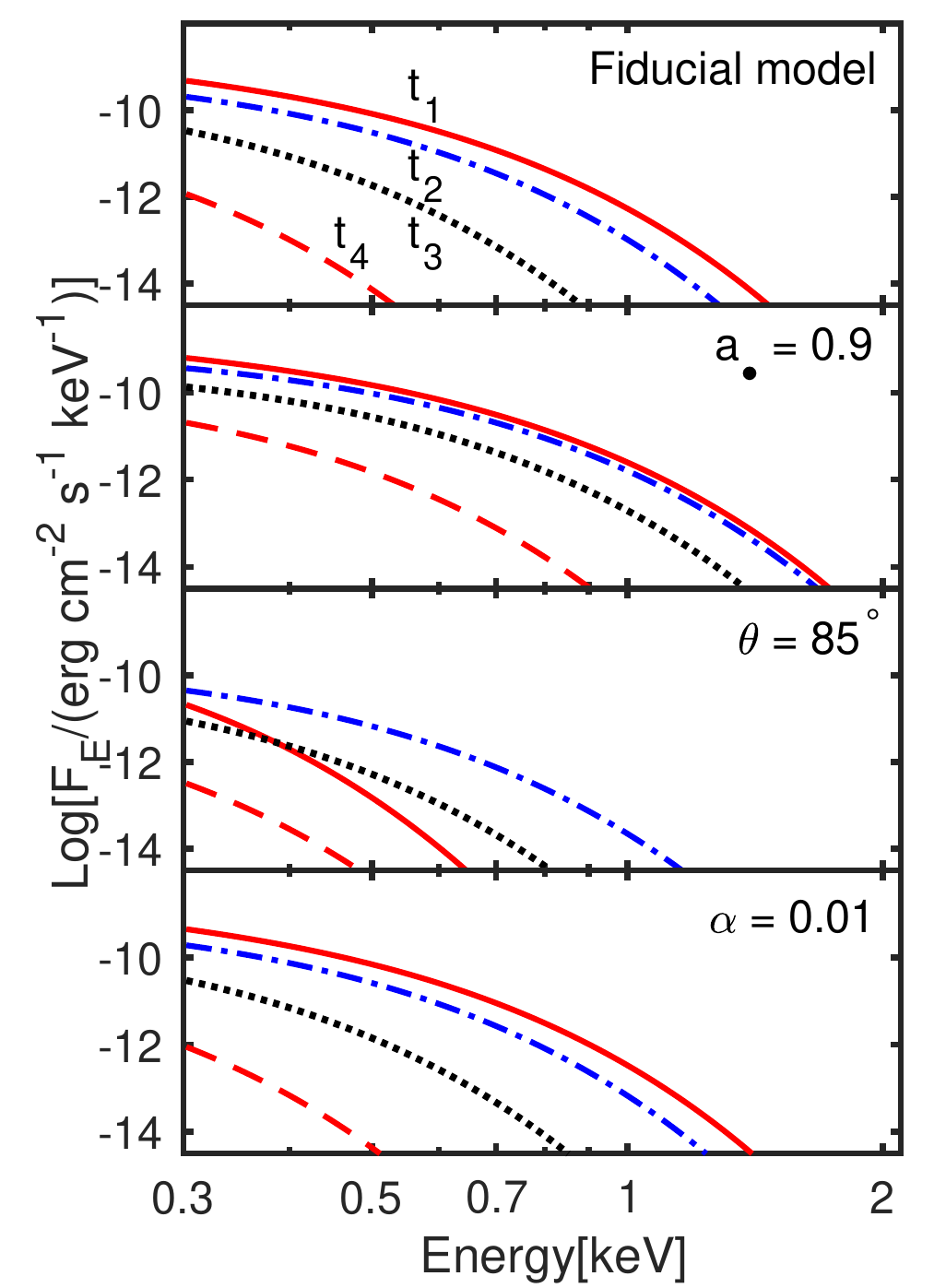}
\caption{Theoretical spectra for our slim disk model seen at different times. {\it Top}:
Fiducial case from Fig. \ref{fig:theorySEDs1} at $t_1=7$ days $(\dot m=100)$, $t_2=204$ days $(\dot m=3.4)$, $t_3=513$ days $(\dot m=0.8)$, and $t_4=1196$ days $(\dot m=0.2)$. 
As time passes, 
the accretion rate decreases, and the disk spectrum monotonically becomes dimmer and softer.  
{\it Upper middle}: Same as the top panel, but now for a rapidly spinning SMBH ($a_\bullet=0.9$).  The results are qualitatively similar to the fiducial case, although the initial rate of spectral softening (and dimming) is slower.  
{\it Bottom middle}: Same as the top panel, but now for a nearly edge-on observer ($\theta=85^\circ$). 
The time evolution here is different than in the other panels. Here the 
synthetic spectrum
from $t=t_1$ (red solid line)
to $t_2$ (blue dot-dashed) becomes both brighter and harder.
Later, at $t_3$ (black dotted) and $t_4$ (red dashed), the spectrum fades and 
softens.
%
The intermediate-time hardening and brightening
arises from ``disk slimming,''  
where the disk height declines as the accretion rate decreases in time, allowing many more null geodesics from the X-ray bright inner disk to escape to the observer.
%
{\it Bottom}: Same as the top panel, but now for low disk viscosity ($\alpha=0.01$). The results are very similar to those in the top panel for the fiducial case.}
\label{fig:theorySEDs2}
\end{figure}

\begin{figure}
\includegraphics[height=0.5\textheight]{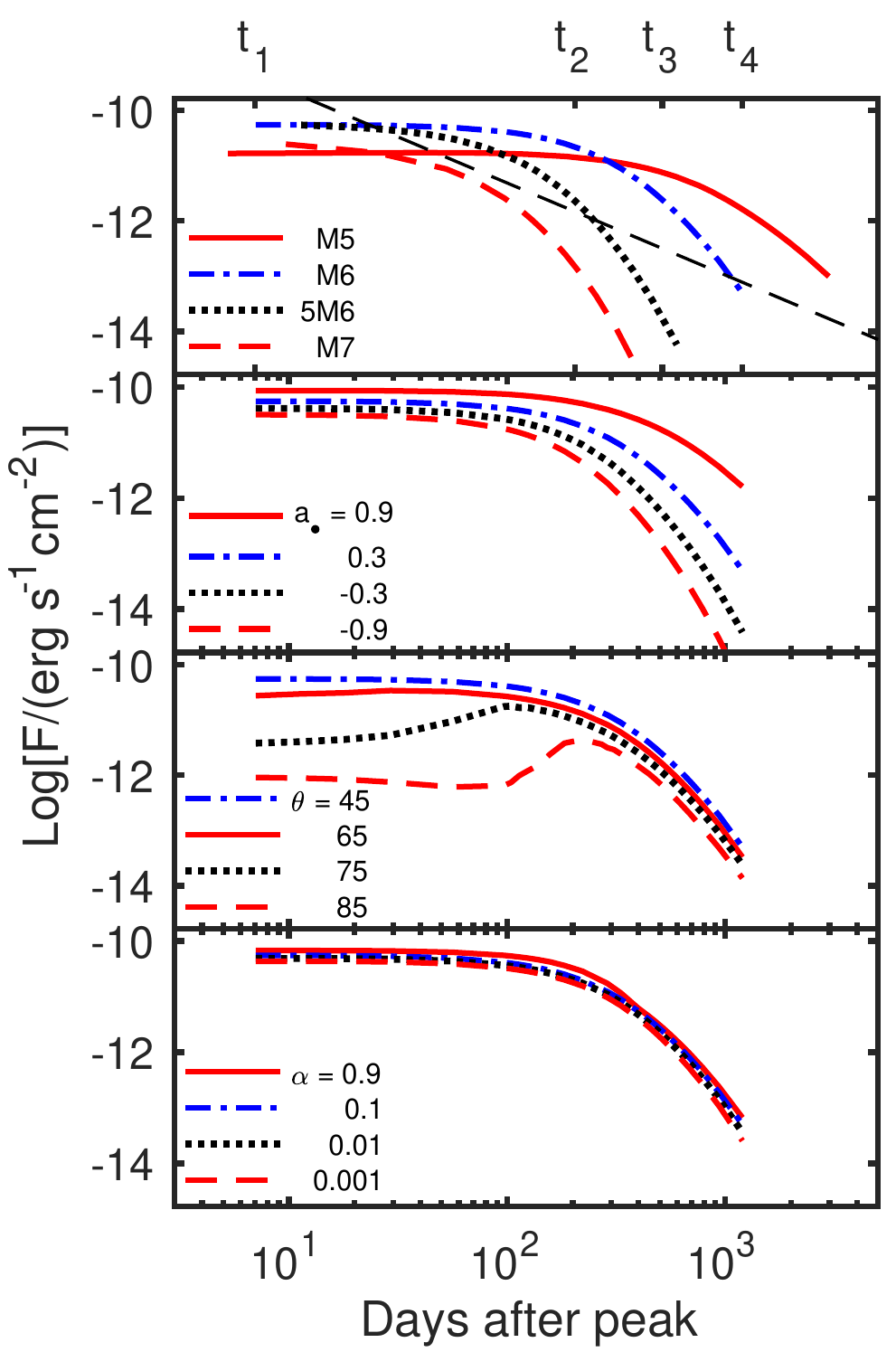}
\caption{Theoretical X-ray light curves for our slim disk model, showing the 0.3-$10~{\rm keV}$ X-ray flux 
versus time since the peak of the mass fallback rate.  
As in Fig. \ref{fig:theorySEDs1}, the four panels explore the effects of changing individual parameters from the fiducial case ($M_\bullet=10^6 M_\odot$, $a_\bullet = 0.3$, $\alpha=0.1$, and $\theta=45^\circ$; blue dot-dashed line). 
Most of the light curves have a ``plateau'' phase when the X-ray emission is nearly constant, corresponding to super-Eddington accretion rates.  Once the accretion rate becomes significantly sub-Eddington, the X-ray luminosity falls off exponentially with time, reflecting its origin in a quasi-thermal Wien tail from the (cooling) innermost disk annuli.  In no cases does the X-ray luminosity trace the $t^{-5/3}$ power law (black dashed line) that describes the asymptotic mass fallback rate.  
{\it Top}: Compared with the fiducial case, lower SMBH mass 
($10^5 M_\odot$) produces a longer plateau phase and lower initial luminosity, whereas higher SMBH masses 
($\geq 5\times10^6 M_\odot$) 
produce a shorter plateau phase.
{\it Upper middle}: Larger and more prograde SMBH spins extend the duration and increase the normalization of the plateau.  
{\it Lower middle}: Varying the observer's inclination angle 
dramatically affects
the X-ray light curve, for reasons also visible in Fig. \ref{fig:theorySEDs2}.  Nearly edge-on viewing angles produce a delayed rise in the luminosity, corresponding to the time when the disk becomes geometrically thin enough to stop self-shielding the observer from the inner disk's X-rays.  
{\it Bottom}: Varying $\alpha$ has a negligible effect on the light curve in comparison to the other parameters explored here.}
\label{fig:theoryLCs}
\end{figure}

The sensitivity of the synthetic spectrum and light curve to various model parameters is shown in Figures \ref{fig:theorySEDs1}-\ref{fig:theoryLCs}, where we 
assume that
the mass accretion rate tracks the fallback rate and that the spectral hardening factor $f_{\rm c}$ is well-described by Eq. \eqref{fc2}. Figure \ref{fig:theorySEDs1} shows the theoretical X-ray spectra.
As is generally the case for steady-state $\alpha$ disks, lower SMBH masses imply smaller disk sizes and higher disk temperatures, although the bolometric luminosity from smaller SMBHs is increasingly Eddington-limited\footnote{As noted earlier, our slim disk models are not strictly Eddington-limited, but the highly super-Eddington fallback rates characteristic of TDEs around SMBHs with $M_\bullet \lesssim 10^6 M_\odot$ only increase their bolometric luminosities over the Eddington limit by factors $\sim \ln(\dot{M}_{\rm t}/\dot{M}_{\rm Edd}).$}. Consequently, our calculation shows that the highest total soft X-ray luminosities ($0.3-10.0~{\rm keV}$) occur for $M_\bullet \sim 4\times10^6M_\odot$.
Increasing the SMBH spin pushes the prograde ISCO of the SMBH inward, increasing the emitting area of high temperature gas, the disk effective temperature,  
and thus the X-ray flux. 
When the inclination is large (i.e., the disk is nearly edge-on), the observed luminosity decreases dramatically, because X-rays from the inner region are mostly shielded by the outer annuli of the disk, in which case the low temperature edge is the primary contributor to the observed flux. The strength of this effect, which was first predicted for TDE disks by \citet{Ulmer99}, is weakened somewhat by relativistic lensing, in a way that we have quantified here for the first time.

The evolution of the synthetic spectra over four epochs
is shown in Figure \ref{fig:theorySEDs2}. The top panel indicates how the fiducial case changes with time. As the accretion rate decreases, the disk spectrum dims and spectrally softens. A higher SMBH spin (upper middle panel) results in a slower decay of the disk luminosity, because higher radiative efficiencies delay the transition to the sub-Eddington regime.  
The results for a lower viscosity parameter (bottom panel) are similar to the fiducial case.

For a more inclined, nearly edge-on disk (bottom middle panel), the total luminosity is initially dominated by the disk edge, as X-ray photons from the inner disk are blocked by the edge. As the accretion rate decreases, the height of the disk also decreases, exposing more photons from the inner disk to the observer's line of sight. As a consequence of this ``disk slimming,'' the disk spectrum initially becomes brighter and harder. At later times, the disk luminosity enters the usual regime of exponential decay, and the spectrum fades and softens. Because the effect of disk slimming is not instantaneous, it introduces some degeneracy in single-epoch spectral fitting; the same overall X-ray luminosity can be produced by two different stages of edge-on disk evolution, e.g., $t_1$ and $t_3$,
with two very different accretion rates. 
The X-ray spectral shape is somewhat harder at the later epoch, however.

The sensitivity of the light curve to various model parameters is shown in Figure \ref{fig:theoryLCs}.
Lower mass SMBHs have a longer super-Eddington emission plateau, transitioning later to exponential X-ray flux decay (see \citealt{lin2018} for a possible observational example). Increasing the spin $a_\bullet$ increases the normalization of the disk luminosity and extends the super-Eddington plateau phase. 
In all of these synthetic light curves, the disk accretion rate decays as Eq. \eqref{Mf}, and the disk temperature and luminosity also decrease.
However, the luminosity does not scale linearly with the accretion rate in a slim disk, as the hot, radially inflowing gas will advect away some fraction of the heat that is generated locally. This effect becomes notable when $\dot m > 0.3$ and comes to dominate over radiative cooling near the Eddington limit \citep{Sadowski2009}. 

In the slim disk model, continuously increasing the accretion rate increases the fraction of disk power advected as heat into the horizon. As a result, at  accretion rates well above the Eddington rate, the total luminosity of the disk increases only slowly with accretion rate \citep{Abramowicz1988}. 
The X-ray luminosity here actually 
saturates as the accretion rate becomes highly super-Eddington, because $f_{\rm c}$ decreases due to an increase in the surface density, which more than counters the slowly increasing total luminosity. 
As a result, slim disk models predict a nearly flat early-time X-ray light curve.

At late times in TDEs, once the accretion rate becomes significantly sub-Eddington, advective cooling is no longer important. The X-ray light curve decays faster than the (power-law) accretion rate at late times, because 1) X-rays are generally emitted from the Wien tail of each annulus, and 2) the spectral hardening factor decreases with the accretion rate. 
The former is generally more important for setting the rate at which the X-ray light curve declines, as the luminosity decays exponentially, i.e., $\nu L_\nu \propto \exp{(-bt^{5/12})}$, where $b$ is a normalization constant, as the accretion rate falls.

The light curve behaves interestingly for large (edge-on) inclinations, exhibiting a late, steep rise at $t \sim100$ days. This behavior is due to the same ``disk slimming'' effect seen earlier in Fig.~\ref{fig:theorySEDs2}.  At early times, X-ray emission is blocked by the 3D structure of the 
disk, but once the disk height decreases sufficiently, X-rays from the hot inner disk annuli can escape along the observer's line of sight, increasing the observed luminosity. This result suggests that the disk vertical structure provides a strong constraint on the inclinations of some TDEs. Given that $H/r\sim0.6$ during the super-Eddington phase, the peak of the X-ray light curve can be delayed in nearly half of TDEs with an initially super-Eddington accretion rate (see also the discussion in \citealt{Jonker+19} on the delay between the observed optical and X-ray peak luminosities). 

Unlike in a standard thin disk model, where the disk luminosity is independent of the viscosity parameter (for fixed accretion rate), the innermost regions of our slim disk have a higher inner temperature 
(for fixed accretion rate)
when the viscosity parameter is larger \cite{Sadowski2009}. The ultimate effect of viscosity on bolometric luminosity is relatively small, so we fix the viscosity parameter to $\alpha=0.1$
for the rest of our analysis.
There is a possible secondary effect; 
holding all else constant, different $\alpha$ values produce different surface density profiles, which in turn can alter the spectral hardening factor $f_{\rm c}$.  We defer a detailed investigation of this effect to future work.

\subsection{Fitting ASASSN-14li's X-ray spectra}


\begin{figure*}
\includegraphics[height=0.65\textheight]{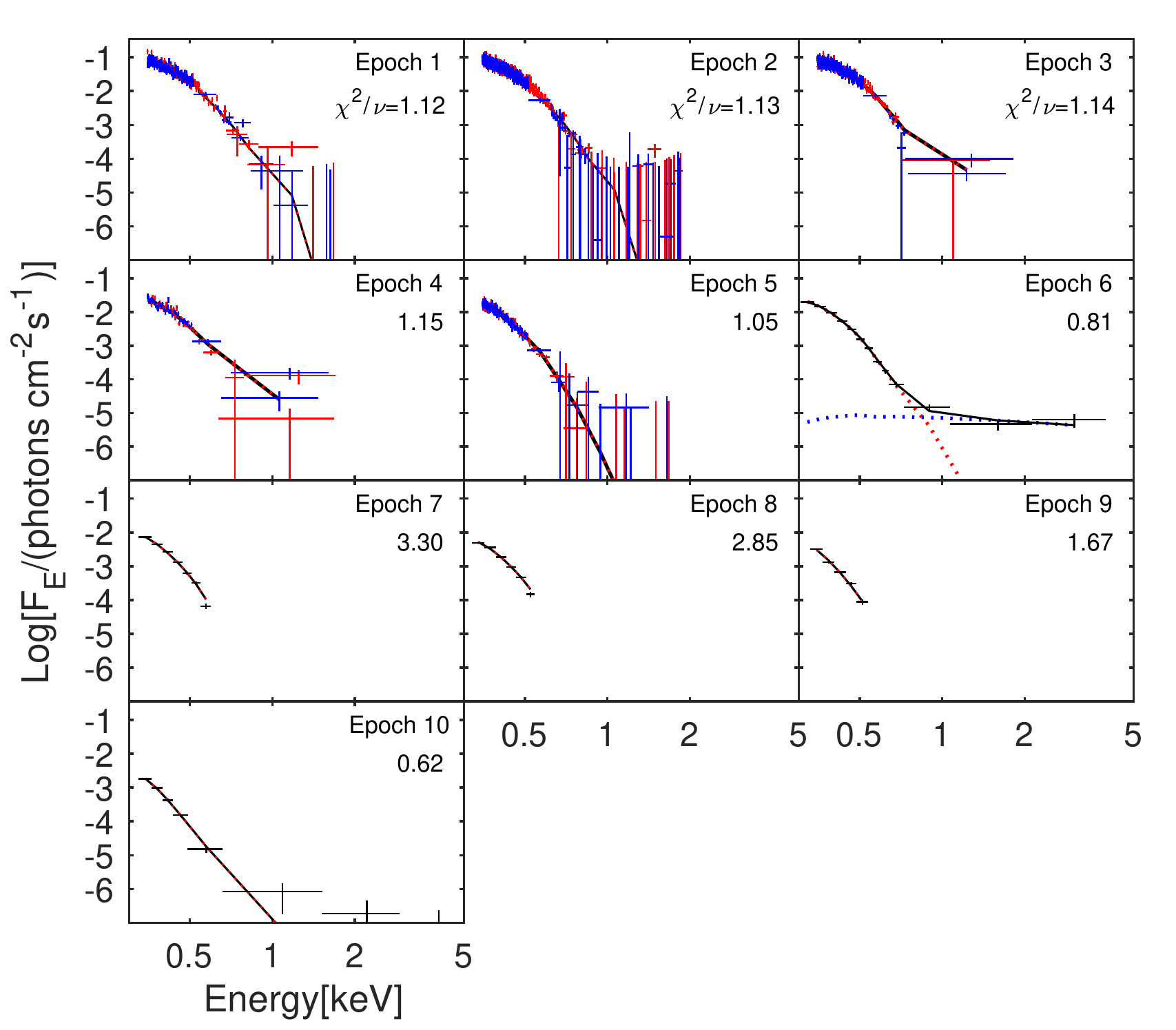}
\caption{Simultaneous slim disk fits to XMM--{\it Newton} spectra for ASASSN-14li from early (Epoch 1) to late (Epoch 10) times over a three-year period. The first five spectra are obtained using the two RGS detectors (red for RGS1, blue for RGS2),
which provide data over the energy range 0.35--1.9 keV. The last five spectra are obtained using the 
pn detector, which is sensitive over 0.3--10 keV. All spectra are background-subtracted, and the data are binned so that there are at least 30 counts per bin. Each panel shows the best-fit model as a solid black line.  In nine of the epochs, only the quasi-thermal disk model (red dotted line) is needed to fit observations. In Epoch 6, however, we need an additional power-law component to fit the hard emission (blue dotted line). The accretion rate is allowed to float between epochs.
The fitting results are in Table~\ref{14liresults}. }
\label{14sp}
\end{figure*}
\begin{deluxetable*}{ccccccccc}
\tablecaption{Best-fit results for ASASSN-14li, with $M_\bullet=10^7M_{\odot}$ and $a_\bullet=0.998$.}
\tablewidth{0pt}
\tablehead{
\colhead{Epoch} & & \colhead{$N_{\rm H}$}  & & \colhead{$\theta$} & & \colhead{$\dot m^a$}  & & \colhead{$\chi^2/v$} \\  
\colhead{} & & \colhead{$[10^{20}{\rm cm}^{-2}]$}  & & \colhead{$[{^\circ}]$} & & \colhead{$\rm [Edd]$}  & & \colhead{}  
}
\startdata
$1$
           &&$5.3\pm0.3$   &&$76\pm3$  &&$3.0\pm0.2$   
           &&629.79/562 \\
$2$ 
           && $5.0\pm0.2$   && ...  && $2.8\pm0.2$ 
           && 2321.45/2057     \\
$3$   &&$5.1\pm0.3$  &&... &&$2.9\pm0.2$ 
           &&591.98/521                \\         
$4$ &&$5.3\pm0.4$   &&...  &&$1.08\pm0.03$
           &&  178.92/155      \\
$5$ &&$4.0\pm0.3$ &&... &&$0.81\pm0.02$ 
           && 601.7/572  \\
$6^b$ &&  $4.3\pm0.3$ &&...  &&$0.79\pm0.02$   && 7.31/9          \\
$7$  && $4.0\pm0.4$ && ...  &&$0.53\pm0.02$   && 16.52/5                        \\
$8$    &&$4.9\pm0.5$ && ...   &&$0.48\pm0.02$     &&     11.4/4                     \\
$9$   &&$2.5\pm0.8$ && ... &&  $0.34\pm0.02$  && 5.02/3             \\
$10$ &&$3.4\pm0.8$&& ... &&$0.30\pm0.01$  && 4.31/7              \\
\enddata
\tablecomments{In our simultaneous fitting, the SMBH spin and mass are held fixed, inclination is required to be the same for all epochs, and other parameters float. The total $\chi^2_{dof}$ is $4368.42/3895=1.12$. $^{\rm a}$The accretion rate (in dimensionless Eddington units) has been corrected for the radiative efficiency $\eta$. $^{\rm b}$The power-law parameters of epoch 6 are: $\Gamma=1.5\pm0.7$ and $A_{\rm pl}=8.0\pm2.7\times 10^{-6}$ $\rm {photons ~s^{-1}~cm^{-2}~keV^{-1}}$. }
\label{14liresults}
\end{deluxetable*}

The quasi-thermal slim disk model of the previous subsection forms the basis for our X-ray spectral fitting and parameter estimation.  However, X-ray spectra may be affected by absorption local to the source, on larger scales in the host galaxy, and/or from our own Galaxy. Here we fit the
multi-epoch spectra of ASASSN-14li by combining our slim disk model with a free-floating absorption parameter, $N_{\rm H}$. The general absorption model  {\sc phabs} in XSPEC \citep{Arnaud1996} is added as a multiplication model to our slim disk model. For simplicity, we combine absorption from material local to the TDE, from the host galaxy ($N_{\rm H}=0.026\times 10^{22}{\rm cm}^{-2}$ at $z=0.0206$),
and Galactic absorption ($N_{\rm H}=0.016\times 10^{22}{\rm cm}^{-2}$)
into one parameter \citep{Miller2015}, neglecting (small) redshift effects.   

For the first five epochs of XMM--{\it Newton} RGS spectra, we find no evidence for a spectral component at energies higher than 1.0 keV. As a result, we use only the absorbed slim disk model to fit those epochs. However, as was reported by \citet{Miller2015}, there is evidence for a disk wind at early epochs, in the form of strong absorption lines. Ignoring these strong absorption lines in our fits would artificially bring down the continuum level, thus affecting the best-fit parameters. To avoid such a bias, we ignore the data bins that contain the strong absorption lines at $\approx 0.4911$, $\approx 0.414$, $\approx 0.3997$, and $\approx 0.3776$ keV. For the last five epochs of XMM--{\it Newton} pn spectra, we fit the spectra over the 0.3--10 keV range. 

The fits to the ten spectral epochs are required to have a constant $M_\bullet$ and $a_\bullet$, but all other parameters can, in principle, vary. Even the observational inclination angle could change from epoch to epoch, as the disk may be subject to Lense-Thirring precession \citep{Stone2012, Liska+18} and/or gradual realignment into the SMBH equatorial plane \citep{Franchini2016, ZanazziLai19}. 
However, because the equations for a tilted relativistic slim disk do not yet exist, 
we assume the same inclination for all epochs, i.e., that the disk was either born with small tilt or has quickly aligned itself into the equatorial plane. 
We fit the ten spectra simultaneously, allowing all ten accretion rates ($\dot m_i$), all ten absorption parameters ($N_{\rm H,i}$) and the one inclination ($\theta$) to float. We minimize the $\chi^2$ for each combination of ($M_\bullet, a_\bullet$). 

\subsubsection{SMBH mass, spin, and other inferred parameters}
\label{14lipi}
The best-fit parameters and their $1\sigma$ errors 
are shown in Table \ref{14liresults}, and the spectral fits are plotted in Fig.~\ref{14sp}. The reduced $\chi^2$ for each spectrum indicates that our model describes the spectra well. The total reduced chi-squared is $\chi^2/v=1.12$. 
The $\chi^2_{\rm dof}$ is larger for epochs 7 and 8, in part due to
a decrease in flux 
at $\approx$0.6 keV. 
The first three spectra are fitted with high Eddington ratio accretion rates, indicating an early, mildly super-Eddington accretion phase. The slim disk model predicts a nearly constant flux for these three epochs.


\begin{figure}[ht!]
\includegraphics[height=.280\textheight]{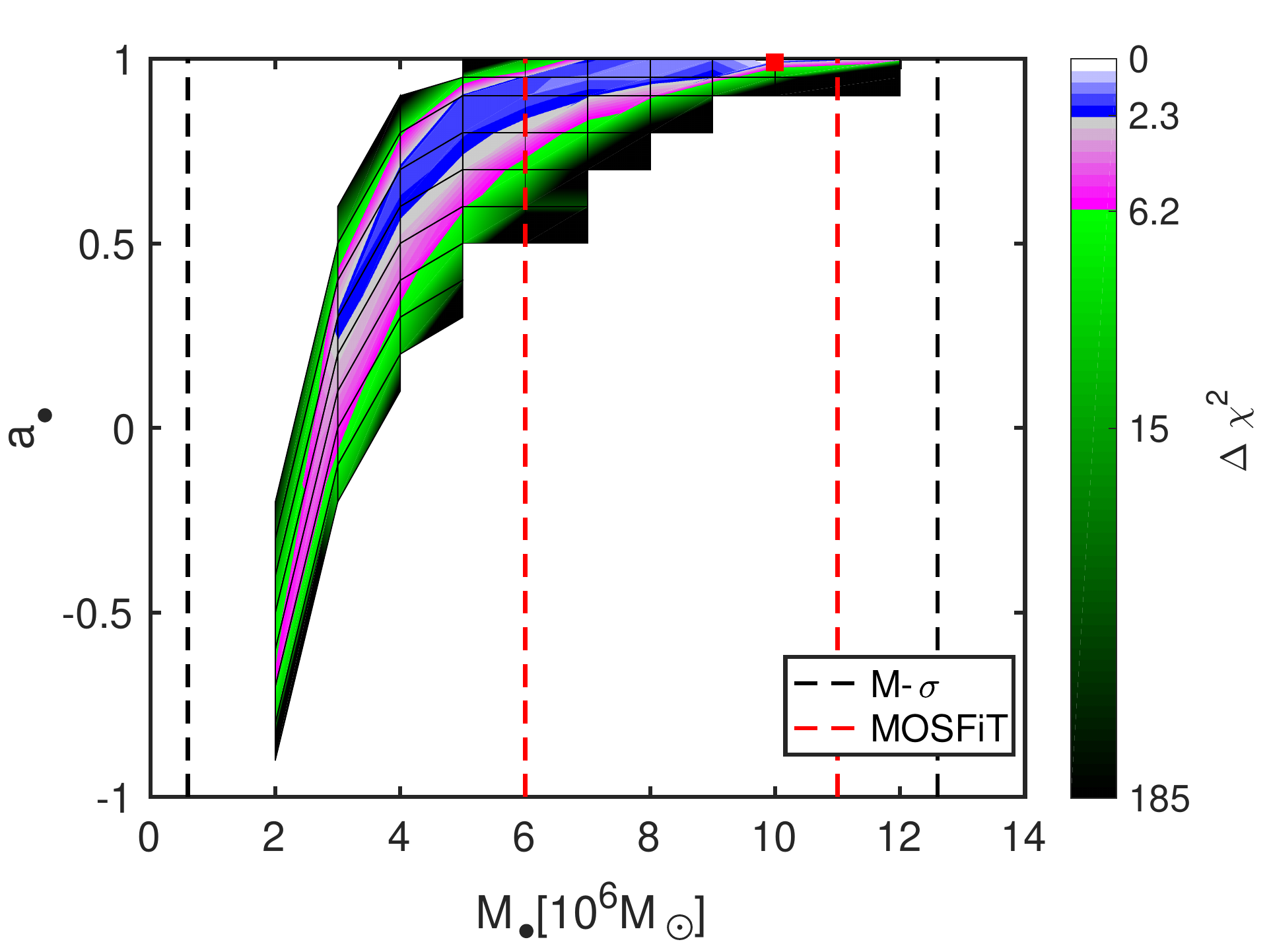}
\caption{Constraints on the SMBH mass ($M_\bullet$) and spin ($a_\bullet$) in ASASSN-14li. We calculate the $\chi^2$ across a model grid in the $(M_\bullet, a_\bullet)$ plane (grid points are indicated by vertices of the black lines). We subsequently fill the intermediate parameter space by linear interpolation. The red square denotes the best fit: $M_\bullet=10^7 M_\odot$ and $a_\bullet=0.998$. The blue region is enclosed by the $1\sigma$ confidence level (CL) contour and the pink region by the $2\sigma$ CL contour.
The best-fit $M_\bullet$ is consistent with independent constraints from the literature, i.e.,
the MOSFiT estimate from optical and UV light curves fitting, $9^{+2}_{-3}\times 10^6M_\odot$ \citep[][red dashed lines]{Mockler2018}, and the values inferred from galaxy scaling relations, $(0.6-12.5) \times 10^6M_\odot$ 
\citep[][black dashed lines]{van2016,Holoien2016a,Wevers2017}. 
At a $1\sigma$ CL, we constrain the mass to  $10^{+1}_{-7}\times 10^6M_\odot$ and the spin to $>0.3$. In combination with the mass constraint
from MOSFiT, our model fits constrain the spin further, to $>0.85$.
}
\label{14lima}
\end{figure}

Figure \ref{14lima} plots the $\Delta\chi^2$ on the ($M_\bullet,a_\bullet$) plane. The best-fit values lie at {\bf ($10^7 M_\odot, 0.998$)}. We compute and plot different confidence level (CL) contours with two-parameter thresholds of $\Delta \chi^2=2.3$ ($1\sigma$ CL, blue) and $\Delta \chi^2 =6.18$ ($2\sigma$ CL, pink).  Within a $1\sigma$ CL, we constrain the SMBH mass and spin to lie in the ranges $10^{+1}_{-7}\times 10^6M_\odot$ and $>0.3$, respectively.  While the marginalized, 1D constraints are fairly broad (e.g., the constraint on $M_\bullet$ has a $1\sigma$ CL comparable to that derived from galaxy scaling relations), the joint, 2D constraint is tighter. To our knowledge, this is the first time simultaneous multi-epoch X-ray spectral fitting has been used to constrain $M_\bullet$ and $a_\bullet$ in a TDE. 

Our constraint on $M_\bullet$ is broadly consistent
with external, independent estimates, including from galaxy scaling relations such as $M_\bullet-M_{\rm bulge}$, which predicts $M_\bullet= 6.3_{-3.1}^{+6.3} \times 10^6M_\odot$ 
\citep{van2016, Holoien2016a}, and $M_\bullet - \sigma$, which predicts $M_\bullet= 1.6_{-1.0}^{+2.4} \times 10^6M_\odot$ 
\citep{Wevers2017}.  Our SMBH mass is also consistent with the MOSFiT estimate from optical and UV light curves fitting: $M_\bullet = 9^{+2}_{-3}\times 10^6M_\odot$ \citep{Mockler2018}. Applying the MOSFiT mass constraint to our Figure \ref{14lima} narrows our $a_\bullet$ constraint to $>0.85$. 

In general, SMBH spin measurements are more challenging to make than SMBH mass estimates, and the only previous
constraint for ASASSN-14li
comes from detection of a 133-second X-ray quasi-periodic oscillation (QPO) by \citet{Dheeraj2019}. While the physical origin of this QPO remains uncertain, if one assumes that the maximum possible QPO frequency is that of fundamental test particle motion at the ISCO, 
the maximum mass 
is $M_\bullet \approx 2\times 10^6 M_\odot$. However, because this mass can only be achieved for near-extremal spins,
it is inconsistent with our parameter inference at the $2\sigma$ CL.  This inconsistency is notable, but, given the questions about
the QPO's origin, a detailed comparison must be deferred for now.


In our exploration of the ($M_\bullet$, $a_\bullet$) plane, $\chi^2$ increases quickly for large $M_\bullet$ and small $a_\bullet$, because these parameter choices produce disks with insufficient X-ray luminosity to fit the first three spectra. 
Thus, observations of super-Eddington spectra offer significant power to constrain parameters of interest and to test accretion astrophysics. For example, the degeneracy of $f_{\rm c}$ with accretion rate would be greatly weakened in the super-Eddington limit, as the bolometric luminosities of super-Eddington disks are insensitive to the accretion rate. Super-Eddington accretion would also push the radius of peak effective temperature inwards, providing more spin information than is available in sub-Eddington accretion disks.

\begin{figure}
\includegraphics[height=0.65\textheight]{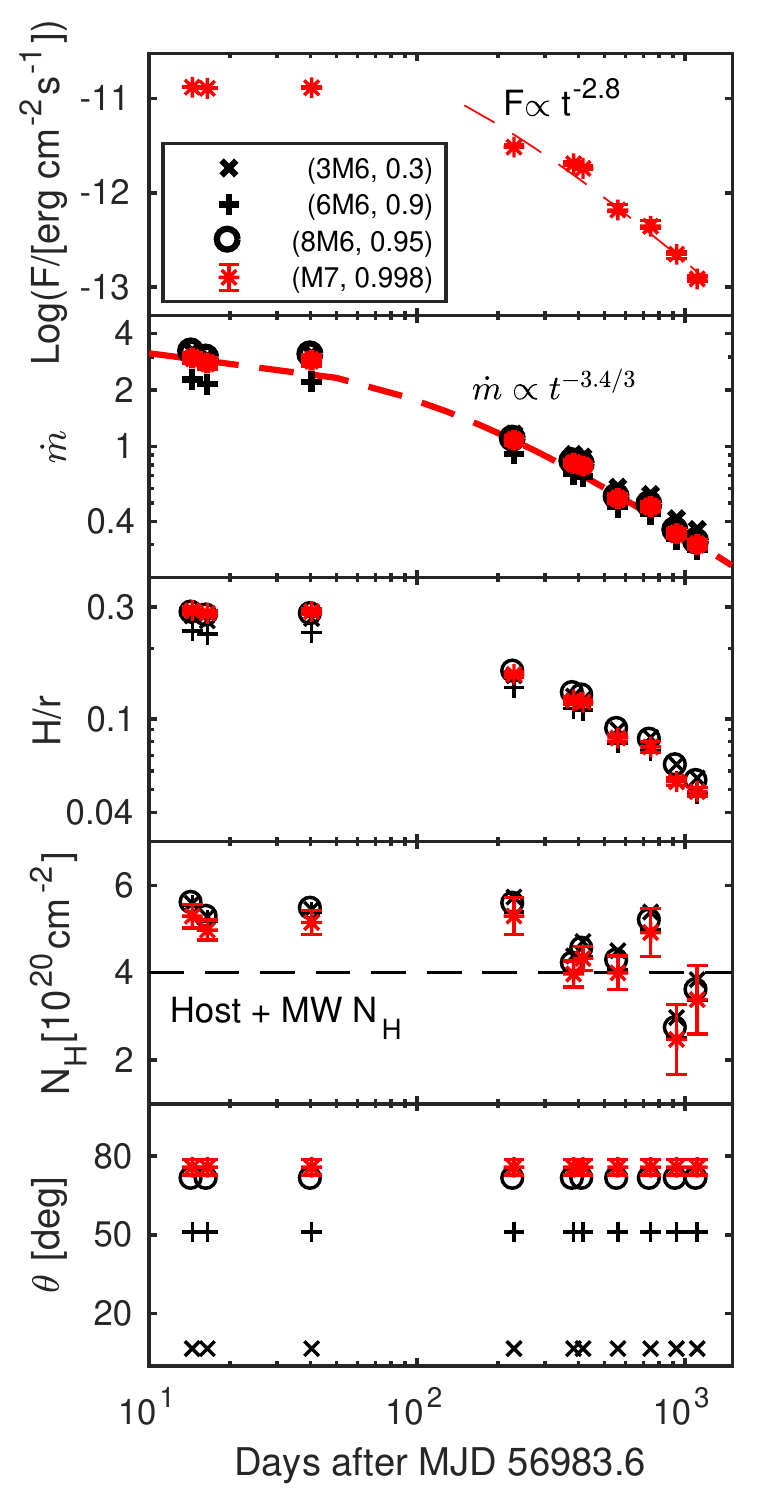}
\caption{Evolution of parameters in ASASSN-14li. The best-fit and $1\sigma$ error bars (red points) for X-ray flux $F$, dimensionless accretion rate $\dot m$ (corrected for radiation efficiency $\eta$), disk aspect ratio $H/r$, absorption $N_{\rm H}$, and inclination $\theta$ are plotted from top to bottom, respectively. 
Fits to the light curve and evolution of the mass accretion rate (red dashed lines) are shown in the top two panels.
Also plotted are the results for three other combinations (black points without error bars, $Mx=10^x M_\odot$) of $M_\bullet$ and $a_\bullet$ within $\pm 1\sigma$ of the best-fit mass and spin.
Here we set the peak date as MJD 56983.6, and $t_{\rm fall}+t_{\rm peak}=150$ days. Our primary conclusions are: 1) The X-ray flux experiences a nearly-constant plateau during the first $\approx 40$ days and then decreases quickly, $\propto t^{-2.8}$, at later epochs. 2) The decay of the accretion rate is slow, going roughly $\propto t^{-1.1}$, suggesting an early delay in disk assembly. 3) The fitted height of the disk decreases by almost an order of magnitude with time. 4) The fitted $N_{\rm H}$ decreases to that of the host galaxy plus Milky Way after $\sim 300$ days, suggesting an early additional contribution from obscuring material near the disk. 5) The fitted $\theta$ depends strongly on the pair ($M_\bullet, a_\bullet$). 6) The decay in the X-ray light curve is dominated by the decay in $\dot{m}$. 
}
\label{14p}
\end{figure}

In Figure \ref{14p}, we plot the time evolution of the fitted TDE parameters. The observed X-ray flux (top panel) is integrated from 0.35 to 1.9 {\rm keV}. The light curve is nearly constant for the first three epochs and then decays quickly, 
as $\propto t^{-2.8}$.
This behavior is similar to the theoretical predictions in Fig. \ref{fig:theoryLCs}, i.e., an early plateau corresponding to the super-Eddington phase and then a rapid decline in X-ray luminosity at sub-Eddington accretion rates. 

The evolution of the mass accretion rate (upper middle panel)
is well constrained, with small error bars for most epochs. The overall uncertainty is dominated by the range of $M_\bullet$ and $a_\bullet$ permitted within the $1\sigma$ CL of Fig. \ref{14lima}. For the best-fit SMBH mass and spin, the accretion rate through the disk
declines from super-Eddington ($\dot m\approx 3.0$) to sub-Eddington ($\dot m\approx 0.3$) over 1100 days. 

The accretion rate decays 
as $t^{-1.1}$, slower than the canonical $t^{-5/3}$ decay of the fallback rate\footnote{Increasing $t_{\rm fall}+t_{\rm peak}$ would yield a steeper 
decay rate, but one still shallower than $t^{-5/3}$.}. 
Figure \ref{14p} also shows that the
specific choice of ($M_\bullet, a_\bullet$) pair has little effect on the 
accretion decay rate. 
However, the decay rate depends on the specific $f_{\rm c}$ parameterization. Given that $f_{\rm c}$ can increase or decrease the 
predicted flux normalization, it is somewhat degenerate with the accretion rate. 

To investigate how our $f_{\rm c}$ assumptions affect the decay of disk accretion rate, we consider a non-fiducial model with constant $f_{\rm c}$, i.e.,
assuming that the last seven
(sub-Eddington) epochs have the same $f_{\rm c}$ as the first three (super-Eddington) epochs.
We then
refit the data by fixing $M_\bullet=10^7M_\odot$ and $a_\bullet=0.998$ and find that the decay of the accretion rate is $t^{-1.2}$, a little steeper than $t^{-1.1}$. 
This result shows that the fitted accretion rate decay is insensitive to the $f_{\rm c}$ parameterization. In agreement with our theoretical analysis, the decay of the light curve is driven by that of the accretion rate and not the modeled time evolution of $f_{\rm c}$. We conclude that the accretion rate decays from $\dot m \approx 3$ to $\dot m\approx 0.3$ roughly as $t^{-1.1}$, which is insensitive to the specific choice of ($M_\bullet, a_\bullet$) pair and $f_{\rm c}$ model.

This analysis is based on
a specific
$f_{\rm c}$ parameterization
DE19, which was originally derived for thin (i.e., sub-Eddington) disks around non-spinning SMBHs. We have already examined the sensitivity of the temporal evolution of $\dot{m}$ to the DE19 $f_{\rm c}$ prescription. In Appendix \ref{app:14li}, we further explore the effect of alternative treatments of $f_{\rm c}$ in the super-Eddington regime and find that the 
DE19
$f_{\rm c}$ prescription can be applied to slim disks. 
Different treatments of $f_{\rm c}$ produce essentially the same best-fit ($M_\bullet, a_\bullet$) pair and have little impact on the broader estimation of $M_\bullet$ and $a_\bullet$\footnote{There is only a minor change in our $2\sigma$ CL constraints in the ($M_\bullet, a_\bullet$) plane when we allow $f_{\rm c}$ to float. Interestingly, the DE19
$f_{\rm c}$ prescription produces substantially {\it weaker} constraints on ($M_\bullet, a_\bullet$) at the $1\sigma$ CL than are produced by looser priors, for reasons analyzed in Appendix \ref{app:14li}. In this sense, our
fits are conservative, particularly for $a_\bullet$.}. By comparing with other treatments of $f_{\rm c}$, we find that the DE19
$f_{\rm c}$ prescription gives a reasonable constraint on the accretion rate; our results are generally insensitive to alternative treatments of super-Eddington spectral hardening.

In all of our fits, the disk height $H/r$ steadily decreases with time
(middle panel). The total absorption $N_{\rm H}$ does the same (lower middle panel).  Because both of these effects will increase the X-ray flux (when all else is equal), the decline of the light curve is driven by that of the accretion rate.  
$N_{\rm H}$ decreases to that of the host galaxy plus Milky Way after $\sim 300$ days, suggesting an early additional contribution from obscuring material near the disk. The fitted (constant) inclination $\theta$ depends strongly on the ($M_\bullet, a_\bullet$) pair (bottom panel).

While we were completing this paper, \citet{Mummery&Balbus20}, hereafter MB20, published an independent analysis of the ASASSN-14li X-ray and UV light curves, using time-dependent models of viscously spreading, general relativistic thin disks. MB20 find
$1.45\times 10^6 < M_\bullet / M_\odot < 2.05 \times 10^6$, with a broad range of permitted SMBH spins. This mass constraint is inconsistent with ours at the $1\sigma$ CL, and marginally consistent at the $2\sigma$ CL. The origin of this tension could arise from differences between the theoretical models,
including our choice to model the local annular emission of the disk as a spectrally hardened blackbody, in contrast to MB20's assumption of a purely thermal spectrum.

Our model allows the accretion rate at each epoch to float, while MB20 compute $\dot m(t)$ deterministically by assuming 
several different initial conditions. Given this difference, it is notable that our accretion rate evolves as a $\dot{m}\propto t^{-1.1}$ power-law; this decay rate is
similar to the classic $t^{-1.2}$ power-law for a viscously spreading disk with zero stress at the ISCO \citep{Cannizzo+90} and somewhat steeper than the $t^{-0.8}$ power-law predicted for finite stresses at the disk inner edge  \citep{Mummery&Balbus19}.  
However, while spreading disk solutions are likely valid at late times in TDE disk evolution \citep{vanVelzen+19}, it is not clear if they apply at early times, when $\dot{m}$ is influenced not just by internal stresses, but also by deposition of tidal debris falling back to pericenter.

The two models have other strengths and weaknesses. Our slim disks can fit the detailed shape of the X-ray spectrum, while their thin disk models are applied to the {\it Swift} X-ray light curve and to the late-time UV light curves.  MB20 explore a wider range of inner boundary conditions and also account for the spreading of the outer disk edge, which we neglect. It is not clear whether this latter effect should change the X-ray spectrum significantly, given the physical origin of the X-rays in the innermost annuli.

\subsubsection{
No evidence for missing energy problem}
By using relativistic slim disk models, we have measured the time evolution of the mass accretion rate.
This task is challenging to perform with optical or NUV light curves due to the lack of a first-principles model for the bolometric corrections. By linearly interpolating between epochs and 
integrating under the $\dot{m}$ curves
in Fig. \ref{14p}, we can directly measure the total mass accreted onto the SMBH during the duration of observations:
$\Delta M=0.17$, $0.20$, $0.22$, and $0.27 M_\odot$ for the SMBH mass-spin pairs drawn from the $1\sigma$ CL in Fig. \ref{14p}, i.e., for masses $10\times 10^7$, 
$8\times 10^6$, $6\times 10^6$, and $3\times 10^6 M_\odot$,  respectively.
In comparison to MB20, our $\Delta M$ estimates are well outside the $0.016 M_\odot$ obtained by assuming finite stress at the ISCO and only partially overlap the $0.26-0.34 M_\odot$ calculated by assuming vanishing ISCO stress. 

Because half of the disrupted star's mass $m_\star$ remains dynamically bound to the SMBH \citep{Rees1988}, we can translate $\Delta M$ into a lower limit on $m_\star$ of $\ge 2\Delta M$. This lower limit 
is similar in magnitude to a cruder estimate from the large fitted peak accretion rate; assuming rapid circularization, i.e.,
$\dot{M}(t)=\dot{M}_{\rm t}(t)$, and our best-fit SMBH mass
$M_\bullet = 10^7 M_\odot$
gives $m_\star \gtrsim 0.5 M_\odot$. All of our lower limits on $m_\star$ have some tension
with the value of $0.2^{+0.1}_{-0.1}M_\odot$
inferred from MOSFiT optical and UV light curves fitting.

TDE rate calculations predict that most disrupted stars come from the lower main sequence \citep{Stone&Metzger16}, with $0.1 \lesssim m_\star/ M_\odot \lesssim 1$.  Our X-ray spectral modeling demonstrates that $\Delta M$ is of the same order of magnitude. 
This result contrasts strongly with the ``missing energy problem'' \citep{Stone&Metzger16, Piran+15}, where, if one integrates the total energy emitted under optical/NUV TDE light curves and assumes radiatively efficient accretion, 
$\Delta M \lesssim 0.01 M_\odot$. 
Indeed, the total optical/NUV energy release seen in the first $\approx 200$ days of the ASASSN-14li light curve can be explained by radiatively efficient ($\eta=0.1$) accretion of $\Delta M = 0.0013 M_\odot$ \citep{Metzger&Stone17}, which is at least 10 times smaller than our estimate. 
Our X-ray continuum fitting 
appears to have solved the missing energy problem, at least for ASASSN-14li, by using a disk model that accounts for both 1) the energy released at unobservable EUV wavelengths and 2) the energy directly advected into the SMBH horizon. The first of these effects dominates
for the accretion rates 
inferred from our fits to ASASSN-14li.

Our result is in agreement with simple theoretical models for TDE disks, which have long predicted a spectrum dominated by EUV emission \citep{Ulmer99, Lodato2011, Lu&Kumar18}, and 
with observations of infrared dust echoes from TDEs selected through optical/NUV emission.  While the infrared emission is not itself particularly bright, the accretion luminosities required to heat circumnuclear dust to the requisite temperatures are substantial. For the TDE PTF-09ge, the observed peak optical/NUV luminosity was $\approx 10^{44.1}~{\rm erg~s}^{-1}$, but modeling of the dust echo implies that the unobserved EUV/X-ray luminosity is $\sim 10^{45}~{\rm erg~s}^{-1}$ and the total energy absorbed by the circumnuclear dust is $E_{\rm abs} \sim 10^{52}~{\rm erg~s}^{-1}$ \citep{vanVelzen+16}.

Such a large energy release corresponds to radiatively efficient ($\eta=0.1$) accretion of $\Delta M\approx 0.1 M_\odot$, similar to that we infer for
ASASSN-14li.  While ASASSN-14li itself has a dust echo \citep{Jiang+16}, with observed infrared luminosity $\sim 10^{41.5}~{\rm erg~s}^{-1}$, 
conversion to the TDE bolometric luminosity is very model-dependent.  Depending on the assumed model for the nuclear dust grain size distribution, the inferred bolometric luminosities vary between $L\sim 10^{43-45}~{\rm erg~s}^{-1}$ \citep{Jiang+16}.  
By using X-ray continuum fitting to measure $\dot{m}(t)$ and infer the bolometric luminosity for a large TDE sample, 
it will be possible to rule out various models for circumnuclear dust in galactic nuclei.

\subsubsection{Implications for circularization efficiency and reprocessing layers}
For ASASSN-14li, we have seen that
the best-fit evolution of the accretion rate is $\dot{m}(t) \propto t^{-1.1}$ (Fig. \ref{14p}), apparently unlike
the theoretical mass fallback curves $\dot{M}_{\rm t}(t)$. 
To quantify this result, we conduct an F-test \citep{James2020}
by imposing the theoretical fallback rate from Eq. \ref{Mf} as a prior for the disk accretion rate $\dot{m}$. The resulting spectra are poor fits, with $\chi^2$ increasing by 147.1 and an F-test 
value of $p < 10^{-16}$ relative to the model in which $\dot{M}$ is allowed to float freely from epoch to epoch.  This small $p$ value indicates that the best-fit accretion rate evolution is inconsistent with the fallback rate.
Delayed circularization thus may be operative. If mass returning near (an unobserved) peak fails to circularize, but is incorporated into the inner disk on later pericenter passages \citep{Shiokawa+15}, the disk accretion rate can exhibit a shallower decline from a lower peak than for the case where
$\dot{M}=\dot{M}_{\rm t}$.

Another possibility is that the bound debris 
does efficiently circularize, but the 
disk launches powerful outflows that carry away most of the circularized material.  Mass-loaded outflows arise naturally in some analytic \citep{King&Begelman99, Lodato2011} and numerical hydrodynamics models \citep{Sadowski+14, Dai+18} of super-Eddington accretion disks; other simulations
predict relatively low mass outflows \citep{Jiang+19a}.
Our fitted $\dot{m}(t) \propto t^{-1.1}$ relation holds even at late times, when the accretion rate is  sub-Eddington and disk winds are unlikely, suggesting that inefficient circularization is responsible for the evolution of the mass accretion rate.  

Whether or not disk winds drive the evolution of $\dot{m}$, they are a potential ``reprocessing layer'' that might explain the observed optical/NUV emission in TDEs like ASASSN-14li \citep{Metzger2016, Dai+18, LuBonnerot19, BonnerotLu19}. A quasi-static layer of loosely bound gas could also reprocess ionizing X-ray/EUV photons from the central disk into longer wavelengths \citep{Loeb+97, Guillochon+14, Roth+16}. Either of these scenarios would entail a time-dependent absorption column $N_{\rm H}$.

As discussed above, $N_{\rm H}$ in 
ASASSN-14li
decreases to that of the host galaxy plus the Milky Way (\citealt{Miller2015}) after $\sim300$ days (Fig. \ref{14p}).
This evolution is insensitive to the specific pair ($M_\bullet, a_\bullet$) chosen from within the $1\sigma$ CL. To test whether the $N_{\rm H}$ decrease is statistically significant, we refit all ten epochs assuming a constant $N_{\rm H}$.
The new fit yields $N_{\rm H}=4.8\pm0.1 \times 10^{20}{\rm cm}^{-2}$, with $\chi^2$ increasing by 49.7 and a small F-test p value of $1.7\times10^{-6}$. 
Thus, a constant $N_{\rm H}$ is strongly disfavored. 

This result has intriguing parallels with late-time 
ultraviolet photometry of ASASSN-14li.  Roughly 300 days after the start of observations, the multi-band ultraviolet 
light curves transition from a steep decline into a 
nearly constant evolution \citep{Brown+17}.  A single-temperature blackbody fit to these light curves suggests a dramatic flattening in the bolometric luminosity of the optical/NUV photosphere \citep{vanVelzen+19} and in the previously declining fitted photospheric radius \citep{Brown+17}.  

These changes may mark
the slow recession and dilution of a reprocessing photosphere \citep{vanVelzen+19}, culminating in the exposure of a bare accretion disk that emits a faint, slowly evolving far-ultraviolet flux from its outer annuli.  Notably, Fig. 5 of \citet{vanVelzen+19} shows the late-time flattening, and posited transition to disk-dominated emission, occurring at a time $\approx 300$ days after the start of the flare.  Our finding that $N_{\rm H}$ 
asymptotes after this time to a constant value, consistent with that of the host galaxy plus Milky Way, further supports the picture that a gaseous reprocessing layer became increasingly optically thin over the course of ASASSN-14li's first year.  

\subsection{Fitting ASASSN-15oi's x-ray spectra}

ASASSN-15oi is notable for its unusual X-ray light curve. The total X-ray flux (0.3--10 keV) stays roughly constant during the first 100 days, and then increases by a factor of 10 \citep{Gezari2017} about 250 days after discovery. So far, there are two hypotheses to explain this phenomenon, both described in \citet{Gezari2017}. The first is the ``delayed accretion'' theory, in which inefficient circularization of bound debris (e.g., \citealt{Piran+15, Shiokawa+15, GuillochonRamirezRuiz15, Hayasaki+16}) prevents the disk mass accretion rate from tracking the mass fallback rate, with little circularization at early times and more at later times.  The second is the ``variable obscuration'' proposal, which posits a time-dependent absorption column between our line of sight and the hot inner disk of the TDE. If the absorbing column density $N_{\rm H}$ diminishes over time, the X-ray flare will appear to brighten, even if the intrinsic luminosity is steady or falling.  
``Variable obscuration'' is disfavored by \citet{Gezari2017}, who find no evidence for declining absorption.

We suggest a third possible explanation.
As we saw in Fig.~\ref{fig:theoryLCs}, 
late-time X-ray 
brightening can also occur due to evolution in the structure of highly inclined disk.
In this ``slimming disk'' scenario, a nearly edge-on disk with initial $\dot{m} \gtrsim 1$ experiences a decline in its accretion rate, which in turn reduces the aspect ratio $H/r$ and exposes hot, X-ray bright inner annuli to the observer's line of sight.  
In this section, we initially fit the ASASSN-15oi spectra with few priors (our ``General Case'' fit), as we did for ASASSN-14li.  We then add additional priors to systematically study and compare the ``Slimming Disk'', ``Delayed Accretion'', and ``Variable Obscuration'' hypotheses.

For ASASSN-15oi,
there are only two epochs with enough X-ray counts for detailed spectrum fitting; both are from XMM--{\it Newton} and were
reported to be well fit with an absorbed blackbody plus power-law model \citep{Holoien16b}.
However, after careful filtering of background flaring events and background subtraction, we find no evidence for a power-law component related to the source spectrum in either epoch. Because there are only 223 counts for epoch 1 in the 0.3-10 keV band, we employ Cash statistics \citep{Cash1979} in fitting the data. We require that there be at least 1 count per bin. 

The background spectra themselves cannot be well fit by a single power-law model in band 0.3-10 keV. A broken power-law model ({\sc bknpow}) in the 0.3-5 keV range
works better; for
epoch 1, ${\rm Cstat}/\nu=117.66/91$, and, for epoch 2, ${\rm Cstat}/\nu=132.98/97$. We use the broken power-law plus our slim disk model multiplied by the {\sc phabs} model in {\sc XSPEC} to account for absorption and to fit the source and background spectra together. The parameters of the broken power-law component are held fixed to the best-fit values from the background-only fit, except for the normalization, which is permitted to float to allow for differences between the background and source extraction region.


\begin{figure}[ht!]
\includegraphics[height=0.19\textheight]{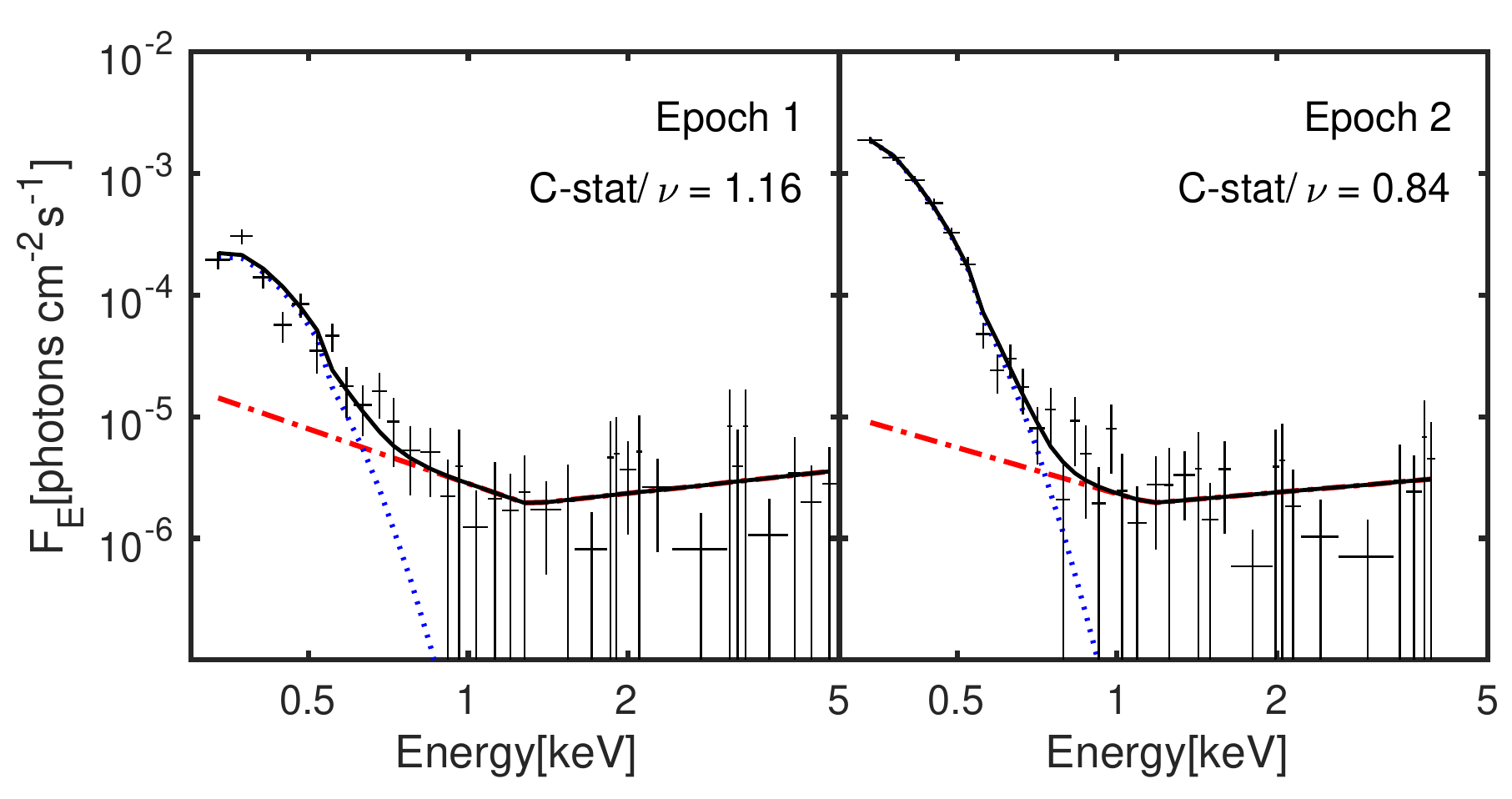}
\caption{XMM--{\it Newton} X-ray spectra of ASASSN-15oi $+$ background light overplotted with our best-fit, General Case model. The observations were obtained on October 29, 2015 and April 4, 2016 (epoch 1 and epoch 2, respectively). As we employ Cash statistics \citep{Cash1979}, we fit the data and the background simultaneously. The blue dotted line displays our disk model, the red dot-dashed line the background broken power-law model, and the solid black line the combined model. The model fits the data $+$ background well, with total ${\rm Cstat}/\nu=62.83/63$. The best-fit ($M_\bullet,a_\bullet$) pair is $(3\times 10^6M_\odot, 0.9)$. The best-fit parameters are listed in the General Case column of Table~\ref{tab:15oi}. 
}
\label{spectrum15oi}
\end{figure}
\begin{deluxetable*}{ccccccccccccc}
\tablecaption{Fitting results for ASASSN-15oi for four scenarios.}
\tablewidth{1pt}
\tablehead{
\colhead{} & \colhead{General} &  \colhead{Slimming} & \colhead{Delayed}& \colhead{Variable}& \\
\colhead{} & \colhead{Case }  & \colhead{Disk} & \colhead{Accretion}& \colhead{Obscuration}&\colhead{} & 
}
\startdata 
$N_{\rm H,1}[10^{20}{\rm cm}^{-2}]^{a}$    & $10.5\pm2.3$  & $14.3\pm5.7$ & $\bm{10.1\pm2.5}$    & $26.1\pm7.9$
& \\ 
$\dot m^b_1[\rm Edd]$         & $0.5\pm0.1$  & $\textbf{4.7}$ & $1.1\pm0.3$       & $\textbf{92}$
& \\
$A_{pl,1}[10^{-6}]$ &$2.8\pm0.5$    &$2.8\pm0.5$  &$3.5\pm0.6$  &$2.8\pm0.5$
& \\
\hline
$N_{\rm H,2}[10^{20}{\rm cm}^{-2}]$  & $8\pm3$   & $3.4\pm1.6$ & $\bm{=N_{\rm H,1}}$    & $10.9\pm1.5$
\\
$\dot m_2[\rm Edd]$                  & $84_{-58}^{+867}$    & $\textbf{1.0}$   & $2.8\pm1.2$         & $\textbf{28}$
 \\
$A_{pl,2}[10^{-6}]$                &$2.3\pm0.4$   &$2.2\pm0.4$   &$2.4\pm0.4$    &$2.6\pm0.4$ 
\\
\hline
$\theta_1[\circ]^{c}$     & $80\pm2$  & $89_{-1.5}^{+1.0}$ & $\bm{60^{+0}_{-55}}$  & $\bm{58.5_{-4.7}^{+1.5}}$ \\
$\beta$           & - & $1.7\pm0.2$& -  & $2.0\pm0.5$
   \\      
$m_{\star}[M_{\odot}]$   & -  & $1.7_{-0.5}^{+5.1}$& -  & $100_{-50}^{+0}$
  \\
$M_\bullet[10^6M_\odot]$    & 3      &4       & 6              & 8
& \\
$a_\bullet$       & 0.9  &0.7 & -0.1          & -0.9
& \\
$Cstat/\nu$                 &62.83/63   &67.1/63 &77.02/64        &75.33/63
&\\
$\Delta AIC$ &0  &4.27 &12.19 & 12.5 \\
\enddata
\tablecomments{The four scenarios considered here are a minimum-prior (``General Case'') scenario and three other idealized scenarios, each detailed in a separate column. Priors, where applied, are marked in {\bf bold}. (a) We set $N_{\rm H, 1}=N_{\rm H, 2}$ for the Delayed Accretion scenario, but allow it to float for the Slimming Disk and Variable Obscuration scenarios. (b) For the Slimming Disk and Variable Obscuration scenarios, we assume that $\dot{m}$ tracks the
theoretical fallback rate $\dot{M}_{\rm t}$ with a peak date of MJD 57248.2; the fitting determines the values of TDE parameters $\beta$ and $m_\star$. In contrast, we allow the accretion rates for the General Case and Delayed Accretion scenarios to float. (c) We set the range of priors for $\theta$ to be $(5^{\circ},60^{\circ})$ and $(5^{\circ},60^{\circ})$ for the Delayed Accretion and Variable Obscuration scenarios, respectively, to prevent disk slimming effects from contributing to their explanatory power. $A_{\rm pl}$ is the normalization of the broken power-law model. The asymmetric errors are calculated using the {\sc XSPEC} {\sc error} command. 
The General Case denotes the best fit under minimal priors from the $\Delta$Cstat minima in the ($M_\bullet$-$a_\bullet$)
plane in Fig. \ref{15oima1}.
The $\Delta AIC$ values are calculated with respect to the General Case, which fits the data best overall. The Slimming Disk scenario also fits the data well.
The Delayed Accretion and Variable Obscuration scenarios fit the data poorly and are disfavored on the basis of their $\Delta AIC$ values. Variable Obscuration is further disfavored by the exceptionally large required $m_\star$.
}\label{tab:15oi}
\end{deluxetable*}

\subsubsection{Minimal-prior, ``General Case'' fit}
\label{sec:generalCase}
We initially fit the two ASASSN-15oi spectra following the same procedure as for ASASSN-14li, with the absorption parameter $N_{\rm H,i}$, accretion rate $\dot m_i$, and inclination $\theta$ (as well as the broken power law normalization) allowed to float freely. In this simultaneous, minimal-prior fit, we require $M_\bullet$, $a_\bullet$, and $\theta$ to be constant between epochs.
We fit the spectra across the ($M_\bullet$, $a_\bullet$) parameter space: $8\times10^5M_\odot\le M_\bullet \le 10^7M_\odot$ and $-0.9 \le a_\bullet \le0.9$. Figure \ref{spectrum15oi} shows the best-fit spectra. The corresponding fitting parameters are in the ``General Case'' column of Table~\ref{tab:15oi}. 

Epoch 1 is fit well with a mildly sub-Eddington accretion rate ($\dot{m}=0.5\pm0.1$), while epoch 2 
prefers a large super-Eddington accretion rate ($\dot{m}=84^{+867}_{-58}$).
The epoch 2 accretion rate implies that $\approx 2 M_\odot$ is accreted during the $\approx 200~{\rm day}$ high-luminosity plateau in the X-ray light curve \citep{Gezari2017}, which requires the disruption of a massive star ($m_\star \gtrsim 4 M_\odot$). 
The best-fit $\theta$ is nearly edge-on, which suggests that the disk height $H/r$ impacts the fitting result when the accretion rate is high. 


Does the accretion rate evolution inferred from the General Case fit make sense? 
The large inferred progenitor star mass
already provides reason to question this fit, as disruptions of high-mass stars that could lead to highly super-Eddington accretion rates are relatively rare. Also troubling is that
our theoretical calculations show that, for the best-fit (nearly edge-on) $\theta$,
the peak X-ray flux in a color-corrected SMBH slim disk is achieved near $\dot{m}\approx 4.5$ (at lower Eddington ratios, the decreasing bolometric luminosity suppresses the X-ray flux, while at higher Eddington ratios, X-rays decline due to the increasing $H/r$). 
If $\dot{m} = 84$ really is the correct description of epoch 2, then at later times, as the accretion rate decreases to $\dot m=4.5$, the X-ray flux should increase to $\approx 1.7\times 10^{-12}$ \flx, i.e., four times higher than for epoch 2. The same peak in X-ray flux should also have been observed between epochs 1 and 2, assuming a continuous increase from $\dot{m}_1 \approx 0.5$ to $\dot{m}_2 \approx 84$.  
Yet this prediction of a double-humped (``Bactrian'') light curve was not seen.

The absence of two peaks in the X-ray light curve, before and after epoch 2, does not definitively
disprove our General Case fit, because of inconvenient gaps in the {\it Swift} record, one of
$\approx 100$ days between epochs 1 and 2, and another of $\approx 200$ days, starting after the
$\approx 150$ day X-ray flux plateau that follows epoch 2 \citep{Gezari2017}.
It is possible that both of our predicted X-ray flux peaks occurred during these two data gaps.
In summary, both the large 
$m_\star$ required by our General Case fit 
and the lack of an observed double peaked light curve
are reasons for doubt,
but a more rigorous test would have been possible with higher-cadence X-ray observations.  
\begin{figure}
\includegraphics[height=.280\textheight]{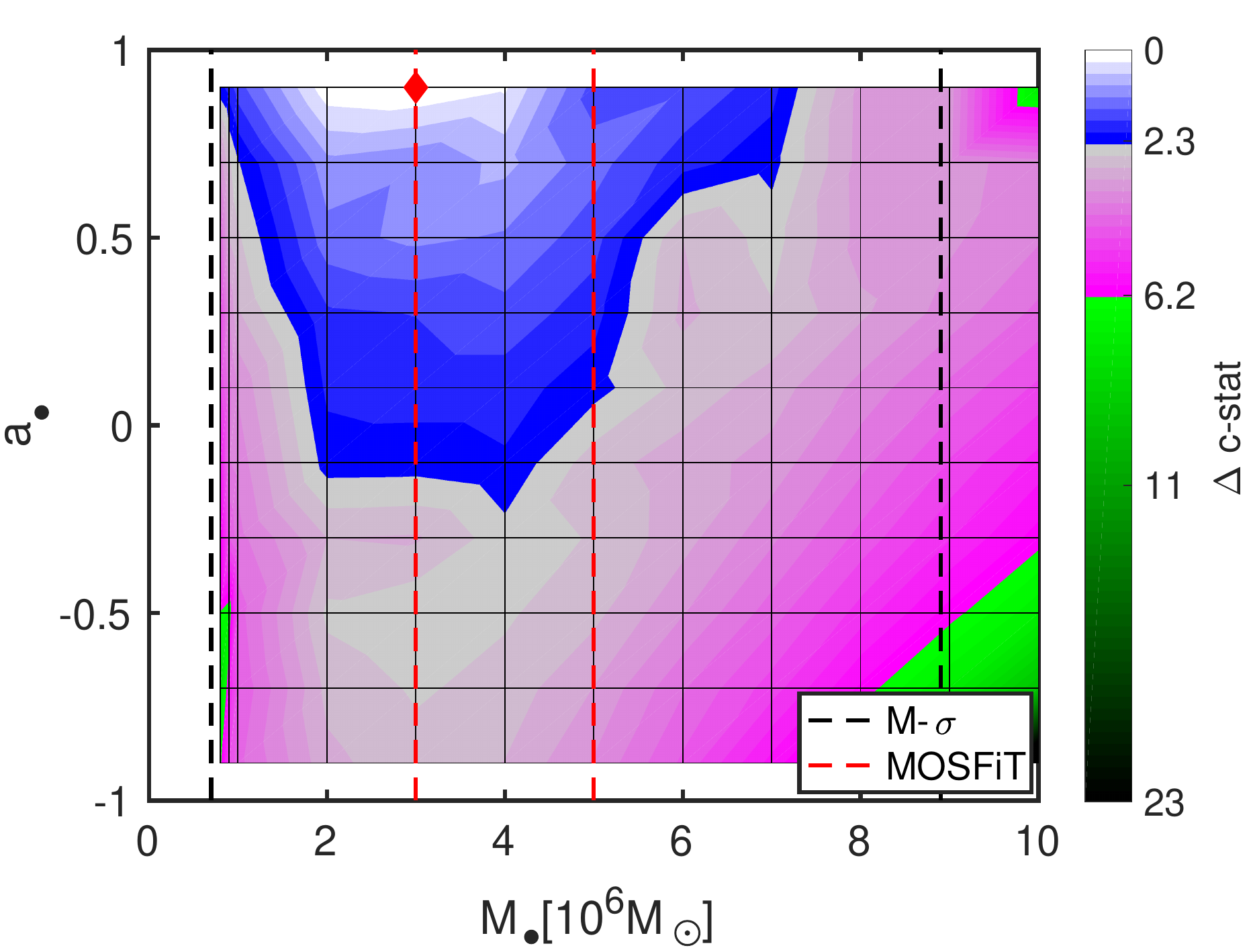}
\caption{Constraints on $M_\bullet$ and $a_\bullet$ in ASASSN-15oi, computed for the minimal-prior, General Case scenario.  Here $\dot{m}$, $N_{\rm H}$, and $\theta$ are allowed to float, but $\theta$ is fixed to the same value between the two XMM-{\it Newton} observing epochs. We calculate $\Delta {\rm Cstat}$ on a grid in the ($M_\bullet, a_\bullet$) plane and then fill in the color contours by linear interpolation. The constraint on the SMBH mass is consistent with 
$M_\bullet= 2.5_{-1.8}^{+6.4}\times 10^6M_\odot$ 
from galaxy scaling relations \citep[][dashed black lines]{Reines2015} and also with $M_\bullet = 4^{+1}_{-1}\times 10^6M_\odot$ from MOSFiT \citep[][dashed red lines]{Mockler2018}. This General Case fit predicts a very high and perhaps problematic second-epoch mass accretion rate (see text).
}
\label{15oima1}
\end{figure}

Figure \ref{15oima1} shows the joint constraints on $M_\bullet$ and $a_\bullet$ for our minimal-prior, General Case fits to ASASSN-15oi. The best fit lies at $(3\times 10^6 M_\odot, 0.9)$. The best-fit parameters are all listed in  Table~\ref{tab:15oi}. 
Past estimates of the SMBH mass for  ASASSN-15oi vary.  
Using a combination of galaxy scaling relations,
\citet{Holoien16b} estimated $M_\bullet \sim 1.3 \times 10^7 M_\odot$. 
Subsequent estimates generally have been lower; 
\citet{Gezari2017} find $M_\bullet=2.5^{+6.4}_{-1.8}\times 10^6M_\odot$, 
and \citet{Wevers+19} get
$M_\bullet=0.5^{+1.4}_{-0.4}\times 10^6M_\odot$.
The range estimated from 
MOSFiT 
is
$4^{+1}_{-1}\times 10^6M_\odot$.
Our marginalized $1\sigma$ CL on $M_\bullet$ is consistent with these independent, external estimates, except for \citet{Holoien16b}, with whom we agree to within $2\sigma$.

\begin{figure}
\includegraphics[height=0.65\textheight]{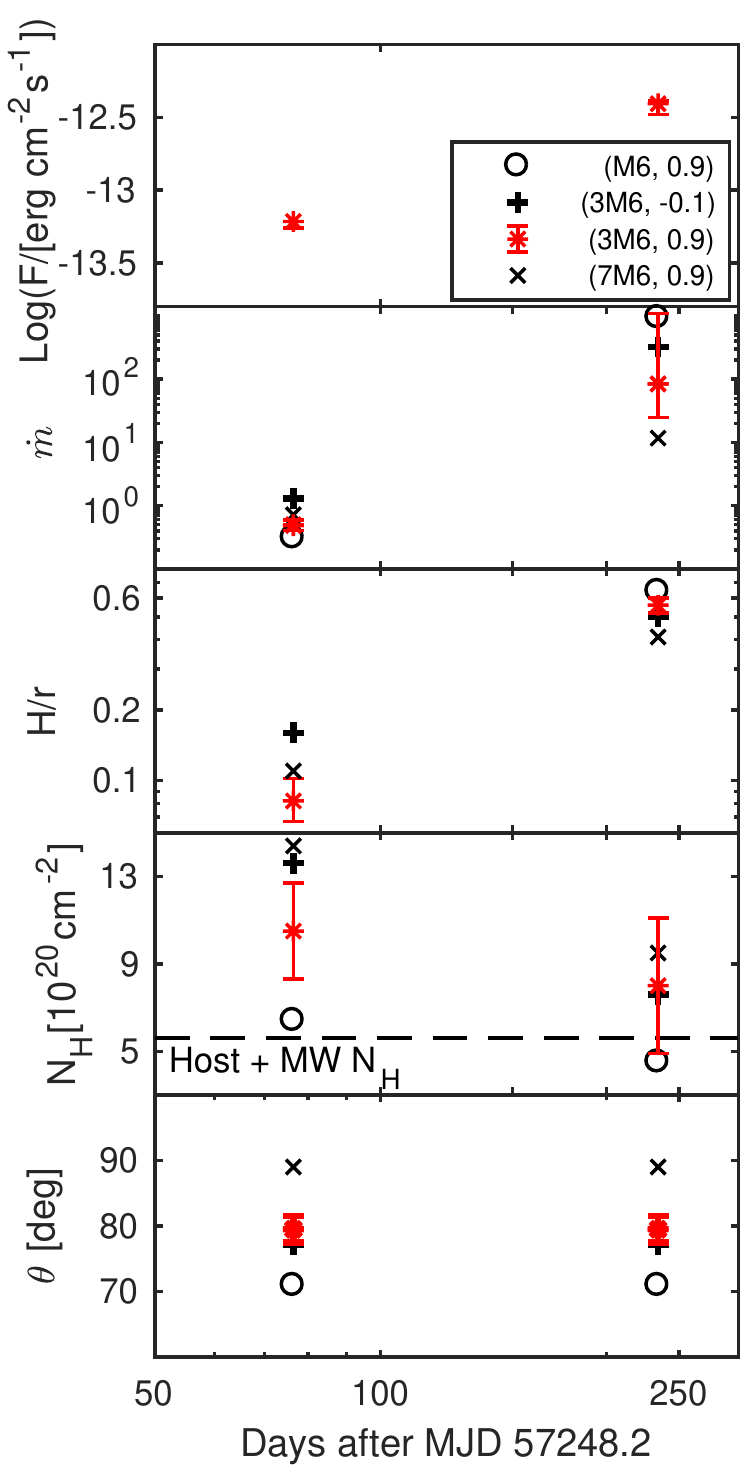}
\caption{Time evolution of best-fit parameters in ASASSN-15oi for the minimal-prior (General Case) scenario. Here the best-fit SMBH mass-spin pair is ($M_\bullet=3 \times 10^6 M_\odot$, $a_\bullet =0.9$).
The total flux $F$, integrated from $0.3-1.0$ keV, increases by a factor of six from epoch 1 to epoch 2 (top panel).
In the remaining panels, we plot the best-fit parameters and the results for three other mass-spin pairs that lie within the $\Delta {\rm Cstat}=2.3$ region in Fig. \ref{15oima1}.
The disk accretion rate in Eddington units (second panel from top) increases significantly between the two epochs, which in turn drives the increase in disk scale height seen in the third panel. The fourth panel shows that, for all combinations of ($M_\bullet, a_\bullet$), the absorption declines between epochs, although not significantly. The best-fit $\theta$ (bottom panel) is large, i.e., closer to edge-on than face-on.
}
\label{figpara15oi}
\end{figure}

Figure \ref{figpara15oi} shows the time evolution of the best-fit parameters, 
including the X-ray light curve derived from the spectral fits to both epochs.
Because the background dominates the modeled disk flux above 1 keV, we only integrate the flux from 0.3-1.0 keV\footnote{We fix the accretion rate when calculating errors, because otherwise the {\sc error} command in {\sc XSPEC} returns inaccurate results.}.
The flux increases by a factor of six from epochs 1 to 2, consistent with the result from \citet{Gezari2017}. 

Our fit to the X-ray spectrum shows that the observed flux at epoch 1 is $6.2\times 10^{-14}$ \flx, while the flux calculated from the best-fit accretion rate and inclination is higher, $3.2\times 10^{-13}$ \flx. From 
ray tracing and running our model through {\sc XSPEC} to account for absorption, 
we find that $1.0\times 10^{-14}$ \flx\ is blocked by the disk edge and $2.4\times 10^{-13}$ \flx\ is absorbed by gas. 
Because both epochs 1 and 2 are affected similarly by absorption, i.e., there is no significant evolution in $N_{\rm H}$ in the best-fit, the
$\sim 6$ times lower flux at epoch 1 relative to epoch 2 is driven primarily by the much lower accretion rate then ($\dot{m}_1 \ll \dot{m}_2$). 

We also explore the effects of allowing epoch 2 to have a lower inclination than epoch 1. We obtain a good fit, with a lower ${\rm Cstat}$ (62.1) than for the General Case, and a best-fit pair of SMBH parameters ($2\times 10^6M_\odot,0.1$). This result hints that precession or realignment of the inner accretion disk in ASASSN-15oi may occur.
Future TDE discoveries will offer ample opportunity to explore disk precession and realignment. In a somewhat edge-on disk, with $\pi/2-\theta \approx H/r$, a small variation in inclination or aspect ratio can be easily detected; the X-ray flux will exhibit large fluctuations in response to small changes in self-shielding.
Studying such reorientation effects is beyond this paper, as our model, which does not yet account for a tilted disk nor ray-trace from a non-equatorial disk, 
cannot make self-consistent predictions.

Another important caveat is our simplistic treatment of the disk outer edge. As described previously, we model the edge of the disk as a single-temperature blackbody and assume that photons are absorbed completely when they intersect with the 3D disk structure. In reality, an accretion disk is not a solid body with infinite optical depth, but instead has a tenuous, optically thin atmosphere some distance above its local height.
An accurate treatment of photon propagation through this atmosphere could alter our fits for highly edge-on models.
For now, we assume that the complicated physics of disk atmosphere self-absorption can be encapsulated in the floating absorption parameter $N_{\rm H}$.


\begin{figure}
\includegraphics[height=.280\textheight]{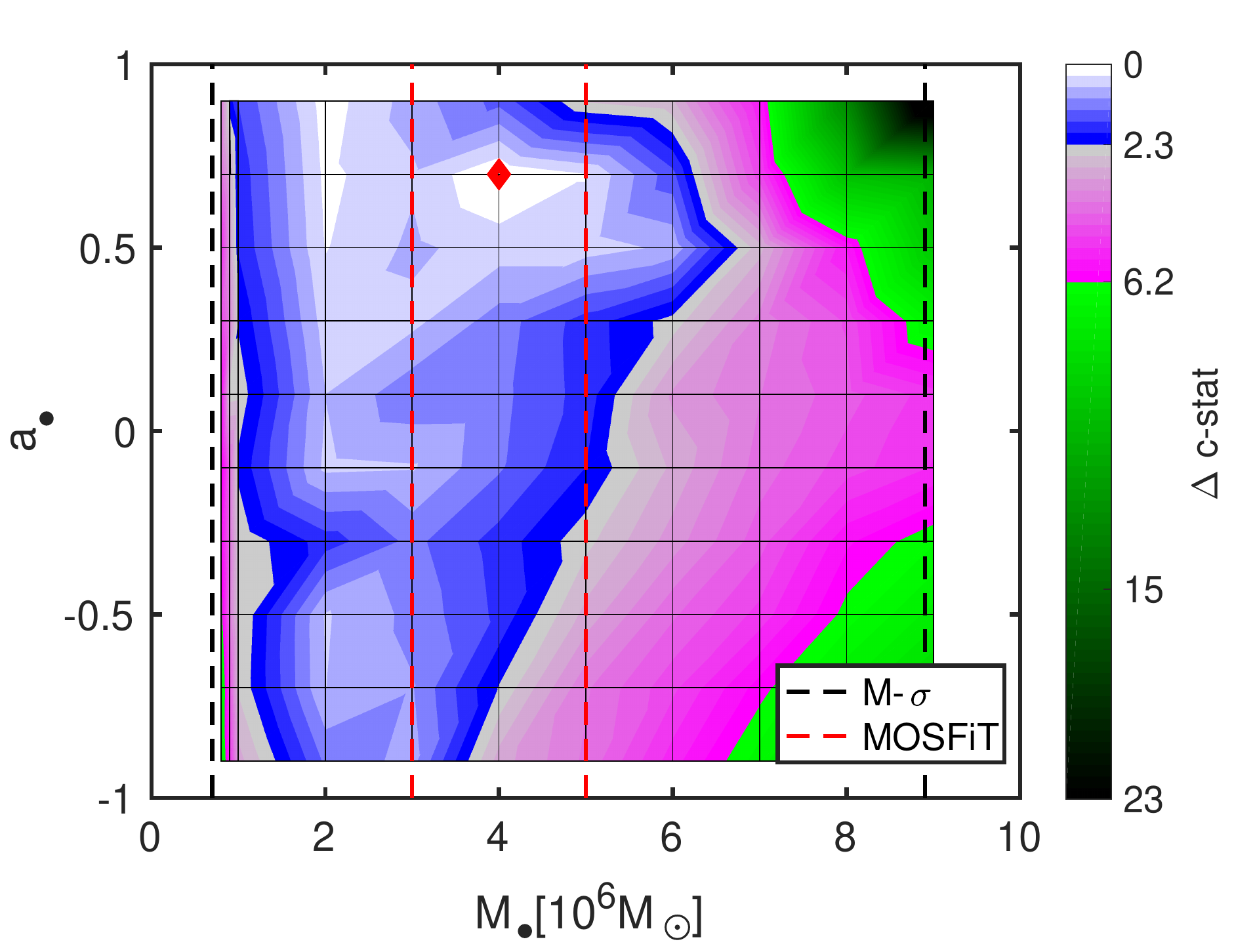}
\caption{Constraints on $M_\bullet$ and $a_\bullet$ for ASASSN-15oi, computed within the ``Slimming Disk'' scenario. Here the accretion rate is forced to follow the theoretical gas fallback rate, and the model responds by selecting a highly edge-on disk configuration. $N_{\rm H}$ is allowed to float between epochs to incorporate the uncertain physics of X-ray absorption by the atmosphere of a nearly edge-on disk. Color contour levels are as in Figure \ref{15oima1}.
The red star denotes the best fit: $M_\bullet=4\times 10^6M_\odot$, $a_\bullet=0.7$. This mass is consistent at the 1$\sigma$ CL with constraints 
from the $M_\bullet-M_{\rm bulge}$ relation (dashed black lines)
and MOSFiT (dashed red lines).
}
\label{15oima}
\end{figure}

\begin{figure}
\includegraphics[height=0.65\textheight]{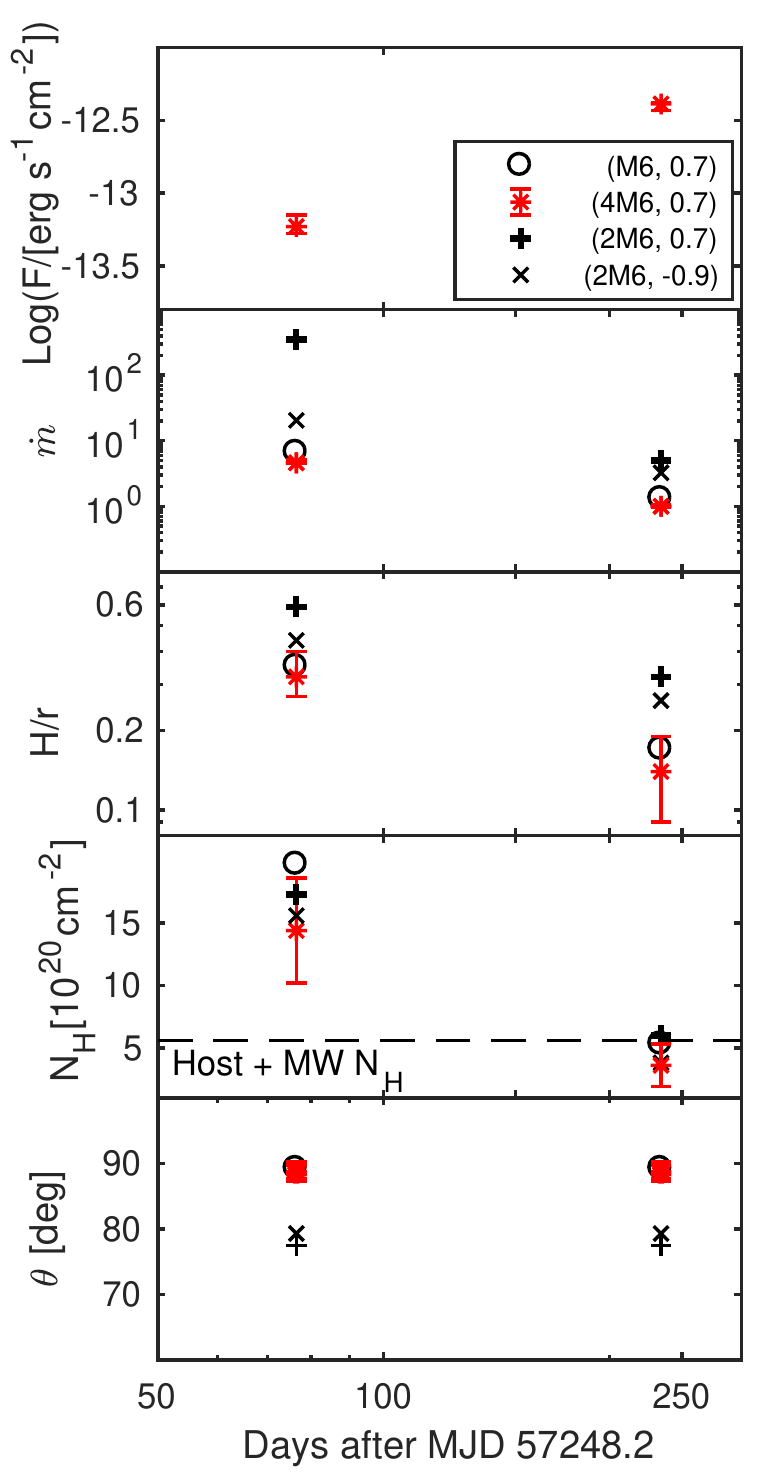}
\caption{Time evolution of best-fit parameters for ASASSN-15oi for the 
Slimming Disk scenario.  
Here the best-fit SMBH mass-spin pair is 
($M_\bullet=4 \times 10^6 M_\odot$, $a_\bullet =0.7$).
The total flux $F$
increases by a factor of six from epoch 1 to epoch 2. 
In the remaining panels, 
we plot the best-fit parameters 
and the results from three other mass-spin pairs that lie within the 
$\Delta {\rm Cstat}=2.3$ region in Fig.~\ref{15oima1}.
In the second panel from the top, the disk accretion rate decays from highly super-Eddington to slightly super-Eddington between the two epochs,
in line with the mass fallback rate prior,
leading to a slimming of the disk (third panel).
For all combinations of ($M_\bullet, a_\bullet$), the Slimming Disk fits prefer a significant decline in absorption between epochs (fourth panel). The best-fit observed inclination is edge-on (bottom panel).
Given the high $\theta$ and decreasing $\dot m$ and $H/r$, the increase in X-ray flux is largely driven by the slimming disk. 
}
\label{figpara15oi2}
\end{figure}

\subsubsection{Testing physically-motivated hypotheses}
\label{sec:hypothesisTesting}

The minimal-prior fit described above
suggests a 
unreasonable accretion rate for epoch 2. 
Therefore, we consider alternative scenarios, with additional priors, to explain the late-time brightening of ASASSN-15oi's unusual light curve:

\begin{enumerate}
\item Slimming Disk: a disk with a declining accretion rate that increases its observed flux through slimming in an edge-on configuration. In this model, 
the mass accretion rate does not float freely, but is forced to follow the mass fallback rate $\dot{M}_{\rm t}$, which requires additional parameters such as the penetration parameter $\beta$ and the progenitor star mass $m_\star$. To account for the uncertain physics of X-rays propagating through an optically thin, edge-on disk atmosphere, we allow $N_{\rm H}$ to float. We do not place a prior on $\theta$, under the expectation that the best fit will involve a high-inclination disk.

\item Delayed Accretion: a disk that increases its luminosity between epochs by increasing its accretion rate, possibly due to delayed circularization of tidal debris \citep{Gezari2017}. Here the mass accretion rate and $N_{\rm H}$ are allowed to float freely as long as $N_{\rm H,1}= N_{\rm H,2}$. We enforce $5 \le \theta \le 60^\circ$ to avoid self-shielding effects associated with an edge-on disk.  

\item Variable Obscuration: a disk that increases its luminosity due to a large decrease in photoelectric absorption, possibly due to the dilution of a reprocessing layer on larger scales \citep{Gezari2017}.  Here we assume that the mass accretion rate follows the mass fallback rate,
$N_{\rm H}$ is allowed to float freely, and $5 \le \theta \le 60^\circ$ is enforced to
eliminate self-shielding.
\end{enumerate}

Two of these idealized models use a theoretical prior for the evolution of the accretion rate, Eq. \eqref{Mf}.  There are three free parameters in this equation, but only two observing epochs, so we fix one parameter, the peak date. We set the time of peak to the discovery date, MJD-57248.2, and allow the other two parameters, $\beta$ and $m_\star$, to float. 
Adopting this prescription for $\dot{M}$ can allow us to test whether the disk accretion rate follows the mass fallback rate, to constrain $\beta$ and $m_\star$, and to tighten constraints on $M_\bullet$ and $a_\bullet$. However, in our exploration of these idealized scenarios, we focus on a simpler question: can the X-ray spectra of ASASSN-15oi be fit by assuming efficient circularization and relying on  evolution in the absorption and/or in the 3D disk structure to produce the observed late-time flux increase?

The Slimming Disk fit answers this question affirmatively.
In Fig.~\ref{15oima}, we plot $\Delta {\rm Cstat}$ values on the ($M_\bullet$, $a_\bullet$) plane. The best-fit ($M_\bullet$, $a_\bullet$) pair is ($4\times 10^6M_\odot$, $0.7$), with ${\rm Cstat}/\nu=67.1/63$. The best-fit results are listed in the Slimming Disk column of Table~\ref{tab:15oi}.  The marginalized, $1\sigma$ CL constraint on $M_\bullet$ is consistent with the range obtained from the  \citealt{Reines2015} relation between $M_\bullet$ and host galaxy luminosity, the $M_\bullet-\sigma$ relationship \citep{Wevers+19}, and 
MOSFiT \citep{Mockler2018}, but is inconsistent with the $M_\bullet-M_{\rm bulge}$ estimate of \citet{Holoien16b}.

This Slimming Disk has more constraining power than our General Case fit and can rule out extreme $M_\bullet$ 
values; e.g., $M_\bullet \gtrsim 6.5\times 10^6 M_\odot$ is excluded at the $1\sigma$ CL.  Unlike for ASASSN-14li, however, 
$a_\bullet$ is unconstrained.
For larger 
SMBH mass, the peak fallback rate (in Eddington units) decreases as $M_\bullet^{-3/2}$, and the radial extent of the disk (in gravitational radii) decreases as $M_\bullet^{-2/3}$. Both scalings decrease the maximum $H/r$ of the disk. 
Consequently, a higher $M_\bullet$ makes it easier for relativistic lensing to propagate X-ray photons to observational sightlines in nearly edge-on disks, thus increasing the X-ray flux from the inner disk.

The Slimming Disk fit also constrains  $\beta=1.7 \pm 0.2$ and $m_\star = 1.7^{+5.1}_{-0.5}M_\odot$. 
Both of these values differ from those inferred from MOSFiT, where $\beta=0.91^{+0.06}_{-0.02}$ and $m_\star=0.11^{+0.04}_{-0.01}M_\odot$. 
This tension is hard to reconcile,
as a smaller $\beta$ and $m_\star$ would drive the accretion rate too low to fit epoch 2. 
Part of this tension may arise from our
assumption that the mass accretion rate follows the mass fallback rate here,
in contrast to the greater number of free parameters in the MOSFiT accretion rates.  Alternatively, the MOSFiT assumptions about the relationship between the accretion rate and the properties of the optical/NUV photosphere may be incorrect.  A joint
analysis of the X-ray and optical/NUV observations will prove valuable in future work.

In Figure~\ref{figpara15oi2}, we plot the evolution of the parameters inferred for the Slimming Disk scenario. Except for the (prior-constrained) behavior of $\dot m$ and $H/r$, the evolution of the other parameters is similar to the General Case. Our calculations show that the observed flux in epoch 1 is $6.1\times10^{-14}$ \flx, while the total 0.3-1 keV flux generated by the best-fit accretion rate and inclination is $3.9\times10^{-12}$ \flx. The bulk of the missing flux, $3.3\times10^{-12}$ \flx, has been blocked by the edge-on disk, while another $5.3\times10^{-13}$ \flx\ has been absorbed by gas (which could be interpreted as the optically thin atmosphere of a nearly edge-on disk). The low flux of epoch 1 is mainly due to the disk edge blocking part of the generated X-rays.

If we assume that the inferred value of $\dot{m}_2 = 1.0$ is constant over
the $\approx 200$ day plateau in the {\it Swift} X-ray light curve, 
we find a lower limit of $\Delta M \approx 0.05 M_\odot$ for the best-fit values of $M_\bullet=4\times 10^6 M_\odot$ and $a_\bullet = 0.7$. Although the lack of temporal coverage makes this estimate quite uncertain, 
it is compatible with the disruption and efficient accretion of a lower main sequence star and appears once again to circumvent the missing energy problem.

Figure~\ref{figpara15oi2} also shows that the fitted absorption parameter $N_{\rm H}$ decreases in time. The fitted observational inclination angle $\theta$ is nearly edge-on. The two results, taken together, agree with the predictions of some accretion reprocessing theories \citep{LuBonnerot19}; for an edge-on configuration, there is more debris in the line of sight at early times. Here, as for
ASASSN-14li, $N_{\rm H}$ 
declines to roughly 
the host galaxy plus Milky Way value after several hundred days, when the accretion becomes sub-Eddington.

We now compare the results of the General Case, Slimming Disk, Delayed Accretion, and Variable Obscuration scenarios. For each scenario, we fit the two spectra through the parameter space $0.8\times 10^5M_\odot\le M_\bullet \le 10^7M_\odot$ and $-0.9 \le a_\bullet\le0.9$. We then use the Akaike information criteria (AIC) \citep{Akaike1974} to test which scenario is favored by the data\footnote{
${\rm AIC}=-2\ln {\cal L}_{\rm max}+2k$,  
where ${\cal L}_{\rm max}$ is the maximum likelihood, and $k$ is the number of free parameters. For Poisson errors, ${\rm Cstat}_{\rm min}=-2\ln{\cal L}_{\rm max}$, while for Gaussian errors, $\chi^2_{\rm min}=-2\ln{\cal L}_{\rm max}$. Generally speaking, two models with $\Delta {\rm AIC}=5$ and $10$ are considered to present strong and very strong evidence, respectively, against the weaker model.}.
Table~\ref{tab:15oi} gives the fitting results and recapitulates the priors. 
The Slimming Disk hypothesis fits the data well, with a ${\rm Cstat}=67.1$ and a $\Delta{\rm AIC}$ of only $4.27$ relative to the General Case fit. On the other hand,
the Variable Obscuration and Delayed Accretion hypotheses, at least in the simple forms tested here, are disfavored. We now discuss these latter two scenarios in greater depth.


For Delayed Accretion, the spectra are best fit with an inclination on the high end of its prior range, suggesting that the fit minimizes ${\rm Cstat}$ (77.02) and $\Delta {\rm AIC}$ (12.19)
only by ingesting some of the underlying physics of the Slimming Disk. The best-fit ${\rm Cstat}$ here is still 7.92 larger than for the Slimming Disk. Unless additional X-ray spectral components are invoked, such as a power law component due to a hot corona, a low accretion rate cannot adequately fit the epoch 1 spectrum at high energies.
The challenge comes from the need to simultaneously fit two spectra with comparable levels of hard ($\approx 0.7$ keV) X-ray emission, but radically different soft X-ray fluxes.  The General Case fit accomplishes this task by using 1) a high $N_{\rm H}$ value to absorb most of the soft X-rays emitted in a low-$\dot{m}$ epoch 1 and 2) an edge-on disk to preferentially obscure hard X-rays in a high-$\dot{m}$ epoch 2.  The Slimming Disk fit succeeds, albeit with a lower quality of fit, by relying on high early-time $N_{\rm H}$, and obscuration by an edge-on disk, to absorb 
X-rays (mostly at softer energies)
from a medium-$\dot{m}$ epoch 1 solution. Later on, a greatly diminished $N_{\rm H}$ and a reduced $H/r$ allow a comparable hard X-ray flux and a much greater soft X-ray flux to emerge from a low-$\dot{m}$ solution in epoch 2.  Without these extra degrees of freedom, the Delayed Accretion scenario fails to fit both epochs satisfactorily.  

This failure does not, however, imply that the accretion rate must track the mass fallback rate precisely. 
Assuming the Delayed Accretion best-fit ($M_\bullet$, $a_\bullet$) pair,
we refit the spectra, allowing $N_{\rm H,1}$ to float freely.  This modification yields a better fit (${\rm Cstat}=67.2$). However, the accretion rate of epoch 1, $\dot m=1.6$, is closer to that of epoch 2, $\dot m=2.7$, and the low flux of epoch 1 is mainly due to absorption.  
The failure of the Delayed Accretion scenario to find a good fit when $\theta\le 60^\circ$ is enforced indicates that its goodness of
fit is being driven mostly by
geometric shielding effects.

The Variable Obscuration scenario is likewise excluded by its large $\Delta{\rm AIC}$ value of 12.5. Another problem is that the best fit selects an extreme progenitor star mass of $m_\star=100^{+0}_{-50} ~M_\odot$, the high bound of the permitted prior range. The heavily absorbed slim disk fits neither the soft spectrum nor the hard spectrum well. 

In summary, our minimum-prior, General Case fits the data best, but with a potentially problematic late-time accretion rate,
followed by our proposed idealized Slimming Disk scenario. The two other simplified scenarios, 
Delayed Accretion and Variable Obscuration,
are disfavored.
Some degree of geometrical change, such as the diminishing $H/r$ that reduces the self-shielding of an edge-on disk in the Slimming Disk scenario, is required for these other scenarios to achieve better fits.

\section{Conclusions}
\label{Conclusions}

We have developed a slim disk model for the accretion disk formed
in a TDE, after a star is tidally disrupted by the central SMBH in a galaxy. We use a sequence of stationary slim disks to model different snapshots in time.
We then calculate the local X-ray flux assuming that a spectrally hardened blackbody spectrum is emitted by each annulus of the inner accretion disk. 
Finally, we employ a general relativistic ray-tracing code to generate synthetic X-ray spectra. 
With these model spectra, we simultaneously fit the multi-epoch XMM-{\it Newton} X-ray spectra of two well-studied TDEs, ASASSN-14li (10 epochs) and ASASSN-15oi (2 epochs)\footnote{The disk solutions and synthetic spectra for 
ASSASN-14li and -15oi are available on GitHub at 
https://github.com/wensx/Continuum-Fitting-the-X-ray-Spectra-of-Tidal-Disruption-Events.}.

Our main findings are:
\begin{enumerate}
\item 
We first use our model to
gain intuition about 
the expected behavior of the X-ray spectra and light curve for different values of SMBH mass, SMBH spin, disk inclination, and disk viscosity. We generate synthetic X-ray data, assuming that the mass accretion rate onto the slim disk is consistent with the mass fallback rate. We find that

a) smaller $M_\bullet$ produces more hard X-ray flux but less soft X-ray flux, as well as a light curve that remains roughly constant in a super-Eddington mode for longer; b) higher 
spin $a_\bullet$ produces a stronger flux in both hard and soft X-rays, in addition to a more slowly decaying light curve; c) a nearly edge-on disk (large $\theta$) screens the
inner X-ray emission, removing higher energy photons from the sightline. The height of this edge-on disk decreases over time due to the declining accretion rate, exposing more inner disk X-rays to the sightline and producing a light curve with a late time flux increase; we call this process the ``Slimming Disk'' scenario.

\item 
Our slim disk model successfully fits the observed X-ray spectra of ASASSN-14li and ASASSN-15oi. 
For ASASSN-14li, we allow the disk accretion rate to float. The best simultaneous fit to the spectra at all 10 epochs indicates that, during the first three epochs, the accretion is super-Eddington. 
This fit yields a SMBH mass and spin of $M_\bullet=10^{+1}_{-7}\times 10^6M_\odot$ and $a_\bullet>0.3$, respectively. This constraint on $M_\bullet$ is consistent with earlier independent determinations: $M_\bullet= 6.3_{-5.7}^{+6.3} \times 10^6M_\odot$ from galaxy scaling relations 
\citep{van2016, Holoien2016a, Wevers2017} 
and $9^{+2}_{-3}\times 10^6M_\odot$ from fitting the optical and UV light curves (MOSFiT; \citealt{Mockler2018}).
Our result is the first direct constraint on $a_\bullet$ for ASASSN-14li. 
When combined with the independent mass constraints from MOSFiT, the spin constraint narrows to $>0.85$.

\item
For ASASSN-15oi, our spectral fits are limited to two epochs. If we assume that the mass accretion rate follows the mass fallback rate, the best simultaneous fit is consistent with the edge-on ``Slimming Disk" scenario described above.
This fit yields $M_\bullet=4.0^{+2.5}_{-3.1}\times 10^6M_\odot$, but fails to constrain $a_\bullet$ due to the low number of observed epochs.  This mass is consistent with  $2.5^{+6.4}_{-1.8}\times 10^6M_\odot$ 
inferred from galaxy scaling relations \citep{Reines2015, Gezari2017, Wevers+19} and with $4^{+1}_{-1}\times 10^6M_\odot$ from MOSFiT.
Our ``Slimming Disk" fit provides
a new explanation for the late time peak in the X-ray light curve.
In contrast, our fits with priors corresponding to the
previously proposed ``Delayed Accretion'' 
and ``Variable Obscuration'' explanations \citep{Gezari2017} fit the data poorly. 


\item
We also explore the time evolution of other important model parameters.  For ASASSN-14li, the disk accretion rate decays from $\dot m=3$ to $\dot m=0.3$ as function of $t^{-1.1}$, which is insensitive to the specific ($M_\bullet, a_\bullet$) pair and $f_{\rm c}$ choice.  Both $\chi^2$ and F-test statistics strongly disfavor theoretical models where the disk accretion rate tracks the mass fallback rate, suggesting inefficient circularization of returning material.  The X-ray light curve does not directly track the accretion rate either, but declines as $\propto t^{-2.8}$ in the later epochs, after about 200 days.   

\item For ASASSN-14li, the equivalent hydrogen column density $N_{\rm H}$ declines to the value of the host galaxy + Milky Way after roughly three hundred days, which is also the time period when the fitted disk accretion rate declines from super- to sub-Eddington. This behavior is reminiscent of the optical and NUV light curves of this flare \citep{Brown+17, vanVelzen+19}, which at first decline steeply, but then flatten at a similar time, asymptoting to a constant fitted blackbody radius. This result hints that we have observed the dilution and recession of a reprocessing layer until it becomes too optically thin to screen the inner accretion disk, a hypothesis that must be explored with future radiative transfer calculations.  For ASASSN-15oi, $N_{\rm H}$ also declines to the host galaxy + Milky Way value after a few hundred days, as the disk goes into the sub-Eddington phase, suggesting that this timescale may be characteristic for obscuring gas depletion or removal in TDEs.

\item We study the effects of a spectral hardening factor $f_{\rm c}$ in the fitting of ASASSN-14li, finding that a) adopting
the semi-empirical $f_{\rm c}$ model from \citet[DE19]{DE2018} can fit the spectra from super- to sub-Eddington accretion phases; b) different $f_{\rm c}$ treatments have no significant impact on the $M_\bullet$ estimation; c) $f_{\rm c}$ is strongly degenerate with inclination,
so it is reasonable to fix $f_{\rm c}$
with the DE19 model.

\item Our model-fitting procedure allows us to directly estimate the time-dependent accretion rate onto the SMBH and thus place a lower limit on the total accreted mass.
Over the $\approx 1100$ days of X-ray observations
for ASASSN-14li, the SMBH has accreted $\Delta M \approx 0.17 M_\odot$, equivalent to the disruption of a $> 0.34 M_\odot$ star,
suggesting that the ``missing energy problem'' \citep{Stone&Metzger16, Piran+15} is not in fact a problem for this TDE.  A similar calculation for ASASSN-15oi is much less accurate due to the lower number of XMM-{\it Newton} observing epochs, but implies $\Delta M \gtrsim 0.04 M_\odot$.

\end{enumerate}

Our model features several idealizations, but the most important is the assumption that the inner, X-ray emitting regions of TDE disks can be modeled with quasi-circular, equatorial, 1D stationary disk solutions.  Real TDEs are 
complex 3D affairs, and the manner in which dynamically cold debris streams dissipate energy to form an accretion flow remains an open problem.
Yet the circularization process is likely to produce quasi-circular accretion flows efficiently in at least some TDEs, if $R_{\rm p} \sim R_{\rm g}$ and relativistic apsidal precession force stream self-intersections deep in the SMBH potential \citep{Hayasaki+13, Dai+15}.  Even if self-intersections occur at great distances and circularization is inefficient, the innermost annuli of the resulting eccentric accretion flow may dissipate excess energy internally, leading to quasi-circular inner annuli well-represented by a slim disk. 
The general success of our slim disk models at fitting multiple TDE X-ray spectral epochs suggests that ASASSN-14li and ASASSN-15oi 
may be well described by one of these explanations.
Theoretical progress on the full problem of tidal debris evolution would aid our interpretation of these flares. 

Other approximations that should be improved in future continuum fitting efforts include treatment of the spectral hardening factor $f_{\rm c}$ (which is theoretically quite uncertain for super-Eddington accretion) and the emission from nearly edge-on disks. Our investigation of different $f_{\rm c}$ treatments is encouraging; the different prescriptions implemented here have no significant impact on $M_\bullet$ estimation. We have approximated the propagation of X-ray photons out of a nearly edge-on disk with a single absorption parameter, $N_{\rm H}$, but it is possible to construct more realistic models for optically thin regions above the disk photosphere.

Even with these idealizations, our model is the most detailed and self-consistent yet applied across the parameter space of TDE X-ray spectral analysis and fitting. (See \citealt{Dai+18} for a hydrodynamics and radiative transfer approach to an individual set of TDE parameters.)  Fully relativistic ray-tracing has allowed us to quantify the novel phenomenon of X-ray brightening due to disk slimming, as well as to constrain masses and spins of the SMBHs that power TDE X-ray emission.  

\section*{Acknowledgements}
We thank the anonymous referee for their helpful comments.
We are also indebted to M.A.~Abramowicz, A.~S{\k a}dowski, I.~Hubeny,
S.~van Velzen, B.~Metzger, W.~Lu, K.~Auchettl, L.~Kuiper, J.~de Plaa, and A.~Mummery for their guidance and suggestions.
SW thanks Steward Observatory and the UA Department of Astronomy for post-doctoral support. During the early part of this work, SW was supported by the International Program for Ph.D. Candidates, Sun Yat-Sen University.
NCS received financial support from NASA, through both Einstein Postdoctoral Fellowship Award Number PF5-160145 and the NASA Astrophysics Theory Research Program (Grant NNX17AK43G; PI B. Metzger). He also received support from the Israel Science Foundation (Individual Research Grant 2565/19). 
PGJ acknowledges support from European Research Council Consolidator Grant 647208.
AIZ thanks the Center for Cosmology and Particle Physics at New York University and the DARK Cosmology Centre at the Niels Bohr Institute, University of Copenhagen, for their support and hospitality. 
Our calculations were carried out at UA on the El Gato and Ocelote supercomputers, which are supported by the National Science Foundation under Grant No.~1228509.
Discussions during the 2020 Yukawa Institute for Theoretical Physics (YITP) workshop on ``Tidal Disruption Events: General Relativistic Transients'' at Kyoto University were critical to the completion of this work.
This paper makes use of data from XMM-{\it Newton}, an ESA science mission with instruments and contributions directly funded by ESA Member States and NASA.

\begin{appendix}

\section{Spectral hardening factor}
\label{app:14li}

\begin{deluxetable*}{ccccc}
\tablecaption{Testing $f_{\rm c}$}
\tablewidth{0pt}
\tablehead{
$f_{\rm c}$   & 1 & 2 & 3 &4}
\startdata
$f_{c1}$   & $f_{ca}\cdot f_{\rm c}^\star$ & (1,  $f_{\rm c}^\star$) & (1,  $f_{\rm c}^\star$) & (1,  $f_{\rm c}^\star$)\\
$f_{c2}$  & {\rm N/A} & {\rm N/A} & 1.7 & (1, 2)  
\enddata
\tablecomments{Priors for four models for the spectral hardening factor that are different than the DE19 $f_{\rm c}$ model assumed in the main text. 
All use $f_{\rm c1}$ to fit the super-Eddington (first three) spectra, and $f_{\rm c2}$ to fit the sub-Eddington (last seven) spectra, of ASASSN-14li. One theoretical upper limit on spectral hardening is $f_{\rm c}^{\star}=(72/T_{\rm keV})^{1/9}$ \citep{Davis2006}, but, in Model 1, we assume this limit can be circumvented and fit $f_{\rm ca}$ as a free parameter. Models 1 and 2 are designed to test whether our fiducial model is robust for super-Eddington phases of accretion; thus, they do
not include sub-Eddington epochs. Models 3 and 4 test the effects of $f_{\rm c}$ on parameter estimation and therefore apply to all epochs.  Parentheses indicate a flat prior in between two limiting values.}
\label{tab:tfc}
\end{deluxetable*}

In this paper, we have assumed 
that the spectral hardening factor $f_{\rm c}$ 
parameterization
from DE19 applies to all observational epochs. This model 
was derived for thin disks (i.e., sub-Eddington accretion) and non-spinning SMBHs, so we must be careful in applying it to TDEs, particularly during the early-time 
super-Eddington accretion phase
and given the possibility of large SMBH spin.
Here we consider
1) if the DE19 $f_{\rm c}$ model 
is valid in the super-Eddington phase, 2) the effect of $f_{\rm c}$ on 
parameter estimation, e.g.,  
$M_\bullet$ and $a_\bullet$, 
and 3) the degeneracy of $f_{\rm c}$ with $\theta$. 

Early studies of X-ray spectral hardening showed that $f_{\rm c}$ is about 1.7 and insensitive to disk parameters \citep{ST93,ST95} for sub-Eddington, geometrically thin accretion disks. Later studies \citep{MFR2000,GD04,DBHT05} indicated that $f_{\rm c}$ may vary from $\sim$1.4 to $\sim$2 and be affected by accretion rate and SMBH mass. Newer theoretical work suggests that $f_{\rm c}$ saturates at a value $f_{\rm c}^\star$ as the effective temperature increases \citep{Davis2006}, even in a super-Eddington phase.
We will now repeat our earlier work, simultaneously fitting the spectra of ASASSN-14li over multiple epochs,
except 
that here we will employ two different prescriptions for $f_{\rm c}$: one for super-Eddington accretion rates ($f_{\rm c1}$) and another for sub-Eddington accretion rates ($f_{\rm c2}$).  

In \S \ref{14lipi}, we found that the first three spectra of ASASSN-14li are best fit by a super-Eddington accretion rate. Thus, to test whether the assumed
$f_{\rm c}$ model
is valid for super-Eddington accretion, we first examine whether 
a better fit is achieved by exceeding
the theoretical saturation value $f_{\rm c}^\star$.  We do this by parameterizing $f_{\rm c1}=f_{\rm ca}f_{\rm c}^\star$ (``Model 1''), where $f_{\rm ca}$ is a free parameter. We then use this model to fit the first three spectra for ($M_\bullet, a_\bullet$) pairs within the $1\sigma$ contour of Fig. \ref{14lima}, thereby disregarding the sub-Eddington epochs. We fix the accretion rate to the best-fit values that we found in \S \ref{14lipi}. Because we fixed the accretion rate, we allow different epochs to adopt different values of $f_{\rm c1}$.  

In a related test (``Model 2''), we allow all the parameters to float, including $N_{\rm H}$, $\theta$ (all the ten epochs have the same inclination), $\dot{m}$, and $f_{\rm c1}$ (over the range 1.0 to $f_{\rm c}^\star$). Then, we re-fit the first three spectra with parameters $M_\bullet$ and $a_\bullet$ taken from within the $1\sigma$ contour of Fig. \ref{14lima}.

To test the effect of $f_{\rm c}$ prescriptions on parameter estimation (specifically, on $M_\bullet$ and $a_\bullet$), we again allow $f_{\rm c1}$ to float within 1.0 and $f_{\rm c}^\star$, but assume two different prescriptions for $f_{\rm c2}$. In ``Model 3,'' $f_{\rm c2}$ is fixed at 1.7. In ``Model 4,'' $f_{\rm c2}$ is allowed to float between 1.0 and 2.0. In analyzing all four of these non-fiducial models, we use the AIC \citep{Akaike1974} to determine which 
prescriptions for $f_{\rm c}$ 
are more favored by the data. 
The priors for Models 1-4 are listed in Table \ref{tab:tfc}.

\begin{figure*}[ht!]
\gridline{\fig{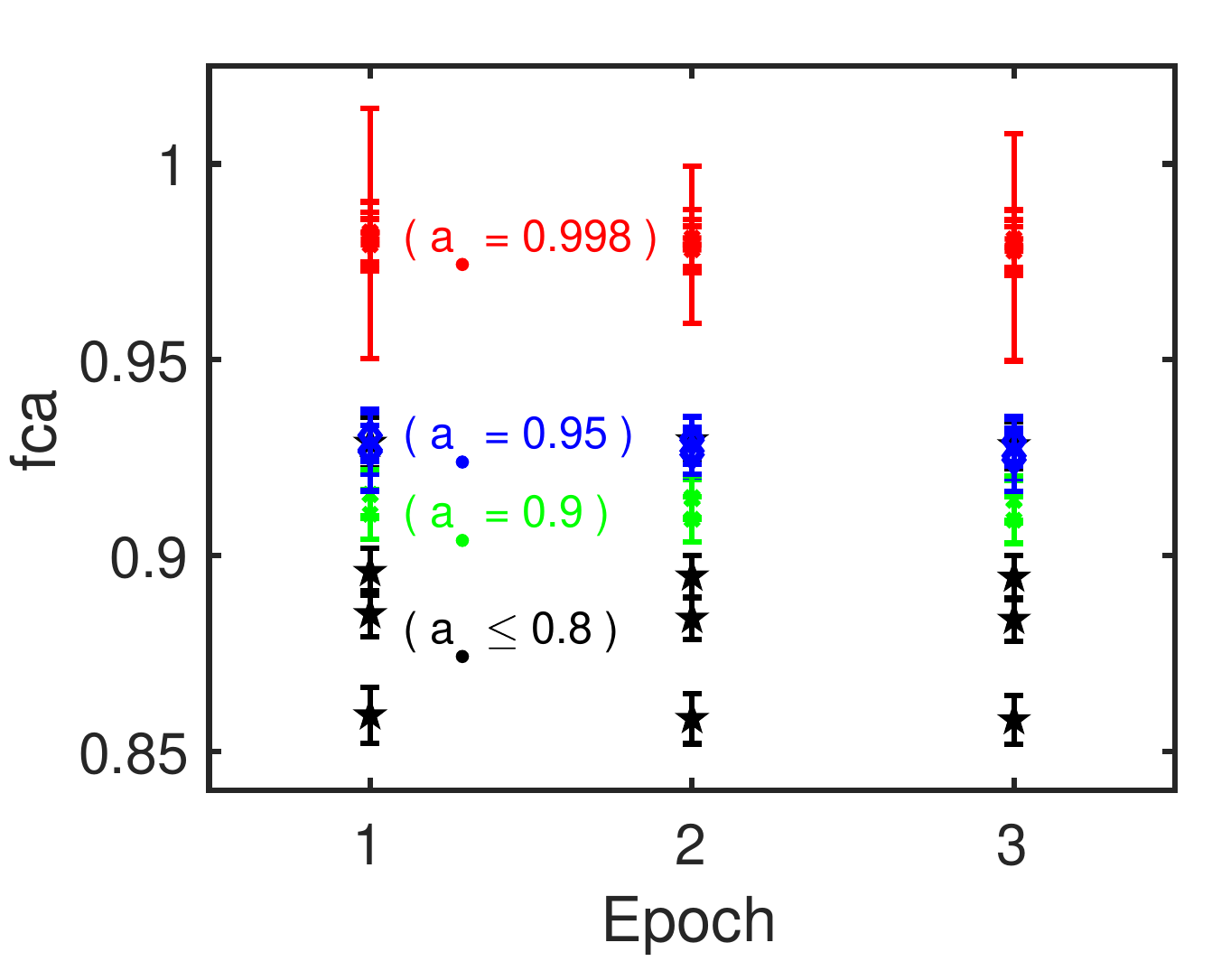}{0.45\textwidth}{Model 1: $f_c=f_{ca} f_c^\star$.}
          \fig{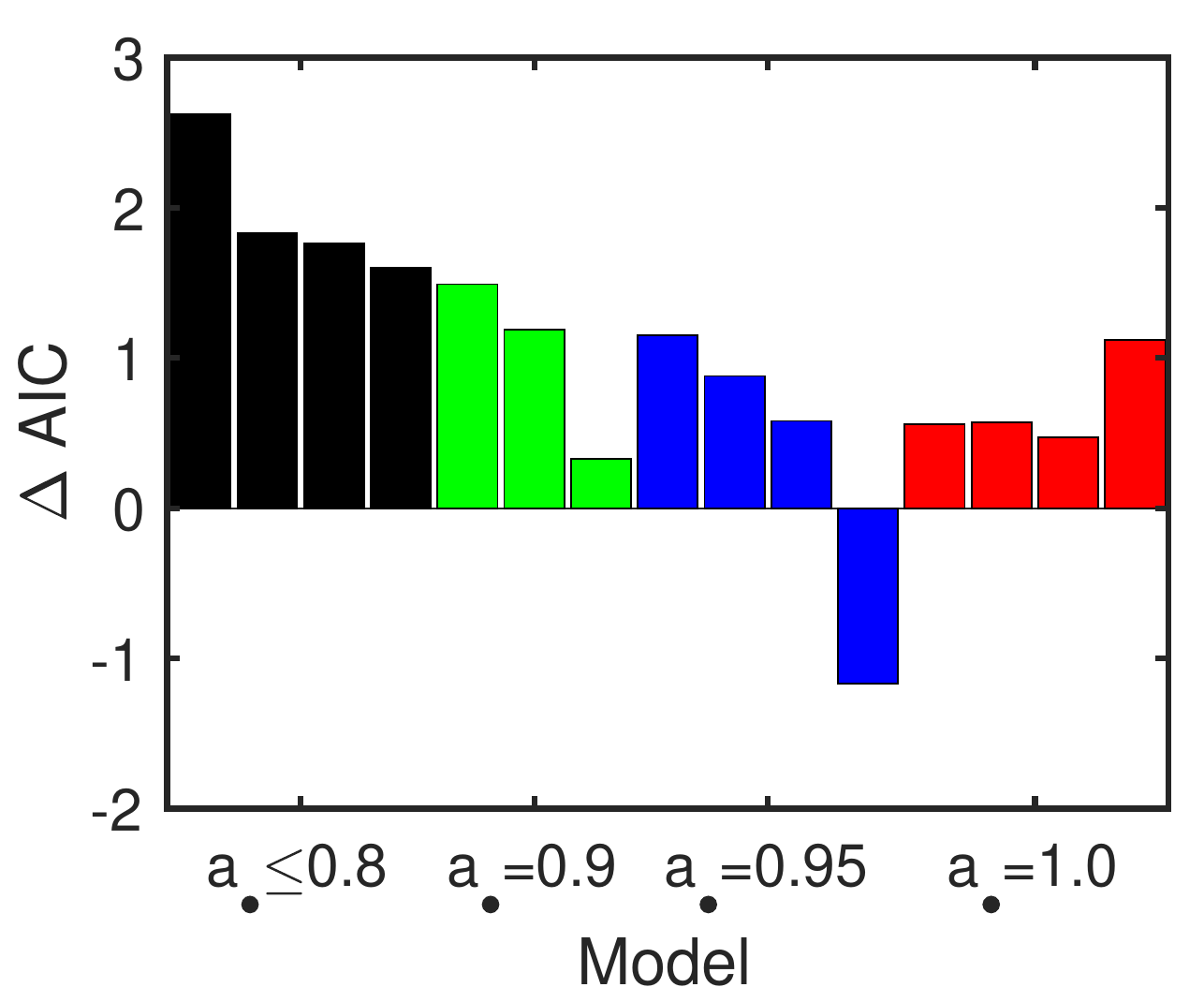}{0.44\textwidth}{Model 2: $f_{c1}$ is allowed to float.}
          }
\caption{Testing $f_{\rm c}$ prescriptions on the super-Eddington (first three) spectra of ASASSN-14li. The left panel tests whether the $f_{\rm c}$ predicted from the DE19 model exceeds the theoretical upper limit $f_{\rm c}^{\star}=(72/T_{kev})^{1/9}$ \citep{Davis2006} (Model 1, see text; Table \ref{tab:tfc}). The right panel compares the fit quality of the DE19 model with Model 2 (see text; Table \ref{tab:tfc}).  We fit the spectra for 
($M_\bullet,a_\bullet$) pairs
within the $1\sigma$ contour 
in Fig. \ref{14lima}.  $\Delta AIC$ is calculated with respect to the  DE19 model. The left panel shows that $f_{\rm c}$ in the DE19 model does not exceed $f_{\rm c}^\star$ for super-Eddington accretion epochs. The right panel shows that there is no significant difference in fit quality between assuming the DE19 model and the less constrained $f_{\rm c}$ of Model 2.}\label{fcTest}
\end{figure*}

The left panel of Fig. \ref{fcTest} shows the results for Model 1. The best-fit $f_{\rm ca}$ is less than 1, which means that $f_{\rm c}$ does not exceed the theoretical upper limit of $f_{\rm c}^\star$, which is about 2.2 for high spin SMBH's of $10^7 M_\odot$. This fitting shows that $f_{\rm c} \sim 2.1$ at the super-Eddington limit for $M_\bullet \sim 10^7 M_\odot$ with the prescriptions of Eq. \ref{fc2}. We also see that $f_{\rm ca}$ decreases with spin, because lower spin yields lower effective temperature at a fixed Eddington level.

The right panel of Fig. \ref{fcTest} shows the result of AIC testing for Model 2 with 
fifteen ($M_\bullet, a_\bullet$) pairs. Here, $\Delta {\rm AIC}$ is calculated with respect to the DE19 model for $f_{\rm c}$. Model 2 yields similar $\chi^2$ values as the DE19 model. As Model 2 has one more free parameter, $\Delta {\rm AIC}$ is worsened accordingly. Fourteen of the ($M_\bullet, a_\bullet$) pairs have $\Delta {\rm AIC} > 0$, 
so allowing $f_{\rm c}$ to float is unnecessary. This test demonstrates that if the spectrum can be well fit by the DE19 $f_{\rm c}$ model, allowing $f_{\rm c}$ to float at super-Eddington epochs would not yield a better fit. However, Model 2 would enable more ($M_\bullet, a_\bullet$) pairs to fit the spectrum, because it allows $f_c$ to adopt a value bigger than the DE19 one. We stress that, although 
the alternate $f_{\rm c}$ priors explored in Models 1 and 2 have no significant impact on $\chi^2$, they can impact the parameter estimation; for example, a bigger $f_{\rm c}$ would result in a smaller $\dot m$.



\begin{figure*}

\gridline{\fig{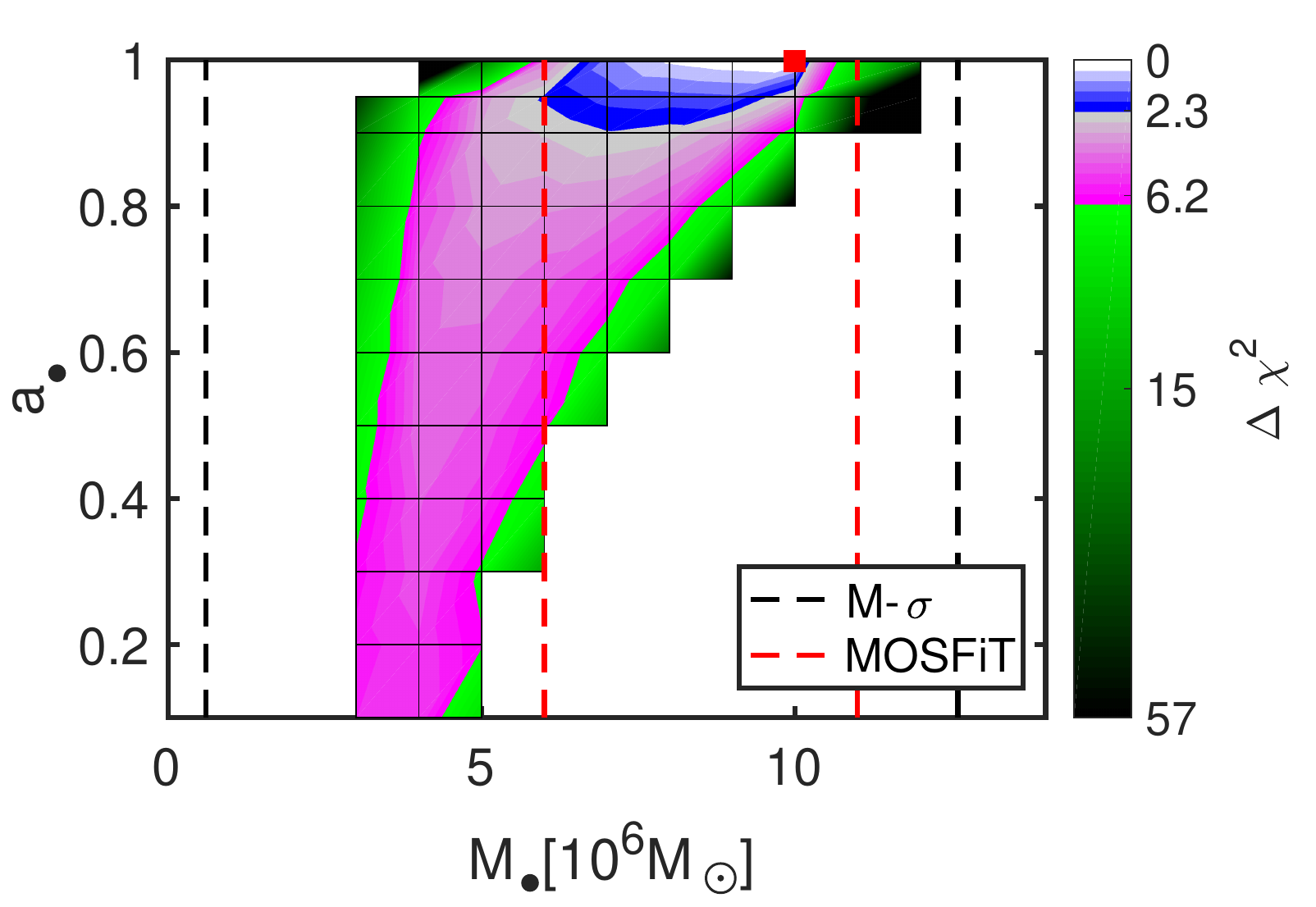}{0.5\textwidth}{Model 3: $f_{c2}=1.7$. }
          \fig{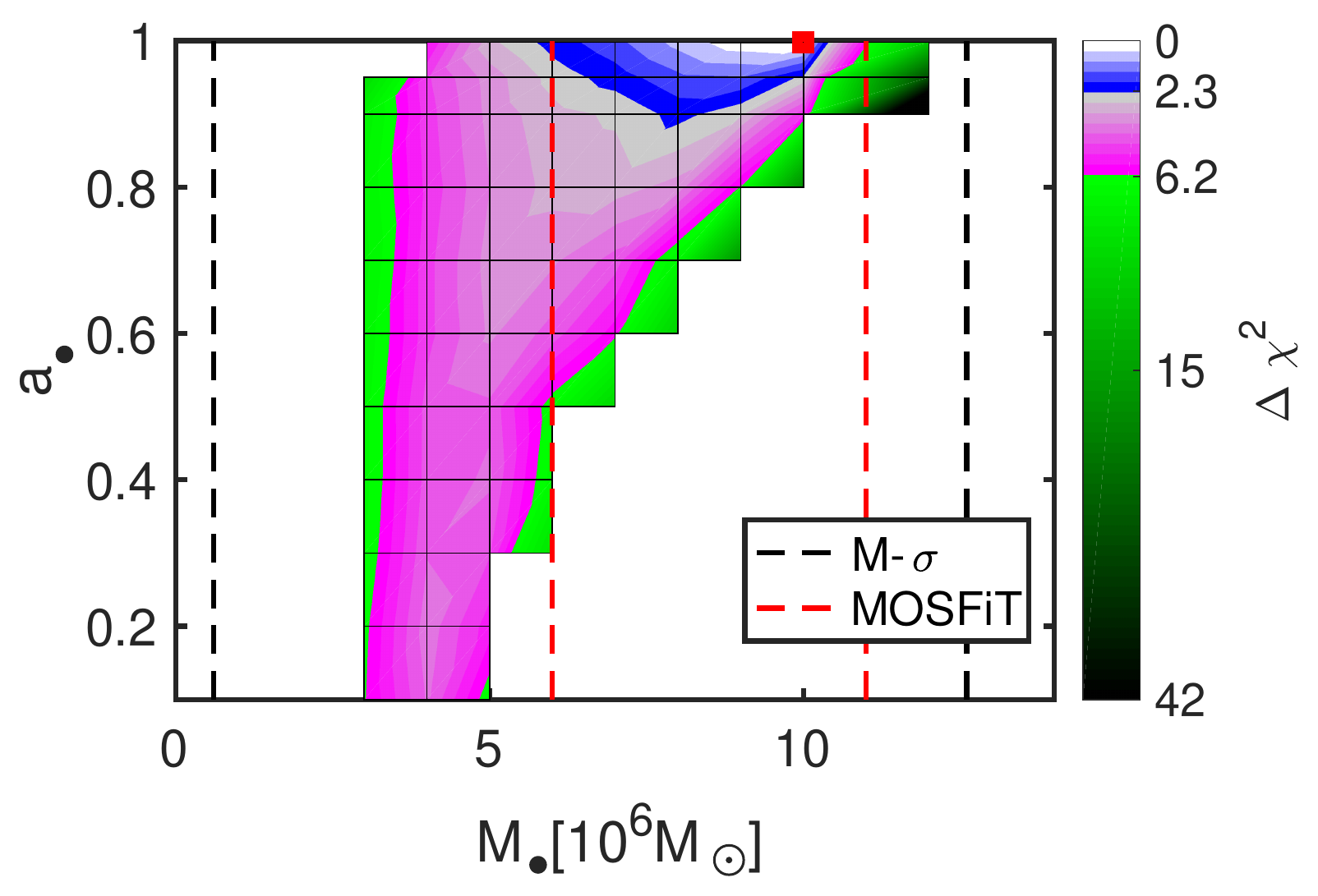}{0.5\textwidth}{Model 4: $f_{c2}<2.0$ is allowed to float.}
          }
\caption{Exploring the impact of $f_{\rm c}$ on estimating $M_\bullet$ and $a_\bullet$ for ASASSN-14li.
Model 3 (left panel) and Model 4 (right), which allow different spectra hardening factors for super- and sub-Eddington accretion, unlike Models 1 and 2 (see text and 
Table \ref{tab:tfc}).  In each panel, the red star denotes the best fit (with $\chi^2$ values of 4360.05 and 4359.81, respectively). The blue region denotes the $1\sigma$ CL contour and the pink the $2\sigma$ CL contour. The constraints on $M_\bullet$ are consistent with both the MOSFiT value of $9^{+2}_{-3} \times 10^6M_\odot$ 
and the range of
estimates from galaxy scaling relations,
$(0.6-12.5)\times 10^6M_\odot$
\citep{van2016,Holoien2016a,Wevers2017}. 
The parametrization of $f_{\rm c}$ 
has little impact on $(M_\bullet, a_\bullet)$ constraints at the 2$\sigma$ level, but these alternate models yield tighter constraints than the DE19 model, particularly on $a_\bullet$ at the $1\sigma$ level.}
\label{14lifc}
\end{figure*}

Figure \ref{14lifc} shows the impact of $f_{\rm c}$ prescriptions on parameter estimation (Models 3 and 4). In these analyses, except for the priors on $f_{\rm c}$, all other parameters are treated the same way as in the General Case (\S \ref{sec:AS14li}; see Fig. \ref{14lima} to make a direct comparison). In Models 3 and 4, the constraint on $M_\bullet$ is still consistent with both the $3^{+7}_{-2}\times 10^6M_\odot$ estimate from galaxy scaling relations \citep{van2016} and the  $9^{+2}_{-3}M_\odot$ range estimated from MOSFiT.
With one more free parameter, the best-fit value of $\chi^2$ for Model 4 is only 0.24 smaller than that of Model 3. 
The AIC comparison shows that the additional parameterization of $f_{\rm c2}$ is unnecessary. 
Models 3 and 4 yield stronger constraints on both $M_\bullet$ and especially $a_\bullet$ at a $1\sigma$ CL, perhaps by releasing a tension forced by overly restrictive priors in the DE19 model. We investigate this issue by studying the degeneracy of $f_{\rm c}$ and $\theta$.


\begin{figure}
\includegraphics[height=.30\textheight]{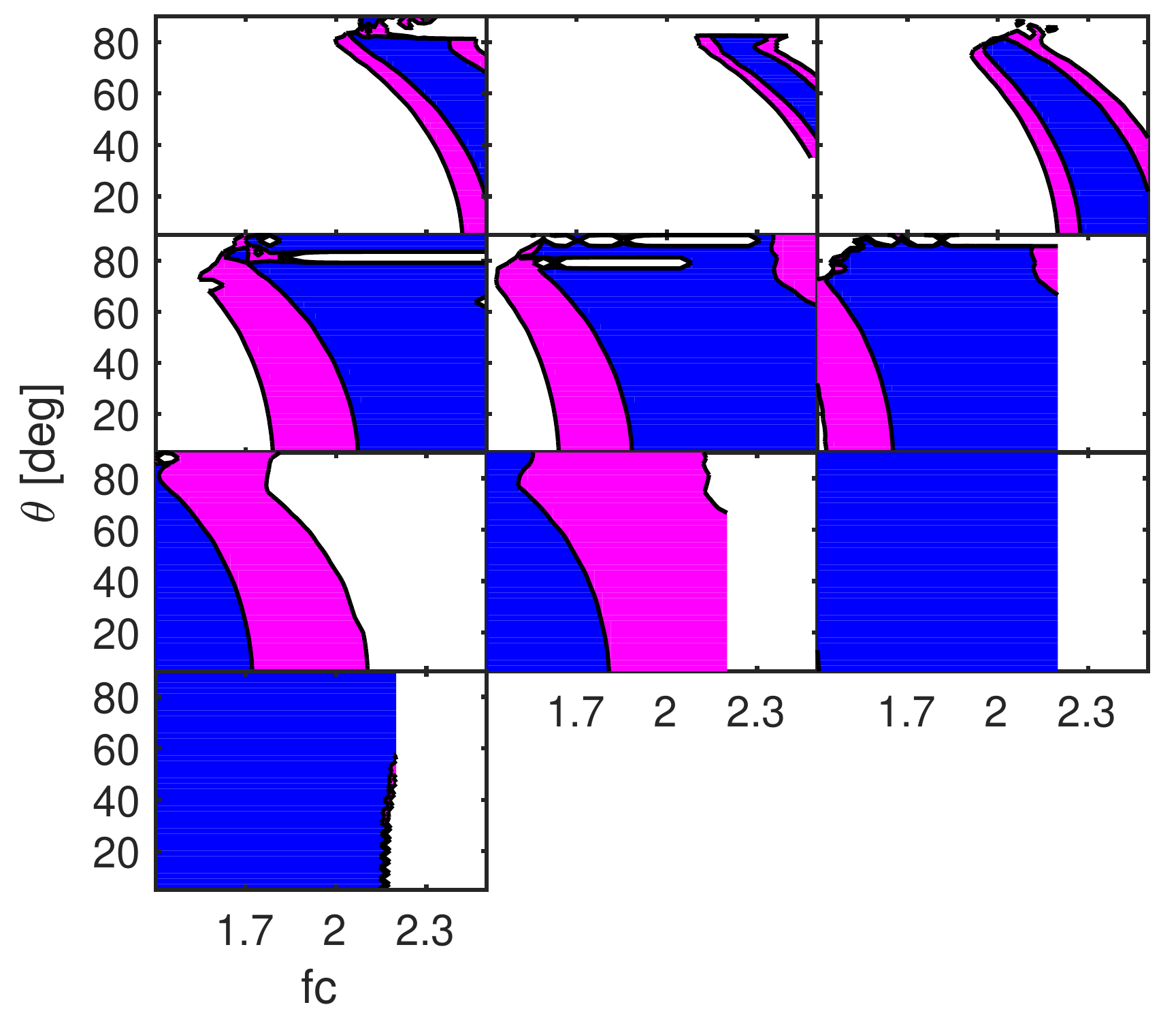}
\caption{Degeneracy of $\theta$ and $f_{\rm c}$ at each of the 10 observed epochs for ASASSN-14li. 
The blue and pink regions correspond to the 1$\sigma$ and 2$\sigma$ CL contours, respectively. Here $M_\bullet$ and $a_\bullet$ are fixed at the best-fit value from the 
DE19 treatment of $f_{\rm c}$: $10^7M_\odot$ and 0.998, respectively. We allow $\theta$, $N_{\rm{H}}$, and $f_{\rm c}$ to float from epoch to epoch. The priors for $f_{\rm c}$ are the ranges (1.4, 2.5) for the first five epochs and (1.4, 2.2) for the last five epochs. We find that $\theta$ is strongly degenerate with $f_{\rm c}$, especially for the later epochs, i.e., for sub-Eddington accretion. The lower limit of $f_{\rm c}$
decreases from $\approx $2.1 to 
$\approx $1.4 from early to late epochs.  This purely empirical result is consistent with theoretical
studies of spectral hardening, where $f_{\rm c}$ decreases from $\approx$2.0 in super-Eddington disks to $\approx$1.4 in the sub-Eddington regime \citep{MFR2000,GD04,DBHT05,DE2018}.}
\label{fci}
\end{figure}
In Figure \ref{fci}, we explore the degeneracy of $\theta$ and $f_{\rm c}$. We allow $f_{\rm c,i}$, $\theta_i$, $\dot m_i$, and $N_{\rm H,i}$ to float for each epoch. The cut-off at high inclination for early epochs arises because larger inclinations shield the observer from X-rays from the inner disk (due to the finite disk height). 
We find that $f_{\rm c}$ is strongly degenerate with $\theta$, especially for the later (sub-Eddington) epochs. Thus, fixing $f_{\rm c}$ in the fitting would not much impact the fit's $\chi^2$ value for each ($M_\bullet, a_\bullet$) pair. Furthermore, the $\theta$ $1\sigma$ CL regions overlap across epochs in Figure \ref{fci}, indicating that the constant $\theta$ prior used throughout the paper
is statistically valid
for this event. 

Figure \ref{fci} also shows that the first three epochs constrain $\theta$ and $f_{\rm c}$ better than the later epochs do. That is because 1) there are more soft and hard photon counts, which helps to break the degeneracy of $N_{\rm H}$ with $\theta$; 2) the effective temperature is insensitive to accretion rate when the accretion is super-Eddington, which works to remove the degeneracy between $f_{\rm c}$ and $\dot m$. 
Epochs 7 and 8 have unusual contours in the $\theta-f_{\rm c}$ plane, due, in part, to a strong decrease in flux near the highest spectral energies ($\sim$0.6 keV). These two epochs would produce an enormous tension in our constant-$\theta$ fitting if $f_{\rm c}$ were fixed to a value $\gtrsim 1.7$.  

While our fitting procedure does not require $f_{\rm c}>1.7$, it does couple $f_{\rm c}$ to the accretion rate, perhaps explaining why we get tighter constraints on ($M_\bullet, a_\bullet$) at a $1\sigma$ CL when we allow $f_{\rm c}$ to float. Letting $f_{\rm c}$ float to smaller values than those provided by the DE19 model releases the tension produced by applying the constant-$\theta$ assumption to Epochs 1-3 {\it and} 7-8.  Alternatively, this tension could arise because the disk is engaged in some level of reorientation over time (either global precession or realignment), which could be addressed later with self-consistent models for relativistic tilted disks.

Figure \ref{fci} demonstrates clear time evolution in the lower limit on a freely-floating $f_{\rm c}$,
which decreases from $\sim$2.1 to $\sim$1.4 from early to late epochs.  This purely empirical constraint on the spectral hardening factor is remarkably consistent with most {\it theoretical} studies, in which
$f_{\rm c}$ is predicted to decrease from $\sim$2.0 (super-Eddington regime) to $\sim$1.4 (sub-Eddington regime) as $\dot{m}$ declines \citep{MFR2000,GD04,DBHT05,DE2018}. 
This result suggests that the DE19 model for $f_{\rm c}$ is consistent with observations of TDE disks.

The analysis of $f_{\rm c}$ in Figure \ref{fci} is based on a specific ($M_\bullet, a_\bullet$) pair. We have shown in Figure \ref{14lifc} that different choices of $f_{\rm c}$ have little impact on the estimated ($M_\bullet, a_\bullet$),
suggesting that 
the conclusions drawn from the $f_{\rm c}-\theta$ plane may apply to other 
($M_\bullet, a_\bullet$) pairs. 
We defer a more thorough study of this degeneracy to future work.

In summary, our DE19 treatment of $f_{\rm c}$ 
does not bias our results significantly nor run afoul of the theoretical upper limit on $f_{\rm c}^\star$.

\section{Stationary slim disk model}
\label{app:slimdisk}

To model accretion disks around spinning BHs, we describe the Kerr metric with the Boyer-Lindquist coordinate system,
\begin{eqnarray}
\label{BL-metric}\nonumber
&&{\rm d}s^2=-\left(1-\frac{2Mr}{\rho}\right)~{\rm d}t^2-\frac{4Mar\sin^2\theta}{\rho}{\rm d}t{\rm d}\phi +\frac{\rho}{\Delta}{\rm d}r^2\\
&& +\rho {\rm d}\theta^2+\left(r^2+a^2+\frac{2Ma^2r\sin^2\theta}{\rho}\right)\sin^2\theta {\rm d}\phi^2\;. \label{eq:BL}
\end{eqnarray}
Here,
\begin{equation}
\Delta\equiv r^2-2Mr+a^2,~~~ \rho\equiv r^2+a^2\cos^2\theta\;.
\label{deltasigma}
\end{equation}
Note that in these formulae and for the remainder of this Appendix, $G=c=1$. 

For convenience, we now follow the procedure of \cite{Novikov1973}. The metric on the equatorial
plane is approximate up to the $(z/r)^0$ terms. Introducing the vertical coordinate $ z = r \cos \theta$,
the line element takes the form,
\begin{equation}
{\rm d}s^2=-\frac{r^2\Delta}{A}{\rm d}t^2-\frac{A}{r^2}({\rm d}\phi-\omega {\rm d}t)^2 +\frac{r^2}{\Delta}~{\rm d}r^2 +{\rm d}z^2, \label{eq:cylindricalMetric}
\end{equation}
with 
\begin{eqnarray}
\label{e.relcorr}
\omega=2Mar/A,~~~~ A=(r^{4}+r^2a^2+2Mra^2). 
\end{eqnarray}
Note that for the remainder of this Appendix, we use $r$ to denote the cylindrical radius (not the spherical radius of Boyer-Lindquist coordinates).  Eq. \eqref{eq:cylindricalMetric} describes the Kerr metric tensor $g_{\mu \nu}$ that shapes the geodesic motion of particles with four-velocities $u$.

The total (local) pressure for the slim disk can be calculated by
\begin{eqnarray}
\label{Pt}
&&p=p_{\rm gas}+p_{\rm rad},\\
&&p_{\rm gas}=k\rho T/(\mu m_{\rm p}), ~~ p_{\rm rad}= a_cT^4/3,
\end{eqnarray}
where $k$ is the Boltzmann constant, $m_{\rm p}$ is the proton mass,
and $a_c$ is the radiation constant. The mean molecular weight is
taken to be $\mu=0.62$. Here we define $\beta_{\rm p}\equiv p_{\rm gas}/p$.

The vertically averaged density and pressure is defined as
\begin{eqnarray}
 \Sigma=\int_{-h}^{+h}\rho\,{\rm d}z=2H\rho,   ~~
 P=\int_{-h}^{+h}p\,{\rm d}z=2Hp,
\label{mu}
\end{eqnarray}
where $H$ is the half of disk thickness. With these basic definitions, we now present the set of disk equations we solve.

(i) The equation of state:
\begin{equation}
\label{p}
P=\frac{k}{\mu m_{\rm p}}\Sigma T_c+\frac23Ha_cT_c^4.
\end{equation}

(ii) Vertical hydrostatic equilibrium  \citep{Abramowicz1997}:
\begin{equation}
 \frac{P}{\Sigma H^2}=\frac{{\cal L}^2-a^2(\epsilon^2-1)}{2 r^4}\equiv{\cal G}
\label{eq_vert}
\end{equation}
with $\epsilon=u_t$ being the conserved energy for test particle motion. 

(iii) Mass conservation:
\begin{equation}
 \dot M=-2\pi \Sigma\Delta^{1/2}\frac{V}{\sqrt{1-V^2}},
\label{sigma}
\end{equation}
where $V$ is measured by the observer co-rotating with the
fluid at a fixed value of $r$, and is given by,
\begin{equation}
\label{vasmeasured}
V/\sqrt{1-V^2}=u^r g_{rr}^{1/2}=\frac{r u^r}{\Delta^{1/2}}.
\end{equation}

(iv) Angular momentum conservation:
\begin{equation}
 \frac{\dot{M}}{2\pi}({\cal L}-{\cal L}_{\rm in})=\frac{A^{1/2}\Delta^{1/2}\gamma}{r}\alpha P,
\label{L}
\end{equation}
where ${\cal L}=u_\phi$ is the $z$-component of angular momentum, ${\cal L}_{\rm in}$ is this angular momentum component at the disk
inner edge, and
\begin{equation}
\gamma=\left(\frac1{1-V^2}+\frac{{\cal L}^2r^2}{A}\right)^{1/2},
\end{equation} 
is the Lorentz factor.

(v) Radial momentum conservation:
\begin{equation}
\frac{V^2}{1-V^2}\frac{{\rm d}\ln V}{{\rm d}\ln r}={\cal A}-\frac{P}{\Sigma}\frac{{\rm d}\ln P}{{\rm d}\ln r},
\label{dvdr}
\end{equation}
where
\begin{equation}
{\cal A}=-\frac{MA}{r^3\Delta\Omega_k^+\Omega_k^-}\frac{
(\Omega-\Omega_k^+)(\Omega-\Omega_k^-)}{1-\tilde\Omega^2\tilde R^2}
\label{eq_rad4}
\end{equation}
and $\Omega=u^\phi /u^t$, $\tilde\Omega=\Omega-\omega$, $\omega=2Mar/A$, $\Omega_k^\pm=\pm M^{1/2}/(r^{3/2}\pm aM^{1/2})$, $\tilde R=A/(r^2\Delta^{1/2})$.

(vi) Energy conservation:
\begin{equation}
\label{Qadv}
Q^{\rm adv}=Q^{\rm vis}-Q^{\rm rad}=-\alpha P\frac{A\gamma^2}{r^3}\frac{{\rm d}\Omega}{{\rm d}r }- \frac{64\sigma T_c^4}{3\Sigma\kappa},
\end{equation}
here, the $Q^{\rm rad}$ is computed by the diffusive approximation \citep{Sadowski2011}. We note that $Q^{\rm rad}$ is two times bigger than the original one in Eq. (6) of \cite{Sadowski2009}. 
The local opacity $\kappa$ is the sum of the electron scattering opacity $\kappa_{\rm es}$ and free-free opacity $\kappa_{\rm ff}$, which we estimate using Kramer's approximation:
\begin{equation}
\label{e.kramers}
\kappa=\kappa_{\rm es}+\kappa_{\rm ff}=0.34+3.2\times 10^{22} T^{-3.5}\Sigma/H.
\end{equation}
In this (areal) energy equation, we have set the difference between viscous heating $Q^{\rm vis}$ and radiative cooling $Q^{\rm rad}$ equal to the advected heat:
\begin{equation}
\label{dqadr}
Q^{\rm \rm adv}=-\frac{\dot M}{2\pi r^2}\frac{P}{\Sigma} \left(\frac{4-3\beta_{\rm p}}{\Gamma_3-1}\frac{{\rm d}\ln T_c}{{\rm d}\ln r}-(4-3\beta_{\rm p})\frac{{\rm d}\ln \Sigma}{{\rm d}\ln r}\right)
\end{equation}
where $\Gamma_3=(8-6\beta_{\rm p})/(24-21\beta_{\rm p})$.

Eq. \eqref{dvdr} and Eq. \eqref{dqadr} can be simplified to:
\begin{eqnarray} 
\label{2D1}
&&V_1\frac{{\rm d}\ln V}{{\rm d}\ln r}+\frac{P(8-6\beta_{\rm p})}{\Sigma(1+\beta_{\rm p})}\frac{{\rm d}\ln T_c}{{\rm d}\ln r} ={\cal A}+V_2,\\ 
\label{2D2}
&&\frac{\Gamma_3-1}{1-V^2}\frac{{\rm d}\ln V}{{\rm d}\ln r}+\frac{{\rm d}\ln T_{\rm c}}{{\rm d}\ln r}=V_3, 
\end{eqnarray}
where
\begin{eqnarray}
\nonumber
&&V_1=\frac{V^2}{1-V^2}-\frac{P(3\beta_{\rm p}-1)}{\Sigma(1+\beta_{\rm p})(1-V^2)}),\\ \nonumber
&&V_2=\frac{P(1-\beta_{\rm p})}{\Sigma(1+\beta_{\rm p})}\frac{{\rm d} \ln {\cal G}}{{\rm d}\ln r}+\frac{P(3\beta_{\rm p}-1)}{\Sigma(1+\beta_{\rm p})}B,\\\nonumber
&&V_3=(1-\Gamma_3)\left(B+\frac{2\pi r^2 Q^{\rm adv} \Sigma}{\dot M P(4-3\beta_{\rm p})}\right).
\end{eqnarray}
Here $B=r(r-M)/\Delta$. Note that $Q^{\rm adv}$ and $\cal G$ are functions of $V$, $T_{\rm c}$ and $r$. \cite{Sadowski2009} calculated $Q^{\rm adv}$ and ${\rm d} \ln{\cal G}/d \ln r$ numerically. Here, we use the following relationship \citep{Sadowski11} to solve this problem analytically,
\begin{equation}
\Omega=\omega+\frac{r^3\Delta^{0.5} \cal L}{\gamma A^{3/2}}.
\end{equation}
 
Thus, Eq. \eqref{2D1} and Eq. \eqref{2D2} can be simplified to:
\begin{eqnarray}
\label{2DD1}
&&\frac{ \rm d\ln V}{{\rm d}\ln r}=\frac{N}{D}=\frac{c_1b_2-c_2b_1}{a_1b_2-a_2b_1}\\
\label{2DD2}
&&\frac{{\rm d}\ln T_{\rm c}}{{\rm d}\ln r}=\frac{c_2}{b_2}-\frac{a_2N}{b_2D},
\end{eqnarray}
where
\begin{eqnarray} 
\nonumber
&&a_1=\left( \frac{V^2\Sigma}{P}+\frac{1-3\beta_{\rm p}}{1+\beta_{\rm p}}\right)\frac{1}{1-V^2}-\frac{1-\beta_{\rm p}}{1+\beta_{\rm p}}\frac{\partial \ln{\cal G}}{\partial \ln V}\\ \nonumber
&&b_1=\frac{14}{1+\beta_{\rm p}}-\frac{1-\beta_{\rm p}}{1+\beta_{\rm p}}\frac{\partial \ln{\cal G}}{\partial \ln T_{\rm c}}-6,\\ \nonumber
&&c_1=\frac{ \Sigma\cal A}{P}+\frac{1-\beta_{\rm p}}{1+\beta_{\rm p}}\frac{\partial \ln {\cal G}}{\partial \ln r}+(3-\frac{4}{1+\beta_{\rm p}})B,\\ \nonumber
&&a_2=\frac{\Gamma_3}{1-V^2}-\frac{2\pi r^2\Gamma_3\Sigma}{(4-3\beta_{\rm p})\dot M P}\frac{\alpha PA\gamma^2}{r^3}\frac{\partial \Omega}{\partial \ln V},\\\nonumber
&&b_2=1-\frac{2\pi r^2\Gamma_3\Sigma}{(4-3\beta_{\rm p})\dot M P}\frac{\alpha PA\gamma^2}{r^3}\frac{\partial \Omega}{\partial T_c},\\\nonumber
&&c_2=\frac{2\pi r^2\Gamma_3\Sigma}{(4-3\beta_{\rm p})\dot M P}\left(Q^{rad}+\frac{\alpha PA\gamma^2}{r^3}\frac{\partial \Omega}{\partial r}\right)-\Gamma_3B,\\ \nonumber
\end{eqnarray}

There are only four free parameters for the two-dimensional differential equations in Eq. \eqref{2DD1} and Eq. \eqref{2DD2}: $M$, $a$, $\dot M$ and $\alpha$. ${\cal L}_{in}$ is the eigenvalue of the problem. It must be chosen properly to ensure that $N=0$ and $D=0$ at the sonic point. Given $M$, $a$, $\dot M$ and $\alpha$, we estimate the initial condition by assuming the Novikov-Thorne disk ($\Omega=\Omega_k^+$ and $Q^{\rm adv}=0$). We use the Runge-Kutta method of the fourth order to integrate the two-dimensional differential equations, and we use the shooting technique to search for the sonic point. If ${\cal L}_{in}$ is greater than the true value, then $D$ decreases to 0 before it reaches the real sonic point, and if ${\cal L}_{in}$ is smaller than the true value, then $D$ will decrease to 0 only near the horizon. The ${\cal L}_{in}$ estimation is updated iteratively, until $\Delta {\cal L}_{in}/{\cal L}_{in}$ is less than $10^{-6}$. We iteratively integrate the equations a radius near the sonic point with the latest ${\cal L}_{in}$ estimate, then take a large step ahead and continue to solve the equations to near-horizon distances.  We find that our solutions are insensitive to the initial conditions.

\section{Ray tracing algorithm}
\label{app:rays}
 
In this section, we review the ray tracing code of \cite{JP11}.
For the Kerr BH, we use the Boyer-Lindquist coordinates, as in Eq. \ref{BL-metric}.
The four null geodesic equations can be written as
\begin{equation}
\frac{{\rm d}^2x^i}{{\rm d}\lambda^2}+\Gamma^i_{kl}\frac{{\rm d}x^k}{{\rm d}\lambda}\frac{{\rm d}x^l}{d\lambda}=0, \label{eq:geodesics}
\end{equation}
where $\Gamma^i_{kl}$ are the various Christoffel
symbols for the metric, and $\lambda$ is an affine parameter.

Using the conservation of energy and angular momentum, the second-order
differential equations for the time and the azimuth of each photon trajectory can be rewritten as,
\begin{eqnarray}
E&=&-g_{tt}\frac{{\rm d}t}{{\rm d}\lambda}-g_{t\phi}\frac{{\rm d}\phi}{{\rm d}\lambda}, \\
L&=&g_{\phi\phi}\frac{{\rm d}\phi}{{\rm d}\lambda}+g_{t\phi}\frac{{\rm d}t}{{\rm d}\lambda}
\label{eq:angmomentum}
\end{eqnarray}
The differential equations for $r$ and $\theta$ are kept in second-order form, as in Eq. \eqref{eq:geodesics}.  Therefore, there are two ($r$ and $\theta$) second-order  geodesic 
equations, and two ($t$ and $\phi$) first-order geodesic equations for each photon.
Note that the norm of the
photon 4-momentum has to vanish, i.e.,
\begin{equation}
g_{kl}\frac{{\rm d}x^k}{{\rm d}\lambda}\frac{{\rm d}x^l}{{\rm d}\lambda}=0\;.
\label{eq:4mom}
\end{equation}
In the code, the four geodesic equations are integrated by the fourth-order Runge-Kutta method, and Eq. \eqref{eq:4mom} is used to monitor the accuracy of the calculation.

A photon from the image plane of an observer at infinity, can be traced back to surface of the accretion disk. Considering a photon with Cartesian coordinate $(x_o, y_o)$ on the image plane at distance $d$ with inclination $\theta_o$, the coordinate $(r_i, \theta_i, \phi_i)$ in the spherical-polar system used for the Boyer-Lindquist metric can be written as
\begin{eqnarray}
r_i&=&(d^2+x_o^2+y_o^2)^{1/2}, \\
\cos\theta_i&=&\frac{1}{r_i}(d\cos\theta_o+y_o\sin\theta_o),\\
\tan\phi_i&=&x_o(d\sin\theta_o-y_o\cos\theta_o)^{-1}.
\label{initiate}
\end{eqnarray}
Three components of the initial four-momentum vector of the photon are set by:
\begin{eqnarray}
&&k^r_i\equiv\frac{{\rm d}r}{{\rm d}\lambda}=\frac{d}{r_i}, \\
&&k^\theta_i\equiv\frac{{\rm d}\theta}{{\rm d}\lambda}=\frac{\frac{d}{r_i^2}(d\cos\theta_o+y_o\sin\theta_o)-\cos\theta_o}{\left(r_i^2-(d\cos\theta_o+y_o\sin\theta_o)^2\right)^{1/2}},\\
&&k^\phi_i\equiv\frac{{\rm d}\phi}{{\rm d}\lambda}=\frac{-x_o\sin\theta_o}{(d\sin\theta_o-y_o\cos\theta_o)^2+x_o^2}.
\label{momentum}
\end{eqnarray}
Note that the photon momentum is perpendicular to the image plane, and the $t$ component can be calculated by Eq. \eqref{eq:4mom}.
The redshift factor $\Upsilon$ of the photon can be calculated by
\begin{equation}
\Upsilon \equiv\frac{E_{obs}}{E_d}=\frac{g_{\mu\nu,obs}k_i^\mu u_i^\nu}{g_{\mu\nu,d}k_d^\mu u_d^\nu}.
\label{eq:redfisht}
\end{equation}
Here the three-velocity of the observer at the image plane is set to zero.

\end{appendix}

\end{document}